\documentclass[3p,times]{elsarticle}

\usepackage{adjustbox}
\usepackage[T1]{fontenc}    
\usepackage{hyperref}       
\usepackage{url}            
\usepackage{booktabs}       
\usepackage{amsfonts}       
\usepackage{nicefrac}       
\usepackage{microtype}      
\usepackage{xcolor}         
\usepackage{algorithm}
\usepackage{algpseudocode}
\usepackage[para,online,flushleft]
{threeparttable}
\usepackage{pifont}

\usepackage{multirow}
\usepackage{graphicx}
\usepackage{amsmath}
\usepackage{amsthm}
\usepackage[textfont=it,labelfont=bf]{caption} 
\usepackage{enumitem}
\usepackage{array}
\usepackage{wrapfig}
\usepackage{colortbl}
\usepackage{tabularray}
\usepackage{graphbox}
\usepackage{times}
\usepackage{helvet}
\usepackage[most]{tcolorbox}
\usepackage{algorithm}
\usepackage{algpseudocode}
\usepackage{subcaption}
\usepackage{xcolor}
\usepackage{mdframed}
\usepackage{pgfplots}
\pgfplotsset{compat=1.17}
\usepackage{xcolor}
\usepackage{shadethm}
\usepackage{thmtools}

\hypersetup{
    breaklinks=true,
    colorlinks,
    linkcolor={blue!80!black},
    citecolor={blue!80!black},
    urlcolor={blue!80!black},
    plainpages=true
}
\newcommand{\cref}[2]{\hyperref[#2]{#1~\ref*{#2}}}
\newcommand{\colref}[3]{\hyperref[#2]{#1~\ref*{#2}{#3}}}
\newcommand{\figref}[1]{\cref{Figure}{#1}}

\newcommand{\secref}[1]{\cref{Section}{#1}}
\newcommand{\eqnref}[1]{\cref{Equation}{#1}}
\newcommand{\tabref}[1]{\cref{Table}{#1}}
\newcommand{\algoref}[1]{\cref{Algorithm}{#1}}

\newcommand{\appendixref}[1]{\cref{}{#1}}

\declaretheoremstyle[%
  spaceabove=-6pt,%
  spacebelow=6pt,%
  headfont=\bfseries\itshape,%
  postheadspace=0.5em,%
  qed=\qedsymbol%
]{mystyle}

\theoremstyle{mystyle}

\definecolor{trueintercepted}{HTML}{FFFF00}
\definecolor{falseintercepted}{HTML}{F08080}
\definecolor{trueboundary}{HTML}{00007f}
\definecolor{notintercepted}{HTML}{008000}

\newcommand{\createVelocityPlot}[1]{
    \includegraphics[width=0.97\linewidth]{#1_diagonal_velocity_plot.pdf}
}

\journal{Computer Methods in Applied Mechanics and Engineering}
\begin{document}

\begin{frontmatter}
\title{Direct Flow Simulations with Implicit Neural Representation of Complex Geometry}

\author[ISU]{Samundra Karki}
\author[ISU]{Mehdi Shadkhah}
\author[ISU]{Cheng-Hau Yang}
\author[ISU]{Aditya Balu}
\author[DU]{Guglielmo Scovazzi}
\author[ISU]{\\Adarsh Krishnamurthy\texorpdfstring{\corref{cor1}}{}}
\author[ISU]{Baskar Ganapathysubramanian\texorpdfstring{\corref{cor1}}{}}
\affiliation[ISU]{organization={Iowa State University}, 
            city={Ames},
            state={Iowa},
            country={USA}}
\affiliation[DU]{organization={Duke University}, 
            city={Raleigh},
            state={North Carolina},
            country={USA}}
\cortext[cor1]{Corresponding Authors}


\begin{abstract}

Implicit neural representations (e.g., neural network-based signed distance fields) have emerged as a powerful approach for encoding complex geometries as continuous functions. These implicit models are widely used in computer vision and 3D content creation, but their integration into scientific computing workflows, such as finite element or finite volume simulations, remains limited. One reason is that conventional simulation pipelines require explicit geometric inputs (meshes), forcing INR-based shapes to be converted to meshes---a step that introduces approximation errors, computational overhead, and significant manual effort. Immersed boundary methods partially alleviate this issue by allowing simulations on background grids without body-fitted meshes. However, they still require an explicit boundary description and can suffer from numerical artifacts, such as sliver cut cells. The shifted boundary method (SBM) eliminates the need for explicit geometry by using grid-aligned surrogate boundaries, making it inherently compatible with implicit shape representations. Here, we present a framework that directly couples neural implicit geometries with SBM to perform high-fidelity fluid flow simulations \textit{without any intermediate mesh generation}. By leveraging neural network inference, our approach computes the surrogate boundary and distance vectors required by SBM on-the-fly directly from the INR, thus completely bypassing traditional geometry processing. We demonstrate this approach on canonical 2D and 3D flow benchmarks (lid-driven cavity flows) and complex geometries (gyroids, the Stanford bunny, and AI-generated shapes), achieving simulation accuracy comparable to conventional mesh-based methods. This work highlights a novel pathway for integrating AI-driven geometric representations into computational physics, establishing INRs as a versatile and scalable tool for simulations and removing a long-standing bottleneck in geometry handling.

\end{abstract}

\end{frontmatter}

\begin{figure}[ht]
    \centering
    \includegraphics[width=0.9\linewidth,clip,trim={0.0in, 2.0in, 0.0in, 2.0in}]{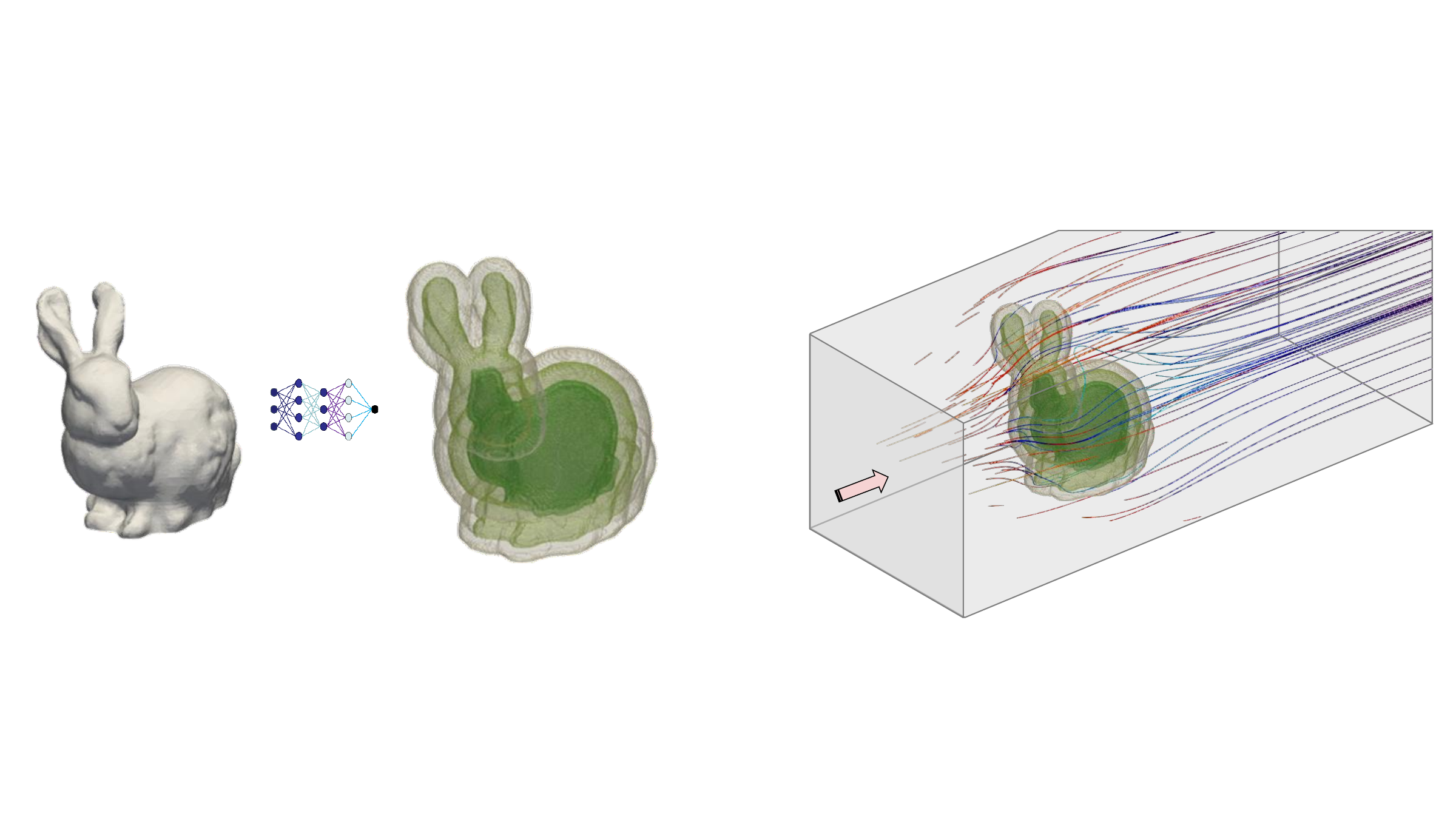}
    \caption{Direct flow simulations with Implicit Neural Representations of complex geometries.}
    \label{fig:teaser_image}
\end{figure}

\section{Introduction}
\label{Sec:Introduction}

Accurately representing complex 3D geometry is fundamental in computational science and engineering, underpinning applications ranging from fluid flow modeling to structural analysis and fluid-structure interactions \citep{nguyen2015isogeometric,liu2022eighty,peskin1972flow}. Traditionally, the choice of a geometry representation is dictated by the simulation at hand, leading to a series of manual translation and conversion steps that are often error-prone and inefficient. In most workflows, explicit geometric models---such as boundary representations (B-reps) and body-fitted polygonal meshes---serve as the foundation for simulations (for instance, in finite element analysis)\citep{mchenry2008overview}. However, generating a high-quality mesh for an arbitrarily complex object is a labor-intensive and computationally expensive process, and it can introduce discretization errors if not done carefully~\citep{CHIBA1998145}.

Implicit neural representations (INRs) have recently gained traction as a transformative paradigm for 3D geometry description. An INR uses a neural network to represent a shape as a continuous field, often as a signed distance function that outputs the distance of any point in space to the surface of the object~\citep{park2019deepsdf, gropp2020implicit, mescheder2019occupancy, chen2019learning}. This yields a compact, smooth, and differentiable representation of geometry that can be learned from data sources such as images, point clouds, or CAD models. INRs have shown great promise in computer vision and graphics, for example, in reconstructing shapes from sparse observations, establishing dense 3D correspondences, or generating novel 3D content~\citep{wang2021neus, ben2022digs, atzmon2020sal, sitzmann2020implicit}. Despite this success, the use of INRs in high-fidelity scientific simulations (e.g., finite element or finite volume analyses) remains largely unexplored.

\begin{figure}[b!]
    \centering
    \begin{subfigure}[b]{0.9\linewidth}
        \centering
        \includegraphics[width=\linewidth, trim=0in 2.8in 0.0in 3.8in, clip]{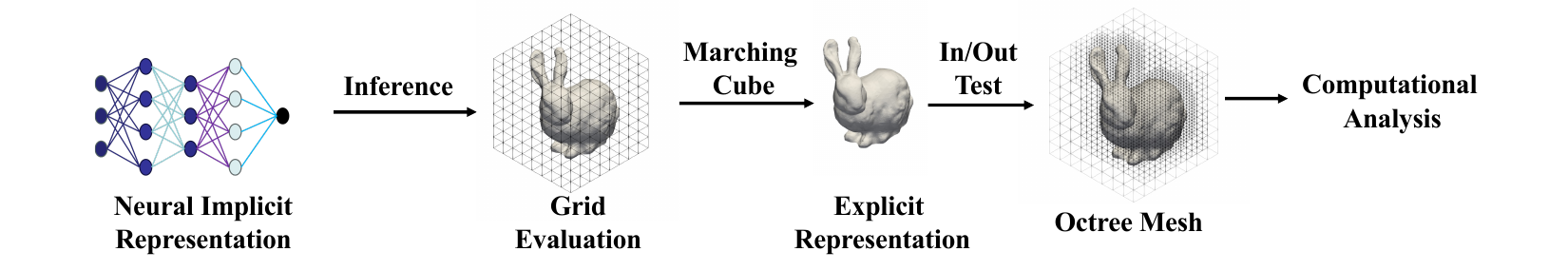}
        \caption{Traditional Approach.}
        \label{fig:traditional_implicit}
    \end{subfigure}
    \begin{subfigure}[b]{0.9\linewidth}
        \centering
        \includegraphics[width=\linewidth, trim=0in 3.2in 0.0in 3.5in, clip]{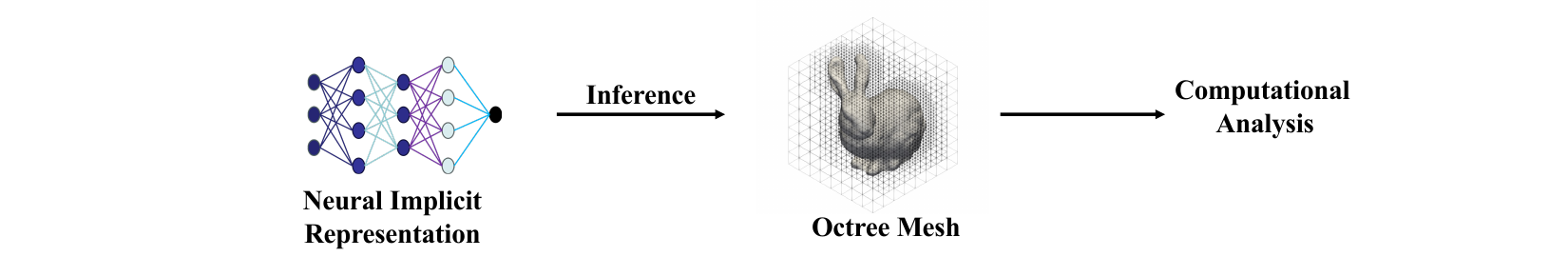}
        \caption{Direct Simulation with INR. }
        \label{fig:our_approach}
    \end{subfigure}
    \caption{In traditional simulations using a boundary-fitted or immersed approach, implicit geometry is converted to explicit representation. In our approach, the Implicit Neural Representations is directly used to obtain the distance vector and the surrogate boundary for simulations using the shifted boundary method (SBM).}
    \label{fig:implicit_vs_our_approach}
\end{figure}

In principle, integrating INRs directly into computational physics solvers could eliminate many of the geometry-processing hurdles. However, the conventional approach to using an INR in a simulation is to first convert it into an explicit form, see \figref{fig:implicit_vs_our_approach}. For instance, one might sample the neural implicit function to produce a polygonal surface (via algorithms like marching cubes) and then generate a mesh suitable for finite element analysis. This conversion step inevitably introduces approximation errors and negates many advantages of the implicit representation (such as resolution independence and easy shape manipulation and editing). Immersed boundary methods (IBMs) ~\citep{peskin1972flow} offer an alternative by allowing simulations on a fixed background grid that does not conform to the object's shape, thereby avoiding the need for a body-fitted mesh. However, standard IBMs still rely on an explicit representation to mark the presence of the object within the grid and often encounter numerical difficulties~\citep{hsu2016direct,burman2015cutfem,burman2012fictitious,saurabh2021industrial}. For example, IBM approaches can produce sliver elements that lead to poorly conditioned matrices and pose challenges for solver stability and parallelization~\citep{burman2015cutfem,saurabh2021industrial,de2019preconditioning}. Additionally, even immersed approaches reintroduce some form of explicit boundary handling, which can become a bottleneck as geometric complexity grows.

The shifted boundary method (SBM) offers a compelling solution to these issues~\citep{main2018shifted1,main2018shifted2}. SBM completely avoids meshing the true boundary by introducing a “surrogate” boundary that conforms to the background grid. The boundary conditions of the problem are imposed on this surrogate boundary, with corrections (via a Taylor-series expansion) accounting for the offset from the true geometry. This eliminates pathological small cut-cells and yields well-behaved linear systems in the simulation. SBM has been demonstrated on a variety of physical simulation ~\citep{colomes2021weighted,chou2023diffusion,huang2020simulation,karatzas2020reduced,atallah2021shifted} problems using octree and Cartesian grids, consistently showing robust performance without the overhead of generating boundary-fitted meshes~\citep{yang2024optimal,yang2024simulating}. Importantly, because SBM only requires distance information from the true boundary to the grid (to construct the surrogate boundary and apply corrections), it is inherently well-suited to implicit geometry definitions like INRs.

Leveraging this compatibility, we propose a framework that directly integrates INRs with the shifted boundary method to perform flow simulations. As summarized in \tabref{tab:geometric_dual}, conventional boundary-fitted and immersed boundary methods rely on explicit geometric representations, whereas the shifted boundary method operates on distance vectors, making it inherently compatible with implicit representations like INRs~\citep{main2018shifted1,main2018shifted2}. In our approach, the neural implicit representation of the geometry is queried during the simulation to provide all necessary geometric information. Specifically, the INR is used to compute distance fields and surrogate boundary locations on the fly as the solver requires them, completely bypassing any explicit surface or mesh extraction (see \algoref{Algorithm ImplicitOctreeGeneration}, \algoref{Algorithm: SurrogateBoundaryIdentificationUsingImplicitNetwork}, \algoref{Algorithm: DistanceFunctionCalculationUsingImplicitNetworkOneGP} for details on the algorithm). This allows us to combine the strengths of modern AI-based shape representations (compactness, flexibility, and data-driven acquisition) with the rigor and accuracy of established finite element/volume solvers in a single seamless workflow.

We validate the efficacy of this framework on both two-dimensional and three-dimensional flow problems. First, we consider the canonical lid-driven cavity flow with an internal circular obstacle represented by an INR and compare the results to a traditional simulation using a body-fitted mesh for the obstacle. The INR-driven simulation closely matches the mesh-based solution across a range of flow conditions (Reynolds numbers), demonstrating that accuracy is preserved. We then extend to a 3D lid-driven cavity scenario with more complex internal geometry and again observe excellent agreement between the implicit and explicit geometry approaches. Beyond these benchmarks, we apply our method to simulate flows around more intricate shapes---such as a porous gyroid, the Stanford bunny, and an airplane model generated by a generative AI algorithm---to showcase its versatility. In all cases, the proposed approach remains stable and accurate, with no manual intervention, indicating its robustness for complex geometrical configurations.

\begin{table}[t!]
    \centering
    \small
    \setlength{\extrarowheight}{2pt}
    \caption{Comparison of different analysis methods and their suitable geometric representations.}
    \label{tab:geometric_dual}
    \begin{tabular}{l|l|l}
        \textbf{Analysis Method} & \textbf{Geometry Information} & \textbf{Suitable Geometric Representation}\\ \hline
        Boundary fitted methods & Boundary representation & Explicit geometry \\
        Immersed boundary methods & Boundary points & Explicit geometry \\
        Shifted boundary method & Distance vector & Implicit geometry \\
    \end{tabular}
\end{table}

This work is a step toward uniting AI-driven geometric representations with high-fidelity physics simulation. It effectively bridges the gap between modern implicit shape encoding techniques and conventional computational analysis, demonstrating that we can achieve state-of-the-art simulation accuracy directly on neural representations of geometry. By removing the long-standing bottleneck of mesh generation, our framework has the potential to simplify and accelerate modeling workflows across a broad range of computational science and engineering problems. Our key contributions include:
\begin{enumerate}[itemsep=0pt,topsep=0pt]
    \item \textbf{Direct INR-based simulation}: Development of a simulation framework that uses a neural implicit representation of geometry directly within the shifted boundary method, eliminating the need for any explicit geometric mesh. A detailed numerical discussion of the framework is provided in the Methods section.
    \item \textbf{Benchmark validation}: Verification of the proposed framework against standard lid-driven cavity flow benchmarks in 2D and 3D, demonstrating that the INR-based approach matches traditional mesh-based results across varying flow conditions.
    \item \textbf{Illustration on complex shapes}: Demonstration of the method’s applicability to complex geometries (e.g., gyroids, the Stanford bunny, and AI-generated aircraft models), illustrating its versatility and robustness for diverse simulation scenarios.
\end{enumerate}

\section{Methods}
\label{sec:Methods}


We first provide the necessary background on implicit representations and the shifted boundary method in \secref{sec:Implicit} and \secref{sec:SBM}, respectively. Subsequently, we discuss various methods for generating Implicit Neural Representations (INRs) in \secref{sec:INR}. Finally, in  \secref{sec:Implementation}, we present the algorithms designed to conduct Shifted Boundary Method (SBM) analyses using INRs.

For clarity, we define a cubic domain $\boldsymbol{\Omega}$, which is composed of two regions: $\boldsymbol{\Omega^-}$, representing the volume enclosed by the surface $\boldsymbol{\Gamma}$, and $\boldsymbol{\Omega^+}$, representing the volume external to the surface $\boldsymbol{\Gamma}$. Thus, $\boldsymbol{\Omega}$ can be expressed as $\boldsymbol{\Omega = \Omega^- \cup \Omega^+}$, with $\boldsymbol{\Omega^- \equiv \Omega \setminus \Omega^+}$. If $\boldsymbol{\Omega^+}$ is a closed set, it includes the boundary, i.e., $\boldsymbol{\Gamma \subset \Omega^+}$. For ease of visualization and explanation, we reduce the cubic domain to a square domain and explain the necessary concepts wherever necessary.

\subsection{Implicit Representation}
\label{sec:Implicit}

Implicit representations encode surfaces implicitly using mathematical functions rather than explicit surface definitions. A commonly used implicit representation is the Signed Distance Field (SDF), a scalar field that assigns the shortest distance from any point in space to the surface of a given geometry. Mathematically, the signed distance field (SDF) for a surface $\mathbf{\Gamma}$ is defined by \eqnref{sdf definition}.
\begin{equation}
f(\mathbf{x}) = 
\begin{cases} 
\phantom{-}\min\limits_{\mathbf{y} \in \Gamma} \|\mathbf{x} - \mathbf{y}\| & \text{if } \mathbf{x} \in \boldsymbol{\Omega^+} \\
-\min\limits_{\mathbf{y} \in \Gamma} \|\mathbf{x} - \mathbf{y}\| & \text{if } \mathbf{x}  \in \boldsymbol{\Omega^-}  \\
0 & \text{if } \mathbf{x} \in \boldsymbol{\Gamma}
\end{cases}
\label{sdf definition}
\end{equation}

\begin{figure}[b!]
    \centering
    \includegraphics[width=0.8\linewidth, trim=2cm 4cm 2cm 4cm, clip]{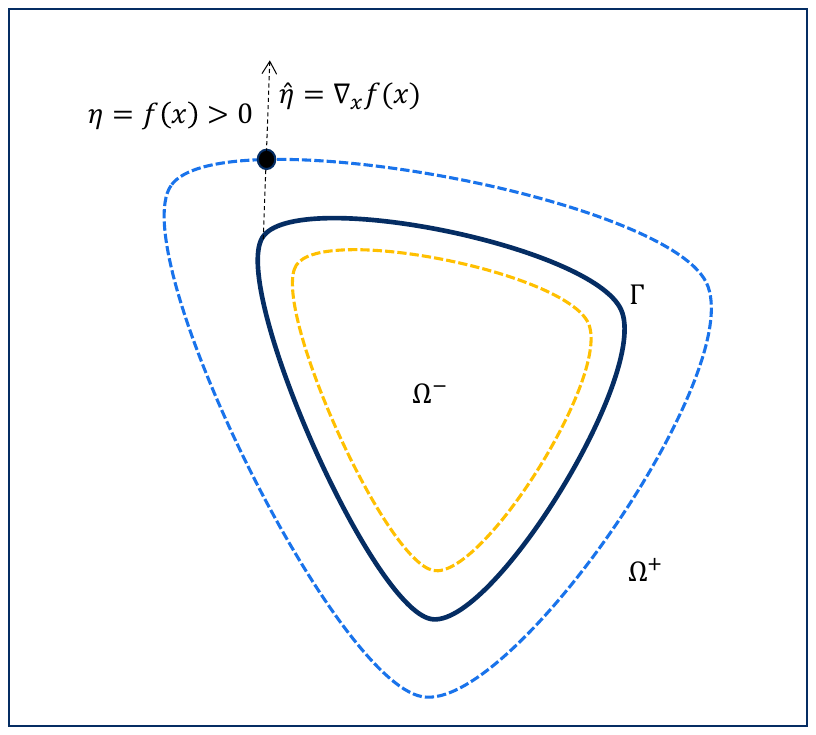}
    \caption{$\Gamma$ is the true boundary with a level set in $\Omega^+$ represented by blue, and a level set in $\Omega^-$ represented in yellow.$\eta$ is the signed distance value at a point in outside point with $\hat{\eta}$ represents the gradient of the signed distance field.}
    \label{fig:gradient-signed distance field}
\end{figure}

The gradient of the signed distance field at a given point, denoted $\hat{\eta} = \nabla_x f$, indicates the direction of increasing signed distance, as illustrated in \figref{fig:gradient-signed distance field}. For a given level-set curve, $\Gamma_{f(x) = c}$, defined as:

\begin{equation}
\Gamma_{f(x)=c} = { x \in \mathbb{R}^n \mid f(x) = c },
\label{equation:level_Set_curve}
\end{equation}

it follows that the gradient vector, $\hat{\eta}$ is always perpendicular to this curve and consequently has no tangential component. Moreover, $\hat{\eta}$ is a unit vector satisfying the Eikonal equation~\citep{crandall1983viscosity}, as given in \eqnref{equation:eikonal_fx}. $-\hat{\eta}$ points towards the closest point on the surface. 
\begin{equation}
    \label{equation:eikonal_fx}
    ||\nabla_x f(x)||=1,
\end{equation}

\begin{figure}[t!]
    \centering
    \begin{subfigure}{0.48\linewidth}
        \centering
        \raisebox{0.22\linewidth}{
        \includegraphics[width=0.15\linewidth]{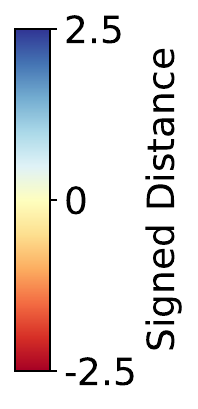}}
        \includegraphics[width=0.8\linewidth]{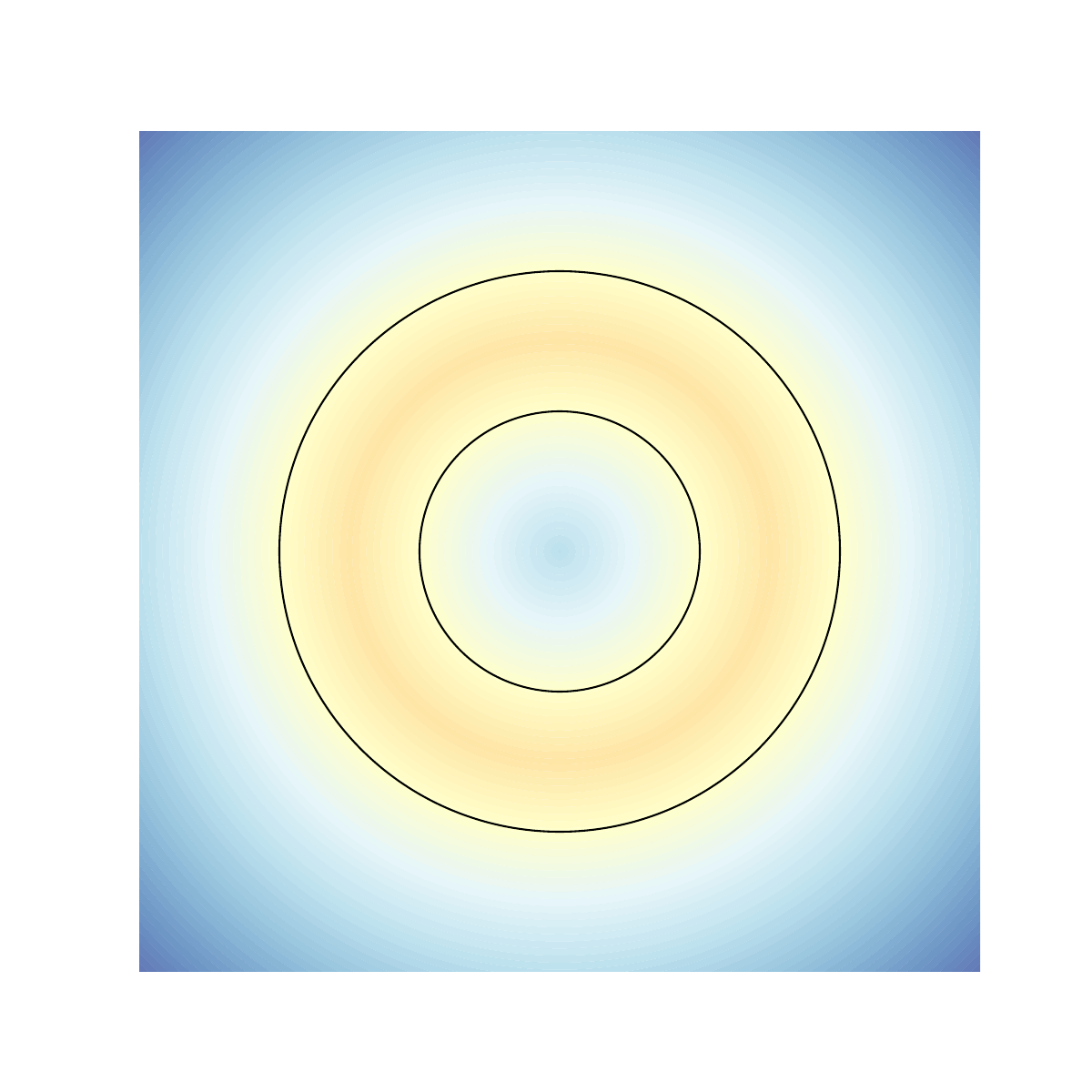}
        \caption{Signed distance field of a ring with inner radius $r_1=1$ and outer radius $r_2=2$.}
        \label{fig:sdf-ring}
    \end{subfigure}
    \hfill
    \begin{subfigure}{0.48\linewidth}
        \centering
        \includegraphics[width=\linewidth]{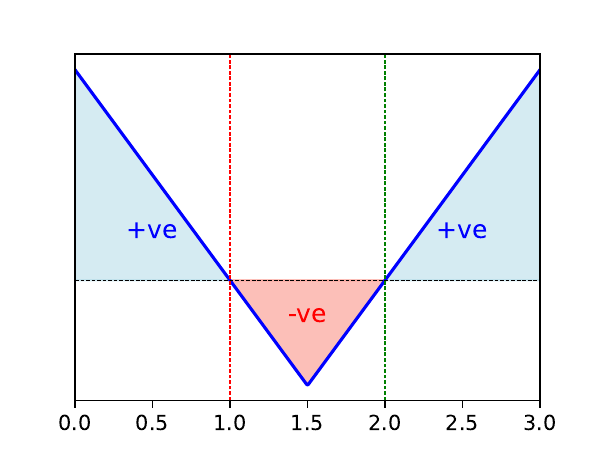}
        \caption{Signed distance values along the x-axis. The red and green line represents the inner and the outer ring, respectively.}
        \label{fig:SDF_RING_X-AXIS}
    \end{subfigure}
    \caption{Visualization of the signed distance field for a ring in the left plot. The right plot has corresponding values of signed distance values along the positive x-axis. }
\end{figure}

\paragraph{Example} \figref{fig:sdf-ring} shows the signed distance field of a ring with inner-ring radius ($r_1=1$) and outer-ring radius ($r_2=2$). The analytical implicit representation of the ring is given in \eqnref{equation:ring}. The contours of the signed distance values are also concentric, as shown in the figure. If we look at the plot of the signed distance for the ring along the positive x-axis in \figref{fig:SDF_RING_X-AXIS}, there appears a kink where the distance between the point from the inner and outer rings are equal. The gradient of the signed distance field blows up and is not deterministic at that point. As we are mostly interested in regions very close to the boundary, such points do not appear in our analysis.

\begin{equation}
f(\mathbf{x}) = 
\begin{cases} 
\phantom{-}(r_1 - \|\mathbf{x}\|) & \text{if } \|\mathbf{x}\| < r_1 \\
\|\mathbf{x}\| - r_2 & \text{if } \|\mathbf{x}\| > r_2 \\
-\min\left(\|\mathbf{x}\| - r_1, r_2 - \|\mathbf{x}\|\right) & \text{if } r_1 \leq \|\mathbf{x}\| \leq r_2 
\end{cases}
\label{equation:ring}
\end{equation}

\subsection{Octree Mesh}
\label{sec:octree_hn}
\figref{fig:SBM_Definition} illustrates a closed region $\mathcal{O}$, representing a complete quadtree, or equivalently, a Cartesian grid consisting of $T_h(\mathcal{O})$ cartesian decompositions of $\mathcal{O}$. Then, the domain of our interest, $\Omega$, is embedded inside $\mathcal{O}$  where $\text{clos}(\Omega) \subseteq \mathcal{O}$ (with $\text{clos}(\Omega)$ denoting the closure of $\Omega$), and with a (true) boundary, $\Gamma$. The union of $T_h$ forms the complete tree but we are only interested in $T \in T_h(\mathcal{O})$ that have a non-empty intersection with the domain of interest $\Omega$.

We define the family of grids as: 
\begin{equation} 
\tilde{T}_h := \left\{ T \in T_h(\mathcal{O}) : \text{meas}(T \cap \Omega) > 0 \right\}
\end{equation}



Now, we can define the surrogate domain: 
\begin{equation} 
\tilde{\Omega}_h := \text{int} \left( \bigcup_{T \in \tilde{T}_h} T \right)
\end{equation}

\begin{figure}

\centering
\begin{tikzpicture}[scale=0.5,every node/.style={scale=2.0} ]


\begin{scope}[shift={(20,0)}]
	\draw[step=4] (0,0) grid +(8,8);
	\draw[step=2] (0,4) grid +(4,4);
	\draw[step=2] (0,0) grid +(4,4);
	\draw[step=2] (4,4) grid +(4,4);
	\draw(2,4) grid +(2,2) rectangle (2,4);
	\draw [fill=red](2,5) circle (0.2cm);
	\draw [fill=red](4,5) circle (0.2cm);
	\draw [fill=red](3,6) circle (0.2cm);
	\draw [fill=red](3,4) circle (0.2cm);
 	\draw [fill=red](4,2) circle (0.2cm);
    \draw [fill=red](6,4) circle (0.2cm);
\end{scope}

\end{tikzpicture}
\caption{2:1 balance octree (quadtree) with hanging nodes (marked in red).} 
\label{fig:hangingnode}
\end{figure}
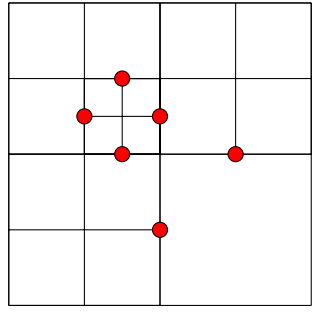

This gives us the surrogate domain, $\tilde{\Omega}_h$, with surrogate boundary $\tilde{\Gamma}_h := \partial \tilde{\Omega}_h$ and outward-oriented unit normal vector $\tilde{n}$ to $\tilde{\Gamma}$ as shown in \figref{fig:dist_vec}. The set of grids $T_h/\tilde{T}_h$ does not contribute towards the analysis of the domain $\Omega$, and these grids (in octree-based terminology leaves) are pruned, which significantly improves memory overhead \citep{saurabh2021scalable}. Although the $T_h$ are presented as regularly uniform grids, each $T \in T_h$ can be further subdivided based on some criteria $F(.)$. This ensures the adaptive refinement of octree-based grids. To preserve the well-posedness of the tree and keep the refinement process gradual, a 2:1 balancing strategy is enforced~\citep{saurabh2021scalable}. Often, 2:1 balanced octree-grids suffer from an ill-constrained node at the interface between regions of higher refinement and lower refinement, which are called \textbf{hanging node} as shown in \figref{fig:hangingnode}. These hanging nodes are not solved for, but rather constrained to respect the interpolation properties of the coarse element. For a detailed discussion of dealing with hanging nodes, see  \citet{saurabh2023scalable}.

\subsection{Shifted Boundary Method}
\label{sec:SBM}

\begin{figure}[h!]
    \centering
    \begin{subfigure}[t]{0.35\linewidth}
        \centering
        \begin{tikzpicture}[scale=0.2] 
            \draw[step=1cm, gray!30] (-10,-10) grid (10,10);

            \draw[blue] (0,0) circle (5cm);

            \draw[red, thick] 
                (-3,-4) -- (-4,-4) -- (-4,-3) -- (-5,-3) -- (-5,3) -- (-4,3) -- (-4,4) -- (-3,4) --
                (-3,5) -- (3,5) -- (3,4) -- (4,4) -- (4,3) -- (5,3) -- 
                (5,-3) -- (4,-3) -- (4,-4) -- (3,-4) -- (3,-5) -- (-3,-5) -- cycle;

            \node[blue] at (3,3) {$\Gamma$};
            \node[red] at (-5,-4) {$\tilde{\Gamma}$};
            \node at (-9.5,-9.5) {$\Omega$};
        \end{tikzpicture}
        \caption{ Surrogate boundary and true boundary.}
        \label{fig:SBM_Definition}
    \end{subfigure}
    \hspace{0.1\linewidth}
    \begin{subfigure}[t]{0.35\linewidth}
        \centering
        \begin{tikzpicture}[scale=0.5]
                \draw [line width = 0.5mm,blue] plot[smooth] coordinates {(1,-0.5) (2.25,2.5) (0.75,6)};
                \draw[line width = 0.5mm,red] (0,0.5) -- (0,5);
                \node[text width=0.5cm] at (0.5,5.5) {\small${\color{red}}$};
                \node[text width=0.5cm] at (1.75,5.5) {\small${\color{blue}}$};
                \node[text width=0.5cm] at (1.55,2.7) {\small$d$};
                \node[text width=0.5cm] at (3,2.9) {\small$n$};
                \draw[->,line width = 0.25mm,-latex] (0,2.5) -- (2.12,3.1);
                \draw[->,line width = 0.25mm,-latex] (2.12,3.1) -- (2.95,3.3);
        \end{tikzpicture}
        \caption{Distance vector $\textbf{d}$ and the true normal $\textbf{n}$.}
        \label{fig:dist_vec}
    \end{subfigure}
    \caption{The domain $\Omega$ is a square grid, with a circle at the center, featuring the true boundary \textcolor{blue}{$\Gamma$} and the surrogate boundary \textcolor{red}{$\tilde{\Gamma}$}. As described in the \textbf{Methods} section, $\Omega^+$ refers to the region outside the true boundary \textcolor{blue}{$\Gamma$}, while the region outside the surrogate boundary \textcolor{red}{$\tilde{\Gamma}$} is the surrogate domain $\tilde{\Omega_h}$.}
\end{figure}
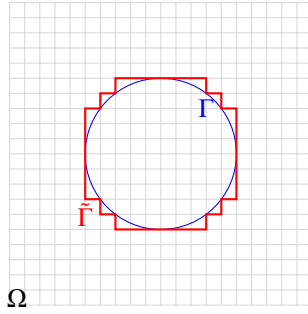
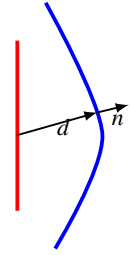 
The shifted boundary method introduced in \citet{main2018shifted1,main2018shifted2} replaces the imposition of the boundary conditions in the true boundary ($\boldsymbol{\Gamma}$) with the imposition of corrected boundary condition on the surrogate boundary ($\boldsymbol{\Tilde{\Gamma}}$). Usually, the distance between the true and surrogate boundaries is small, $\boldsymbol{||\Gamma-\Tilde{\Gamma}||_2 \sim \epsilon}$, where $\boldsymbol{\epsilon}$ scales as the resolution of the background mesh. 
Consider the field variable $\mathbf{\tilde{u}_d}$ at the surrogate boundary, $\tilde{\Gamma}$, Dirichlet boundary condition applied on the true boundary can be computed using Taylor series expansion as: 
\[
u_d = \tilde{u}_d + \nabla \tilde{u}\cdot \mathbf{d}
\]
where
\[u_d=g.\]
The Taylor series expansion gives the corrected Dirichlet boundary condition at the $\tilde{\Gamma}$, which is then applied with Nitsche's method~\citep{main2018shifted1,main2018shifted2}. As shown in \figref{fig:SBM_Definition} does not contain any cut-cells relieving the method with cut-cell-based issues.

\begin{figure}[t!]
    \centering
    \includegraphics[width=0.3\linewidth,trim={1.3in 1.3in 1.3in 1.3in},clip]{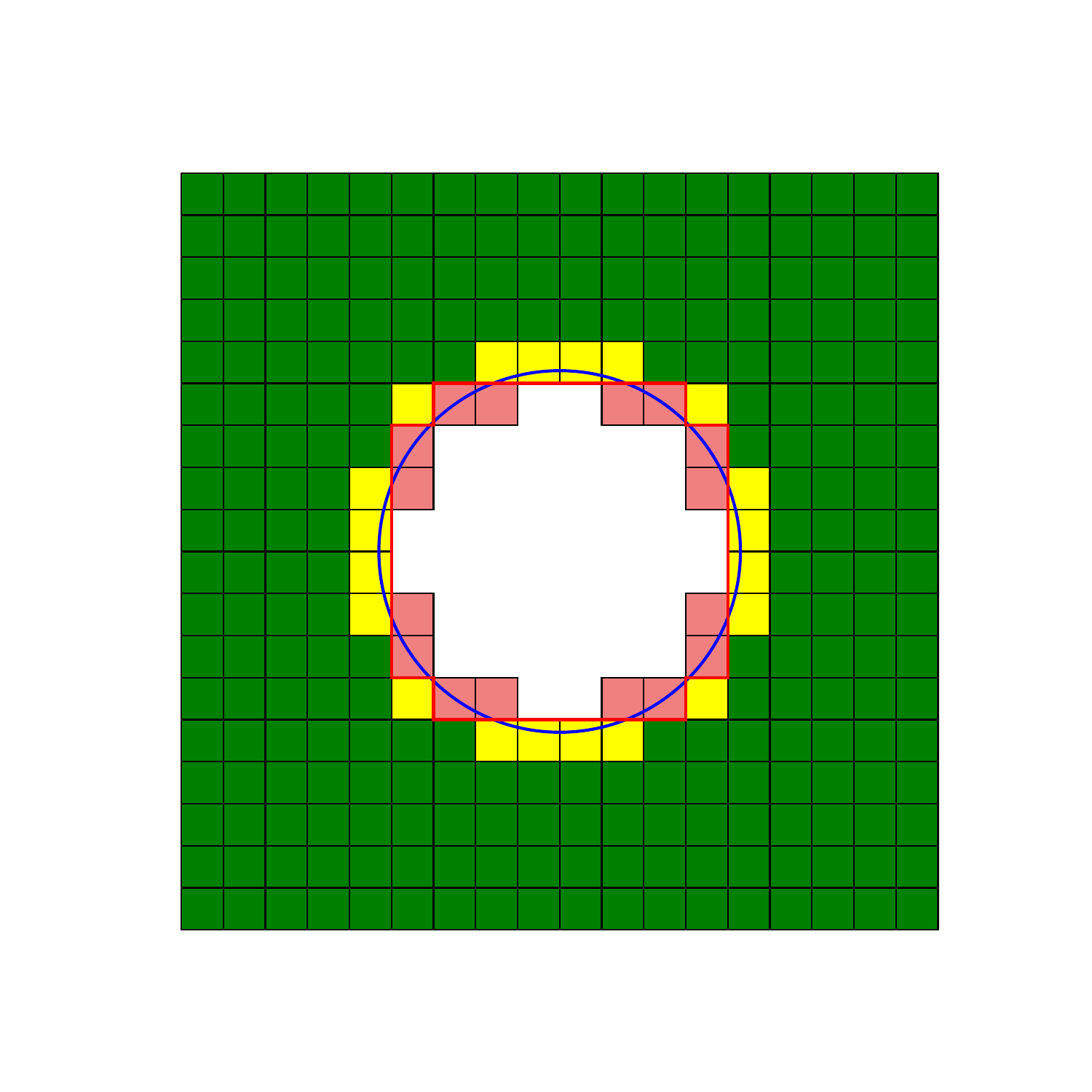}
\caption{
    The elements marked in yellow \colorbox{trueintercepted}{\phantom{\rule{1.5mm}{1.5mm}}} are true intercepted elements, and the elements marked in pink \colorbox{falseintercepted}{\phantom{\rule{1.5mm}{1.5mm}}} are false intercepted elements, with true boundary \textcolor{trueboundary}{$\mathbf{\Gamma}$}, and surrogate boundary \textcolor{red}{$\mathbf{\tilde{\Gamma}}$}. The union of yellow and pink elements (\colorbox{trueintercepted}{\phantom{\rule{1.5mm}{1.5mm}}} $\cup$ \colorbox{falseintercepted}{\phantom{\rule{1.5mm}{1.5mm}}}) forms the intercepted element. The green elements \colorbox{notintercepted}{\phantom{\rule{1.5mm}{1.5mm}}} are interior but not intercepted elements.
}
    \label{figure false intercepted}
\end{figure}

\paragraph{Optimal Surrogate Boundary}
\citet{yang2024optimal} developed a method to compute the optimal surrogate boundary in the case of the octree-based mesh, and the same strategy is used in this work to obtain the surrogate boundary.
\figref{figure false intercepted} presents an incomplete octree; the elements marked in green are not intercepted or exterior and directly contribute to finite element assembly. The elements of interest are intersected by the true boundary, $\Gamma$, defined as the intercepted element. The elements are classified as false intercepted (marked in light red) or true intercepted (marked in yellow) based on the parameter $\lambda$.

\[
\text{Classification} =
\begin{cases}
\text{False Intercepted}, & \text{if } \lambda \geq \lambda_{\text{criteria}} \\
\text{True Intercepted}, & \text{otherwise}
\end{cases}
\]
where:
\[
\lambda = \frac{\text{Count of GP inside } \Gamma}{\text{GP in the element}}.
\]
\citeauthor{yang2024optimal} showed that $\lambda_{\text{criteria}}=0.5$ results in the surrogate boundary with the least mean squared error between the true and surrogate boundary, thereby classifying it as the optimal surrogate boundary. All the analyses performed are based on the $\lambda_{\text{criteria}}=0.5$ in this work.

\paragraph{Navier-Stokes Equations}
This section outlines the finite element shifted boundary method formulation for Navier-Stokes equation. The non-dimensional governing equations for solving incompressible flow can be expressed as:
\noindent \textbf{Momentum Equation:}
\begin{equation}
\label{ns}
    \frac{\partial \mathbf{u}}{\partial t} + (\mathbf{u} \cdot \nabla)\mathbf{u} + \nabla p - \frac{1}{Re} \nabla^2 \mathbf{u} - \mathbf{f} = 0,
\end{equation}
\noindent \textbf{Solenoidality:}
\begin{equation}
\label{contd}
    \nabla \cdot \mathbf{u} = 0,
\end{equation}
where $\mathbf{u}$ is the non-dimensional velocity vector, $p$ is the non-dimensional pressure, and $\mathbf{f}$ represents the non-dimensional forcing term. The Reynolds number $Re$ is given by:
\begin{equation}
    Re = \frac{\rho_r u_r L_r}{\mu_r},
\end{equation}
where $\rho_r$ is the reference density, $u_r$ is the reference velocity, $L_r$ is the reference length, and $\mu_r$ is the reference dynamic viscosity.
The Galerkin discretization of \eqnref{ns} and \eqnref{contd} can be written as:
Find $\{\mathbf{u}_h, p_h\} \in V^h$ such that for all $\{\mathbf{w}_h, q_h\} \in V^h$,
\begin{equation}
\label{galerkin_discretization}
    \left( \mathbf{w}_h, \frac{\partial \mathbf{u}_h}{\partial t} + \mathbf{u}_h \cdot \nabla \mathbf{u}_h + \nabla \cdot \mathbf{w}_h \, p_h - \nabla \cdot \mathbf{w}_h \, p_h + q_h \nabla \cdot \mathbf{u}_h - \mathbf{w}_h \cdot \mathbf{f}_h \right)_\Omega = 0,
\end{equation}
where $\Omega$ is the domain, and terms on the first line represent the standard Galerkin formulation of the Navier-Stokes equations. We use a variational multiscale stabilization strategy in order to utilize equal order basis functions for pressure and velocity within a continuous Galerkin approach. The additional terms introduced by the VMS strategy are:

\begin{equation}
\label{vms_strategy}
    - \sum_{e=1}^{N_{\text{el}}} \left( \mathbf{u}_h \cdot \nabla \mathbf{w}_h, \mathbf{u}' \right)_{\Omega_e} 
    + \sum_{e=1}^{N_{\text{el}}} \left( \mathbf{w}_h, \mathbf{u}' \cdot \nabla \mathbf{u}_h \right)_{\Omega_e} - \sum_{e=1}^{N_{\text{el}}} \left( \nabla \mathbf{w}_h, \mathbf{u}' \otimes \mathbf{u}' \right)_{\Omega_e} 
    - \sum_{e=1}^{N_{\text{el}}} \left( \nabla \cdot \mathbf{w}_h, p' \right)_{\Omega_e} - \sum_{e=1}^{N_{\text{el}}} \left( \nabla \mathbf{q}_h, u' \right)_{\Omega_e}
\end{equation}
where $N_{\text{el}}$ denotes the number of elements, and $\mathbf{u}'$ and $p'$ represent the fine-scale velocity and pressure fields. These terms are further elaborated in \citet{bazilevs2007variational}. The fine-scale variables are defined as:
\begin{align}
    \mathbf{u}' &= -\tau_M \mathbf{r}_M (\mathbf{u}_h, p_h), \\
    p' &= -\tau_C \mathbf{r}_C (\mathbf{u}_h),
\end{align}
where $\mathbf{r}_M$ and $\mathbf{r}_C$ are the residuals of the momentum and continuity equations, respectively. The parameters $\tau_M$ and $\tau_C$ are stabilization parameters, defined as:
\begin{align}
    \tau_M &= \left( \frac{4}{\Delta t^2} + \mathbf{u}_h \cdot G \mathbf{u}_h + \frac{C_M}{Re^2} G : G \right)^{-1/2}, \\
    \tau_C &= \left( \tau_M g \cdot g \right)^{-1}.
\end{align}
Here, $G(G_{ij})$ and $g(g_j)$ represent mappings between physical and isoparametric elements based on the mesh geometry, where $i$ and $j$ refer to spatial coordinates. $C_M$ and $C_E$ are constants chosen as 36.

The Shifted Boundary Method (SBM) introduces surface integration terms to impose the Dirichlet boundary condition $\mathbf{u} = \mathbf{u}_d$ for the non-dimensional Navier-Stokes equations. These surface integration terms can be written as:
\begin{equation}
\label{SBM_NS}
    \begin{aligned}
    \underbrace{- \left\langle \mathbf{w}_h, \left( \frac{2}{Re} \mathbf{\epsilon}(\mathbf{u}_h) - p_h I \right) \cdot \tilde{\mathbf{n}} \right\rangle_{\tilde{\Gamma}_{D,h}}}_{\text{Consistency}} \quad
    & \underbrace{- \frac{1}{Re} \left\langle \left( \nabla_s \mathbf{w}_h + q_h I \right) \cdot \tilde{\mathbf{n}}, \mathbf{u}_h + \nabla \mathbf{u}_h \cdot \mathbf{d} - \mathbf{u}_d \right\rangle_{\tilde{\Gamma}_{D,h}}}_{\text{Adjoint consistency}} \\
    & \underbrace{+ \frac{1}{Re}\frac{\gamma}{h} \left\langle \mathbf{w}_h + \nabla \mathbf{w}_h \cdot \mathbf{d}, \mathbf{u}_h + \nabla \mathbf{u}_h \cdot \mathbf{d} - \mathbf{u}_d \right\rangle_{\tilde{\Gamma}_{D,h}}}_{\text{Penalty}}
    \end{aligned}
\end{equation}

where $\epsilon(\mathbf{u}_h)$ is the non-dimensional strain rate tensor, $\tilde{\mathbf{n}}$ is the outward normal vector at the surrogate boundary, and $\gamma$ is a penalty parameter.

Our final equations are constructed by adding terms in \eqnref{vms_strategy} and \eqnref{SBM_NS} 
 to \eqnref{galerkin_discretization}. We utilize a fully implicit BDF2 scheme with a linear basis functions to solve the non-linear system, see \citet{yang2024simulating} for details.

\subsection{Implicit Neural Representation}
\label{sec:INR}

INRs provide a powerful approach to encoding complex geometries in a continuous and memory-efficient manner. By using neural networks to represent surfaces implicitly, we can capture intricate shapes and topologies that traditional explicit methods often struggle to model accurately. When the function in \eqnref{sdf definition} is represented by a neural network with parameters $\theta$, such a form of representation is called INR. Such representation, if it represents the signed distance field of geometry, follows the Eikonal equation, and the gradient of the field near the surface scaled by the signed distance value gives the distance vector $\mathbf{d}$ given by \eqref{eikonal}. The distance vector $\mathbf{d}$ is the vector pointing towards the closest point in the true surface $\Gamma$. As pointed out, distance vector $\mathbf{d}$ points in $-\hat{n}$ direction with magnitude $|f_{\theta}(x)|$.
\begin{equation}
    ||\nabla_x f_{\theta} (x)||=1,
    \mathbf{d}=-f_{\theta}(x) \cdot \nabla_x f_{\theta} (x)
    \label{eikonal}
\end{equation}

To enable direct simulations using INRs with the Shifted Boundary Method (SBM), the INR has to meet two critical requirements:
\begin{enumerate}
    \item \textbf{Accurate Distance Vector Calculation:} 
    For each point near the region of interest, which is represented by a narrowband width $\delta$, the distance vector must accurately capture both the magnitude and direction. The value of $\delta$ is defined according to the resolution of the mesh intended for use in the final representation, ensuring that the implicit function approximates the true geometry precisely at relevant scales.
    \item \textbf{Precise Classification of Grid Points as Interior or Exterior:} 
    Grid points must be classified as either inside or outside the geometry boundary to enable the construction of a surrogate boundary. This classification step is essential for enforcing the geometric integrity of the Implicit Neural Representation, as it directly affects the accuracy of the boundary region generated during inference.
\end{enumerate}


\begin{figure}[t!]
    \centering
    \includegraphics[width=0.6\linewidth,trim=1.0in 1.0in 0.75in 1.0in, clip]{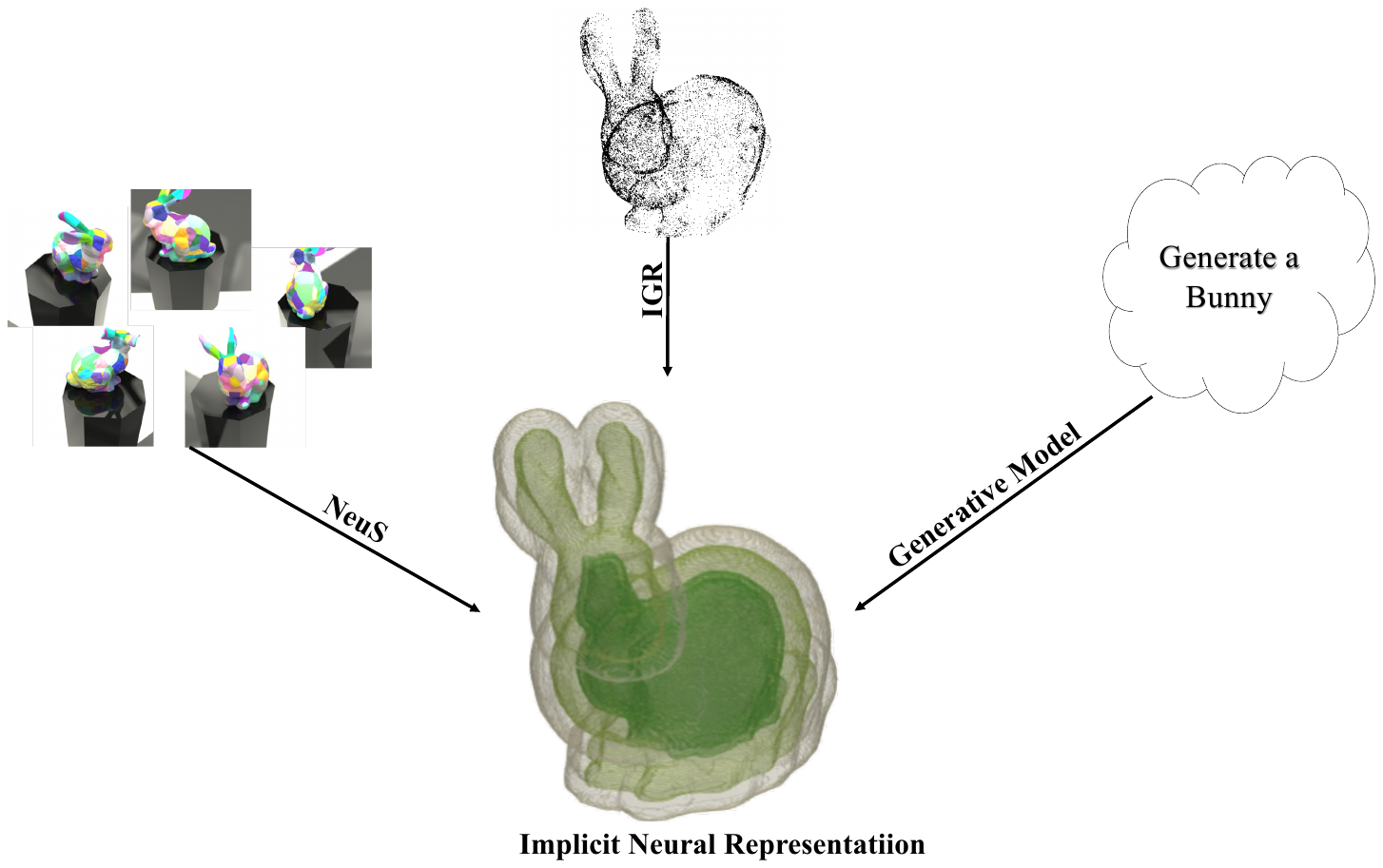}
    \caption{Implicit Neural Representations (INR) can be trained using  Point Clouds and Images for representing 3D geometry. Also, the generative model can directly be used to obtain the INR. The methods pointed out (Neus \citep{wang2021neus}, IGR \citep{gropp2020implicit} and Generative Model \citep{erkocc2023hyperdiffusion}) are few popular methods to go from abstract geometric representation to INRs. These INR obtained from any method then can be used in our framework for computational analysis. }
    \label{fig:INR_generation}
\end{figure}

This section describes three primary methods for generating INR: from point clouds, from images, and through generative AI models, as depicted in \figref{fig:INR_generation}.

\subsubsection{Point Clouds}
A point cloud represents the surface of 3D geometry as points, and several methods exist to obtain point clouds for real-world geometries like LiDAR, photogrammetry, and structured light scanning. A variety of work has been focused on generating Implicit Neural Representations from point cloud data~\citep{gropp2020implicit,ben2022digs,atzmon2020sal,sitzmann2020implicit,ma2020neural}. These works explore several architectural elements of the neural network and take advantage of the eikonal constraint and geometric features of the signed distance field to obtain INRs. 

\subsubsection{Images}
Multi-view 3D reconstruction from images is one of the most studied tasks in terms of 3D reconstruction. Different methods exist utilizing differentiable rendering frameworks to obtain Implicit Neural Representations from Images \citep{yariv2020multiview,wang2021neus,yariv2021volume,niemeyer2020differentiable,oechsle2021unisurf} All these frameworks are based on volume rendering method using INRs to render images. The difference between rendered images and actual images guides the training of INRs.

\subsubsection{Generative AI Models}
Diffusion-based models are gaining prominence in generative tasks \citep{ho2020denoising}. Recent work builds on latent diffusion approaches~ \citet{rombach2022high} to perform diffusion over the neural network weights ~\cite{erkocc2023hyperdiffusion} to generate various INRs, and successfully generate complex shapes like planes, chairs, and cats.

In our work, we start from a point cloud to create the INRs as detailed in \appendixref{section:generating_INR_polygon}. \appendixref{section:Val_Implicit} evaluates the representational accuracy of the INRs specifically for \textbf{SBM} based analysis as presented in \tabref{tab:complexity_objects}. The point cloud data is created from an STL file, which enables us to make comparisons between the analysis of geometry based on INRs and explicit representations as presented in \secref{section:ldc_2d} and \secref{section:ldc_3d}. 

\subsection{Integrating INR with SBM}
\label{sec:Implementation}

In this work, the INR is utilized to obtain an incomplete octree that serves as the surrogate domain. The implicit neural network is employed to selectively refine or discard octree elements, guided by a function \( F() \), which determines the level of refinement needed for a particular location. The algorithm begins by initializing a complete octree \( O \), and the implicit network \( f_{\theta} \) is applied to each octree element. The function \( F() \) encodes refinement criteria and is used to decide whether a given element should be refined further or pruned. The objective is to construct an incomplete octree \( O_{\text{incomplete}} \) that retains only the essential octants, thereby optimizing both storage and computational requirements. During the traversal of the complete octree, each octant \( S \in O \) is evaluated by the implicit network. If the network infers values greater than 0 (not inside the geometry) and the octant satisfies the refinement criteria imposed by \( F() \), the octant is retained. Otherwise, it is pruned. Once this process is complete, the remaining octants form the incomplete octree \( O_{\text{incomplete}} \), which is refined accordingly based on the remaining leaf nodes.

\begin{figure}[b!]
    \centering
    \begin{tikzpicture}[scale=.7]
    \draw[thick, ->] (0.5, 0) -- (6, 0) node[right] {};
    \draw[thick, ->] (0, 0.5) -- (0, 4) node[above, rotate=90] {};

    \draw[thick, green!70!black] (0.5, 2) -- (5.5, 2); 
    \draw[thick, blue] (0.5, 0.5) -- (5, 3.5);          

    \node[rotate=90] at (-0.5, 2.5) {No. of operations};
    \node at (3, -0.5) {Triangles};
    \begin{scope}[shift={(4.5, 3.5)}] 
        \draw[thick, blue] (0.2, 0.9) -- (1, 0.9) node[right, black] {\small Traversal in $\Delta s$};
        \draw[thick, green!70!black] (0.2, 0.4) -- (1, 0.4) node[right, black] {\small Neural Inference};
    \end{scope}
    \end{tikzpicture}
    \caption{Neural Inference Required for computing the distance vector occurs in constant operations. It depends upon the number of Neural Network Layers with other hardware and software constraints. For polygonal meshes, distance vector computation needs traversals across all the triangles. The number of required operations increases with an increase in the number of triangles.}
    \label{fig:neural_inference_against_traversal_in_triangles}
\end{figure}
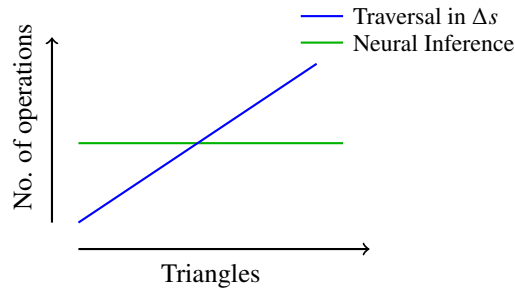

\begin{algorithm}[t!]
  \footnotesize
    \caption{\textsc{ImplicitOctreeGeneration:} Obtain incomplete octree using implicit network}
    \label{Algorithm ImplicitOctreeGeneration}
    \begin{algorithmic}[1]
\Require Complete octree $O$, Implicit network $f_\theta$, Function $F()$
\Ensure Incomplete octree $O_{\text{incomplete}}$
\State Initialize empty set $T$ for storing octree leaf nodes

\State \textbf{Step 1: Apply implicit network to prune octree}
\For{each octant $S \in O$}
    \If{$f_\theta(S) \geq 0$} \Comment{Use implicit network to determine active octants}
        \If{level of $S$ is acceptable based on $F()$}
            \State $T.\texttt{push}(S)$ \Comment{Store the selected octants in $T$}
        \EndIf
    \EndIf
\EndFor

\State \textbf{Step 2: Generate incomplete octree}
\State $O_{\text{incomplete}} \gets$ Refine and prune $O$ based on the leaf nodes in $T$

\State \Return $O_{\text{incomplete}}$
\end{algorithmic}
\end{algorithm}

\algoref{Algorithm ImplicitOctreeGeneration} provides the logic for generating incomplete octree, and is based on prior work in \citet{saurabh2021scalable}. \algoref{Algorithm: SurrogateBoundaryIdentificationUsingImplicitNetwork} generates the surrogate boundary. The algorithm traverses through each element and classifies each Gauss point as outside (if $f_\theta(gp)\geq0$)  or inside (if $f_\theta(gp)<0$). The total count per octant determines $\lambda$ which (along with $\lambda_{criteria}$), is used to classify whether an element is marked as "FalseIntercepted," "Exterior," "Interior," or "TrueIntercepted." Next,  \algoref{alg:boundary}, which is based on \citet{yang2024optimal}, takes the marker \( M \), to generate the optimal surrogate boundary. \algoref{Algorithm: DistanceFunctionCalculationUsingImplicitNetworkOneGP} outlines the procedure for computing the distance vector at the Gauss points located at the surrogate boundary by calculating the gradient of $f_{\theta}$ as presented in  \eqnref{eikonal}. The gradient is computed numerically by using two stencils on each axis using the central difference method. To optimize the process, a mapping mechanism is employed, ensuring that each gradient computation is performed only once.

\begin{algorithm}[t!]
  \footnotesize
    \caption{\textsc{IdentifySurrogateBoundary:} Surrogate Boundary Identification Using Neural Implicit Network}
    \label{Algorithm: SurrogateBoundaryIdentificationUsingImplicitNetwork}
    \begin{algorithmic}[1]
\Require Octree mesh $O$, threshold factor $\lambda$, implicit network $f_\theta$
\Ensure Surrogate boundary $\tilde{\Gamma}$, Element marker $M$
\State Initialize marker $M \gets []$
\For{each element $e \in O$} \Comment{Loop over all elements in the octree mesh}
    \State Initialize $count \gets 0$
    \For{each Gauss point $gp \in \text{GaussPoints}(e)$} \Comment{Loop over Gauss points in element $e$}
        \If{$f_\theta(gp) < 0$} \Comment{Classify Gauss point as interior based on implicit network}
            \State $count \gets count + 1$ \Comment{Increment count for Interior Gauss points}
        \EndIf
    \EndFor
    \State $\lambda_c \gets \frac{\text{count}}{\text{num\_gp}}$ \Comment{Compute fraction of Interior Gauss points}
    \If{$\lambda_c \geq \lambda$} 
        \State $M[e] \gets \text{FalseIntercepted}$ \Comment{Mark element as FalseIntercepted}
    \ElsIf{$count == 0$} 
        \State $M[e] \gets \text{Exterior}$ \Comment{Mark element as Exterior}
    \ElsIf{$count == \text{num\_gp}$} 
        \State $M[e] \gets \text{Interior}$ \Comment{Mark element as Interior}
    \Else
        \State $M[e] \gets \text{TrueIntercepted}$ \Comment{Mark element as TrueIntercepted}
    \EndIf
\EndFor
\State Extract the surrogate boundary $\tilde{\Gamma}$ based on marker $M$ as outlined in \algoref{alg:boundary}
\State \Return $\tilde{\Gamma}$, $M$
\end{algorithmic}
\end{algorithm}

\begin{algorithm}[t!]
  \footnotesize
    \caption{\textsc{ComputeDistanceVector:} Distance Vector Calculation using Neural Implicit Network}
    \label{Algorithm: DistanceFunctionCalculationUsingImplicitNetworkOneGP}
    \begin{algorithmic}[1]
\Require Gauss point position on the surrogate boundary ($Q$), Implicit network $f_\theta$, Mapping $M$
\Ensure Distance vector $(\boldsymbol{d}_{gp})$ for Gauss point $Q$
\If{$Q$ exists in $M$}
    \State Retrieve $(\boldsymbol{d}_{gp})$ from $M(Q)$ \Comment{Retrieve precomputed distance vector if available}
\Else
    \State Compute $\nabla f_\theta(Q)$ \Comment{Calculate gradient of implicit network at $Q$}
    \State Compute signed distance vector $\mathbf{d}_{gp} = \left( \frac{\nabla f_\theta(Q)}{\|\nabla f_\theta(Q)\|} \right) \times f_\theta(Q)$ \Comment{Determine distance via implicit network}
    \State Store $\mathbf{d}_{gp}$ in $M(Q)$ \Comment{Save mapping from $Q$ to $\mathbf{d}_{gp}$ for future reference}
\EndIf
\State \Return Distance vector $(\boldsymbol{d}_{gp})$
\end{algorithmic}
\end{algorithm}

All the algorithms are based on function queries (forward pass across the neural network or Neural Inference) in place of exhaustive traversal through a polygonal mesh~\citep{atallah2021shifted,yang2024optimal}. \figref{fig:neural_inference_against_traversal_in_triangles} represents the linearly increasing number of operations for computing distance vectors through Polygonal Meshes. For each triangle, the shortest distance needs to be computed. As outlined earlier, computing distance vectors through INR only requires Inference calls, which are independent of the geometry. The calls will take a constant number of operations dependent upon the size of the Multi-Layer Perceptron (number of matrix-matrix multiplications or matrix-vector multiplications).

\subsubsection{Comparison of Computational Complexity}
\label{sec:computational_complexity}

We compare the computational cost of three different simulation pipelines: (i) our method combining implicit neural representations with the shifted boundary method (INR-SBM), (ii) the traditional shifted boundary method (SBM), which uses triangle meshes to describe the geometry overlaid on a background mesh \footnote{For specificity, we consider the background mesh to be an octree type mesh}, and (iii) a boundary-fitted method. In all cases, we evaluate only the cost of solving the PDE on the given geometry. Importantly, we do not include the cost of training the INR, just as we typically do not account for the cost of generating STL files in mesh-based approaches. Analogous to how users can select pre-existing STL files from a repository for analysis, we envision a similar library of pre-trained INRs representing common geometries. Thus, the training cost of the INR is considered an offline, amortized step and is excluded from the comparison.


We use the finite element method (FEM) in all three cases. \tabref{tab:comparisonfem} summarizes the computational complexity associated with three key stages of the FEM pipeline: \textit{meshing}, \textit{assembly}, and \textit{solution}. The \textbf{meshing} step is fully automated in both the INR-SBM and standard SBM pipelines. For INR-SBM, the complexity of meshing reduces to $\mathcal{O}(N_{\text{Leaf}})$, where $N_{\text{Leaf}}$ is the number of octree leaves. In the standard SBM, additional processing over the triangle mesh geometry is needed (for in-out tests), leading to complexity $\mathcal{O}(N_{\text{Leaf}} N_{\Delta})$, where $N_{\Delta}$ is the number of surface triangles. In contrast, the boundary-fitted method typically requires a separate meshing stage, often with \textbf{user supervision}. Even in 2D, the best randomized incremental Delaunay triangulation algorithm has complexity $\mathcal{O}(N \log N)$, where $N$ is the number of input points~\cite{devillers1998improved}. In 3D, generating volumetric meshes from surface triangulations is even more expensive; for example, advancing front methods have worst-case complexity between $\mathcal{O}(N^2)$ (and even $\mathcal{O}(N^3)$) depending on geometry and refinement~\cite{lohner2008applied}.

The \textbf{assembly phase} can be divided into computational cost for integrating the \textit{volumetric} and \textit{boundary} contributions. All methods yield comparable volumetric integration times (denoted as ``same''), while the boundary assembly cost varies. INR-SBM retains its implicit representation to compute integrals over surrogate boundaries, again scaling as $\mathcal{O}(N_{\text{Leaf}})$. In contrast, SBM depends on mesh-extracted elements ($\mathcal{O}(N_{\text{Leaf}}N_{\Delta})$). The boundary-fitted approach has no boundary integration cost beyond its manual mesh specification, since the boundary conditions can be directly enforced.

Finally, the \textbf{solver} computational time depends primarily on the number of degrees of freedom (DoFs) in the system. For direct solvers (e.g., multifrontal LU or Cholesky), the computational complexity typically scales as $\mathcal{O}(N_{\text{DOF}}^3)$ in the worst case. In practice, however, we often employ iterative solvers such as GMRES or conjugate gradients, which offer significantly reduced complexity.\footnote{Iterative solvers—particularly Krylov subspace methods like GMRES or conjugate gradients—can achieve complexities as low as $\mathcal{O}(N_{\text{DOF}}^2)$ (in the worst case) or even $\mathcal{O}(N_{\text{DOF}}\sqrt{\texttt{condition number}})$ when combined with optimal preconditioning. However, their performance is sensitive to the condition number of the system matrix. In our experience, immersed methods such as SBM or INR-SBM tend to exhibit higher condition numbers compared to body-fitted meshes, which can slow down convergence and offset the theoretical gains in asymptotic complexity. This trade-off must be considered when evaluating solver efficiency.} The number of DoFs across INR-SBM, standard SBM, and boundary-fitted approaches is comparable when similar background mesh resolutions are used.


\begin{table}[t!]
    \caption{Comparison of Computational Complexity across methods}
    \centering
    \label{tab:comparisonfem}
    \begin{tabular}{|c|c|c|c|}
    \hline
    \textbf{Method} & \textbf{INR-SBM} & \textbf{SBM} & \textbf{Boundary Fitted} \\
    \hline
    \textbf{Meshing} 
    & $\mathcal{O}(N_{\text{Leaf}})$ & $\mathcal{O}(N_{\text{Leaf}}N_{\Delta})$ & Manual \\
    \hline
    \multirow{4}{*}{\textbf{Assembly}} 
    & \multicolumn{3}{c|}{\textbf{Volumetric}} \\
    \cline{2-4}
    & Same & Same & Same \\
    \cline{2-4}
    & \multicolumn{3}{c|}{\textbf{Boundary}} \\
    \cline{2-4}
    & $\mathcal{O}(N_{\text{Leaf}})$ & $\mathcal{O}(N_{\text{Leaf}}N_{\Delta})$ & None \\
    \hline
    \textbf{Solver} 
    & $\mathcal{O}(N_{DOF}^3)$ &$\mathcal{O}(N_{DOF}^3)$ & $\mathcal{O}(N_{DOF}^3)$ \\
    \hline
    \end{tabular}
\end{table}

\section{Flow Simulation Results}
\label{Sec:Results}

To evaluate the accuracy and robustness of our INR-based approach, we first apply it to a 2D lid-driven cavity (LDC) flow with a circular obstacle represented using an implicit neural representation (see \secref{section:ldc_2d}). The LDC problem is a well-established benchmark in computational fluid dynamics (CFD), often used to assess the performance of numerical solvers in handling boundary-driven flows and internal obstacles.

In our setup, we compare INR-based flow simulations against conventional body-fitted mesh-based simulations across a range of Reynolds numbers (Re = 100, 400, 1000, 3000, 5000). We analyze velocity profiles along the vertical and horizontal centerlines to assess deviations between the two approaches. The results show that INR-based simulations closely match the traditional body-fitted approach, demonstrating that our framework accurately captures key flow features, including forming primary and secondary vortices. Moreover, the INR representation eliminates the need for explicit mesh generation, simplifying the simulation pipeline while maintaining high accuracy. These findings confirm that INR-driven simulations provide a robust and reliable alternative to mesh-based approaches, even in scenarios with internal obstacles and complex flow interactions.

Building on the 2D validation, we extend our methodology to a 3D lid-driven cavity flow, incorporating more complex geometries to assess the scalability and robustness of INR-based simulations in higher dimensions (see \secref{section:ldc_3d}). This extension is critical for verifying the framework’s applicability to real-world CFD problems involving three-dimensional structures. We compare INR-based results against polygonal mesh-based simulations---using geometries such as spheres, cones, and cylinders placed inside the cavity. The simulations span multiple Reynolds numbers, ensuring that our approach remains consistent and stable across different flow regimes. Velocity magnitude profiles along diagonal sections of the domain confirm excellent agreement between INR and traditional approaches. One of the key advantages observed in the 3D case is that INR-based simulations avoid the computational overhead associated with meshing complex objects, particularly for geometries that would require extensive manual preprocessing in conventional methods. By leveraging neural implicit representations, our framework preserves geometric accuracy while enabling seamless integration into high-fidelity CFD solvers.

To further demonstrate the versatility of our framework, we apply it to more intricate and non-traditional geometries, including gyroids, the Stanford bunny, and an AI-generated airplane model. These geometries pose significant challenges for conventional mesh-based simulations due to their high surface complexity, intricate topology, and fine details.
\begin{itemize}[itemsep=0pt,topsep=0pt]
\item Gyroid Structure (\secref{section:gyroid_flow}): We simulate internal flow within a gyroid, a periodic minimal surface geometry relevant to porous media and metamaterial applications. The INR-based simulation successfully captures complex internal flow behavior while maintaining numerical stability.
\item Stanford Bunny in Pipe Flow (\secref{section:bunny_pipeflow}): The Stanford bunny, a widely used test object in graphics and geometry processing, is placed in a pipe flow scenario to assess the method’s ability to handle realistic, highly detailed shapes. Our results reveal intricate flow separations, stagnation zones, and velocity profiles, all computed directly from the neural representation without mesh extraction.
\item Generative-AI Airplane (\secref{sec:GenAI_plane}): We integrate an aircraft geometry output from a generative AI model with a CFD simulation, demonstrating compatibility with generative design workflows. This highlights the potential of using INR-based simulation as a rapid evaluation tool for AI-generated geometries, accelerating the design-to-simulation pipeline.
\end{itemize}

These experiments illustrate that our INR-driven framework generalizes across diverse and complex geometries, offering a scalable and flexible solution for computational fluid dynamics. By removing the bottleneck of explicit meshing, our approach streamlines CFD workflows and enables simulations on previously challenging or computationally expensive geometries.


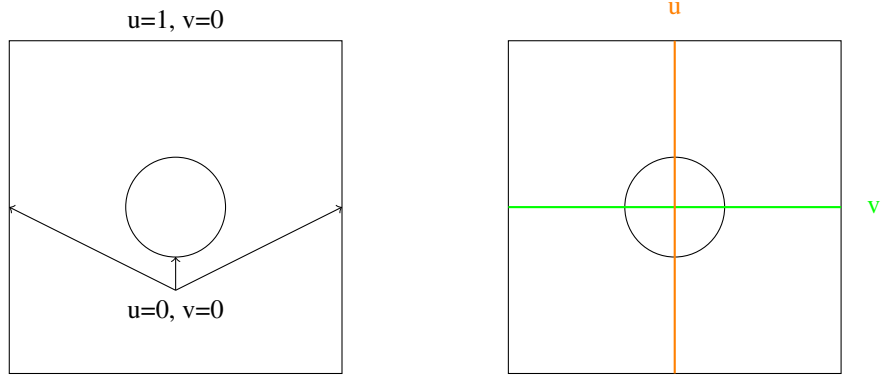
\begin{figure}[!t]
    \centering
    \begin{tikzpicture}[scale=2.2]
    \begin{scope}
        \draw (0,0) rectangle (2,2);  
        \draw (1,1) circle (0.3);      
        \draw[->] (1,0.5) -- (0,1);    
        \draw[->] (1,0.5) -- (2,1);    
        \draw[->] (1,0.5) -- (1,0.7);  
        \node[anchor=south] at (1,2) {u=1, v=0};
        \node[anchor=north] at (1,0.5) {u=0, v=0};
    \end{scope}

    \begin{scope}[xshift=3cm] 
        \draw (0,0) rectangle (2,2);  
        \draw (1,1) circle (0.3);      

        \draw[green, thick] (0,1) -- (2,1);  

        \draw[orange, thick] (1,0) -- (1,2); 

        \node[anchor=south] at (1,2.1) {\textcolor{orange}{u}};

        \node[anchor=west] at (2.1,1) {\textcolor{green}{v}};
    \end{scope}
    \end{tikzpicture}
    \caption{Illustration of the square domain $\Omega$ with a circle at the center with no slip boundaries everywhere except the top, as shown in the figure at the right. The figure on the left shows the location where velocity components \( u \) and \( v \) are plotted.}
    \label{fig:LDC_Circle_Schematic}
\end{figure}

\label{section:CFD}
\subsection{Lid driven cavity flow in 2D with circular obstacle}
\label{section:ldc_2d}

\begin{figure}[b!]
    \centering
    \begin{subfigure}[b]{0.3\linewidth}
        \centering
        \includegraphics[width=\linewidth]{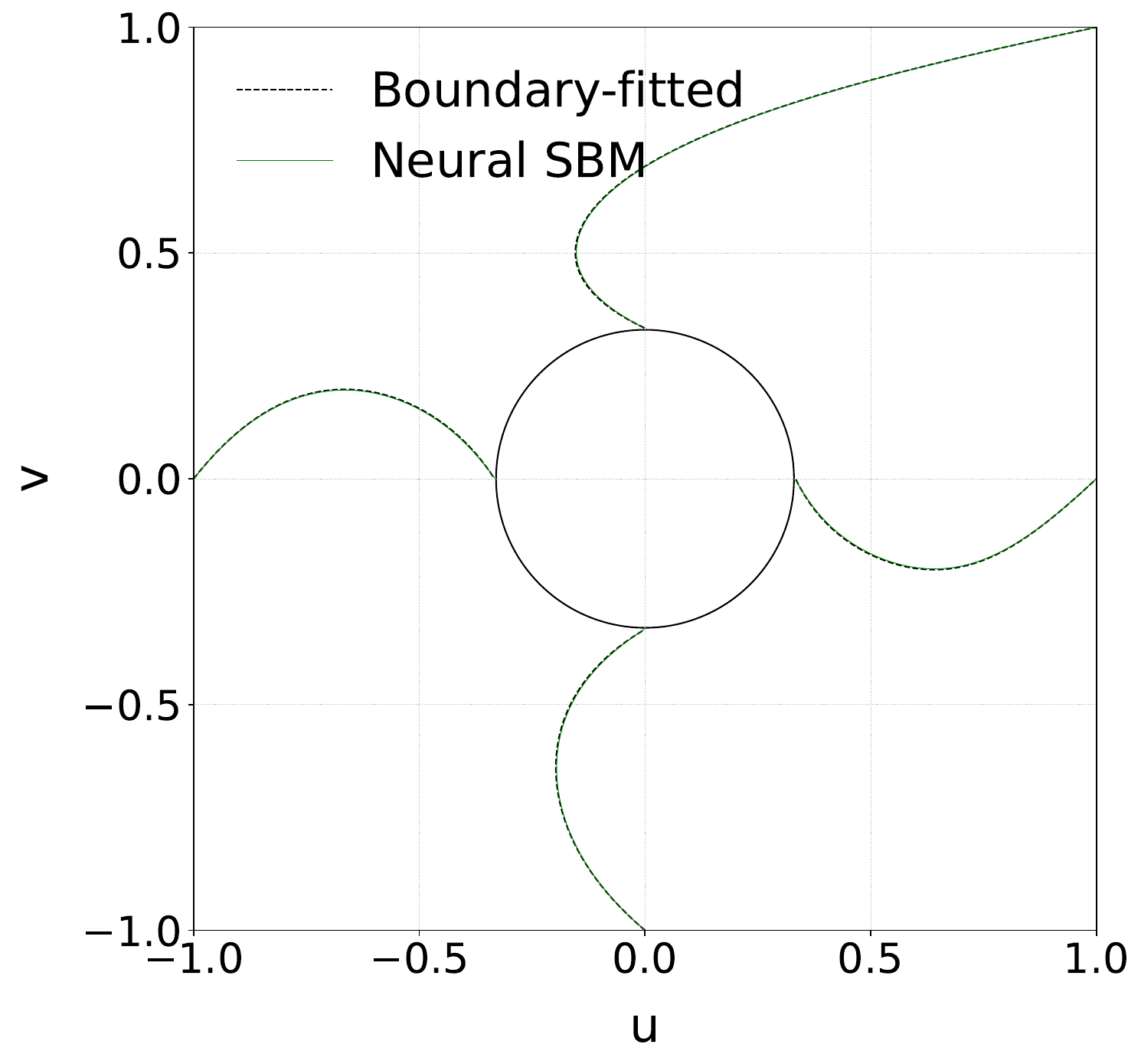}
        \caption{$\mathbf{Re=100}$}
        \label{fig:Re_100_LDC}
    \end{subfigure}
    \begin{subfigure}[b]{0.3\linewidth}
        \centering
        \includegraphics[width=\linewidth]{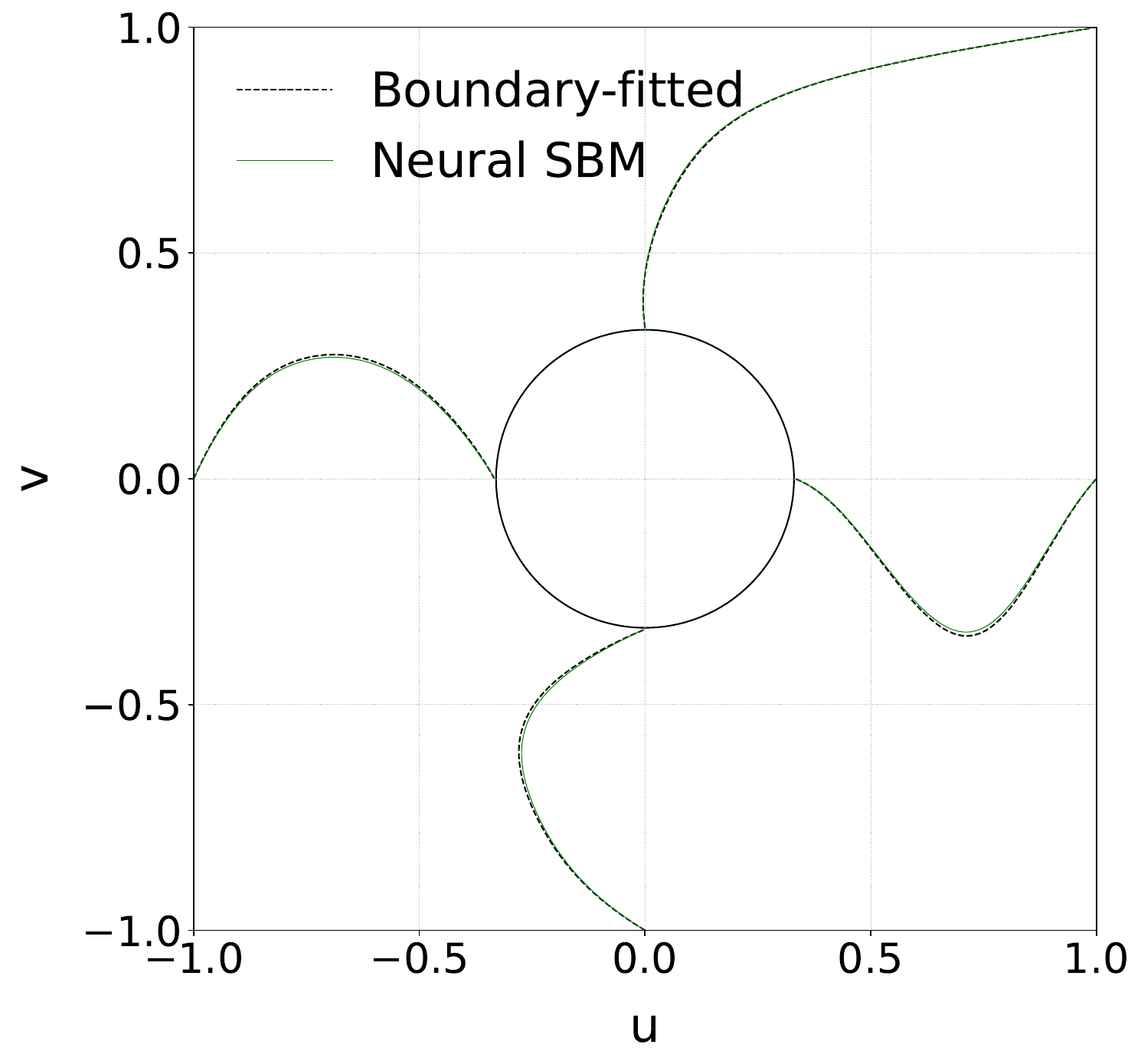}
        \caption{$\mathbf{Re=400}$}
        \label{fig:Re_400_LDC}
    \end{subfigure}
    \begin{subfigure}[b]{0.3\linewidth}
        \centering
        \includegraphics[width=\linewidth]{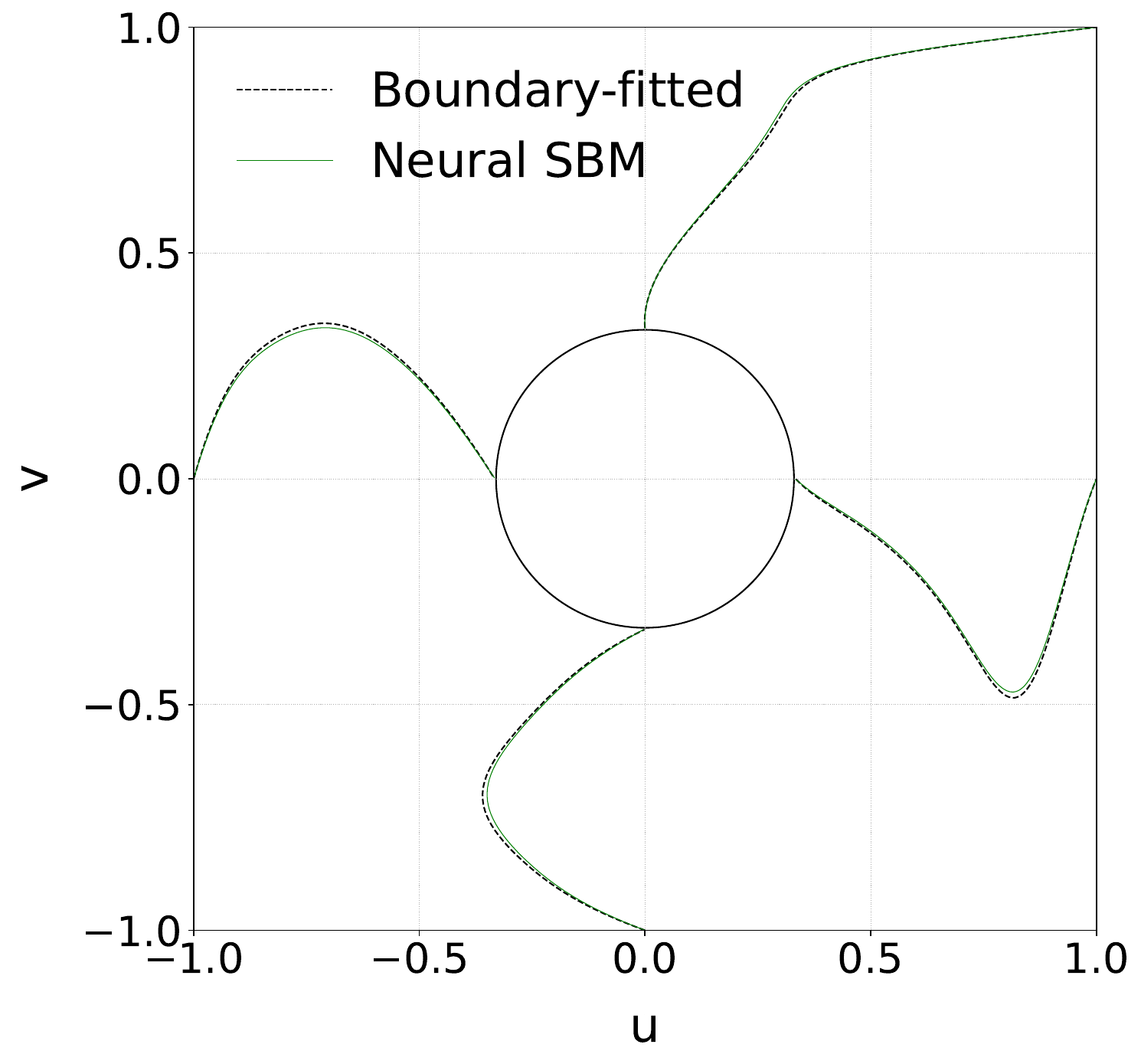}
        \caption{$\mathbf{Re=1000}$}
        \label{fig:Re_1000_LDC}
    \end{subfigure}
    \begin{subfigure}[b]{0.3\linewidth}
        \centering
        \includegraphics[width=\linewidth]{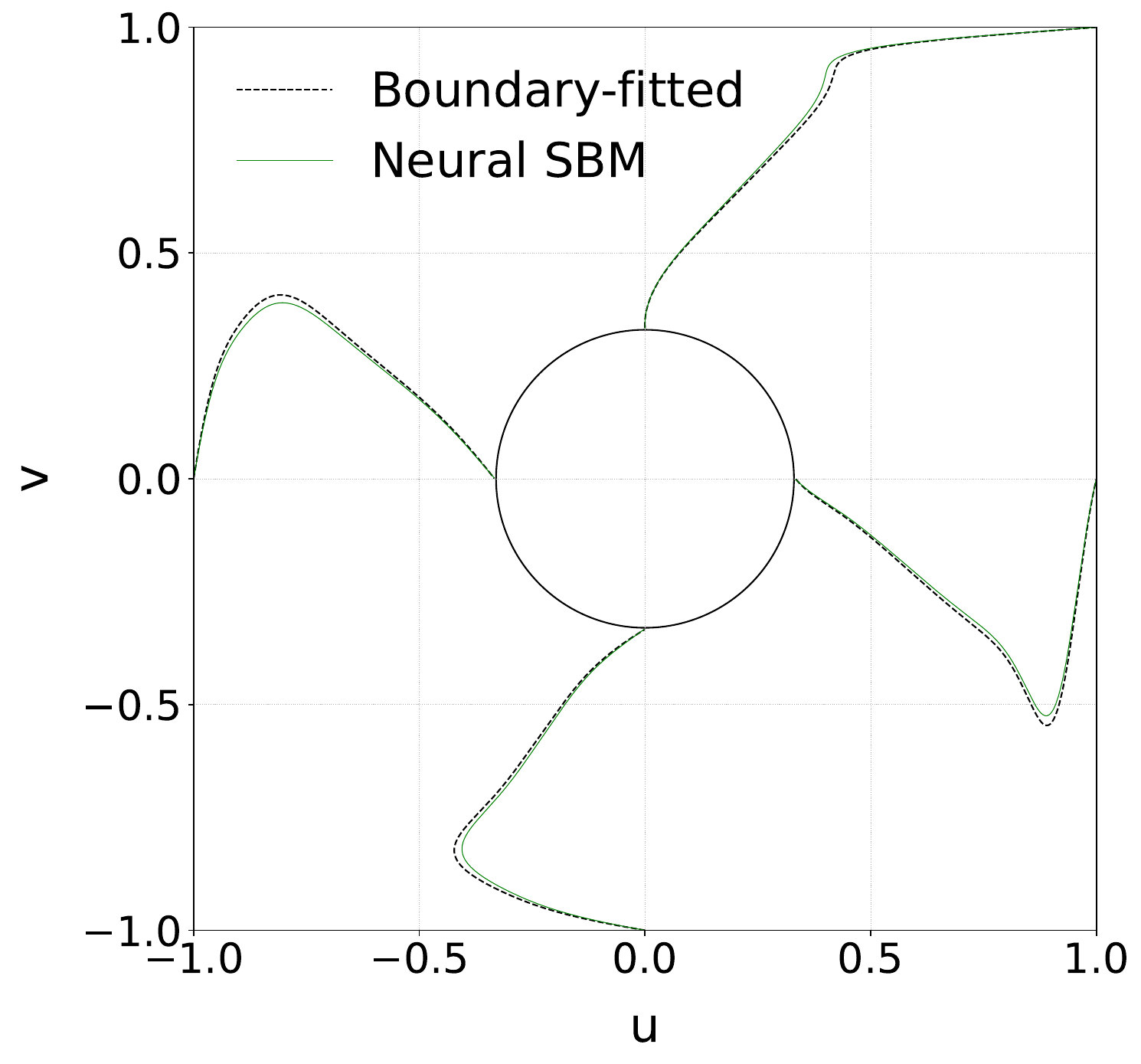}
        \caption{$\mathbf{Re=3000}$}
        \label{fig:Re_3000_LDC}
    \end{subfigure}
    \begin{subfigure}[b]{0.3\linewidth}
        \centering
        \includegraphics[width=\linewidth]{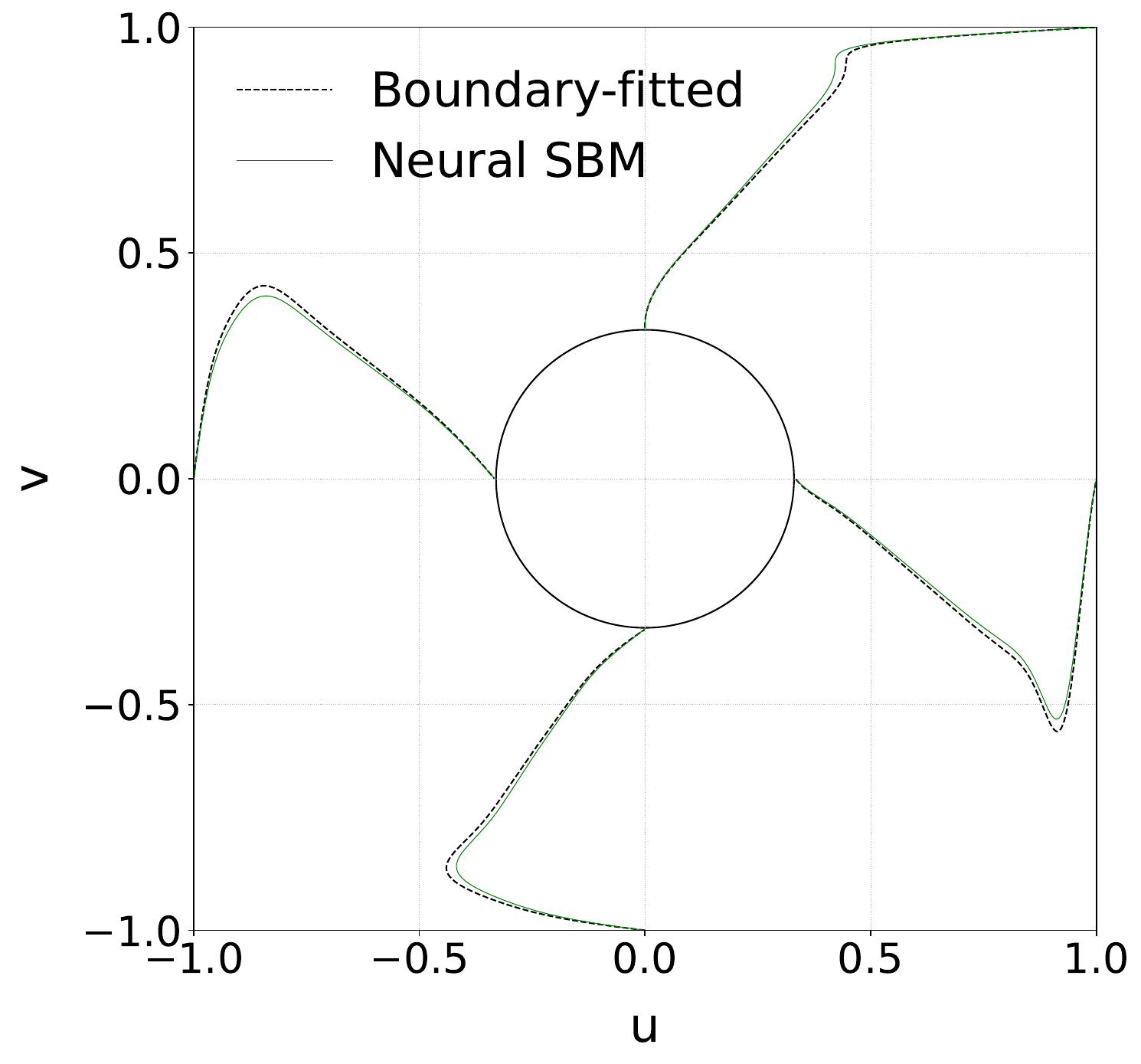}
        \caption{$\mathbf{Re=5000}$}
        \label{fig:Re_5000_LDC}
    \end{subfigure}
    \caption{Validation for 2D Lid-Driven Cavity with a circular obstacle at various Reynolds numbers using boundary-fitted mesh.}
    \label{fig:CFD_Validation_LDC}
\end{figure}

\begin{figure}[t!]
    \centering
    \begin{subfigure}[b]{0.3\linewidth}
        \centering
        \includegraphics[width=\linewidth, trim=150 100 100 100, clip]{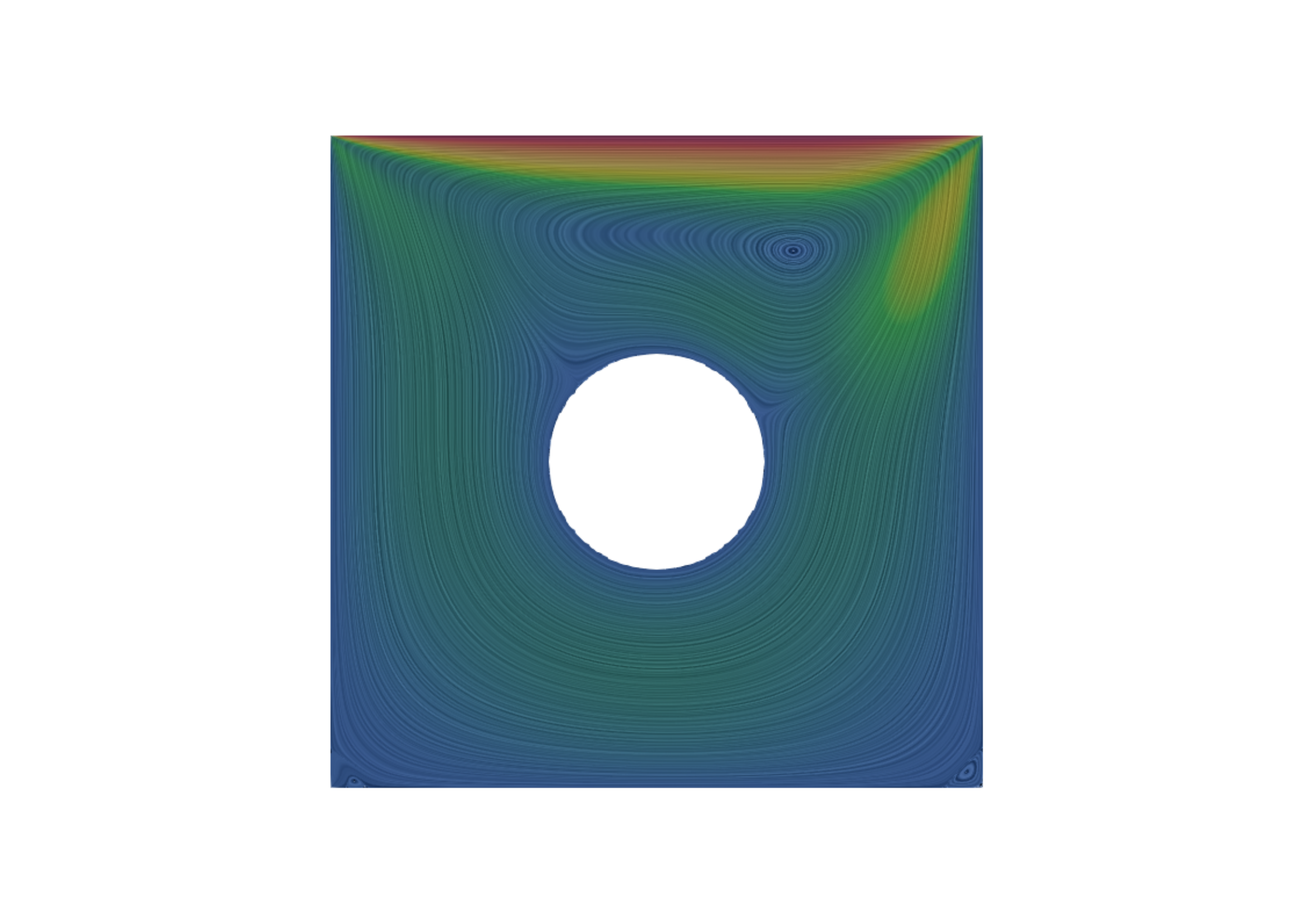}
        \caption{$\mathbf{Re=100}$}
        \label{fig:Re_100_LDC_LIC}
    \end{subfigure}
    \begin{subfigure}[b]{0.3\linewidth}
        \centering
        \includegraphics[width=\linewidth, trim=150 100 100 100, clip]{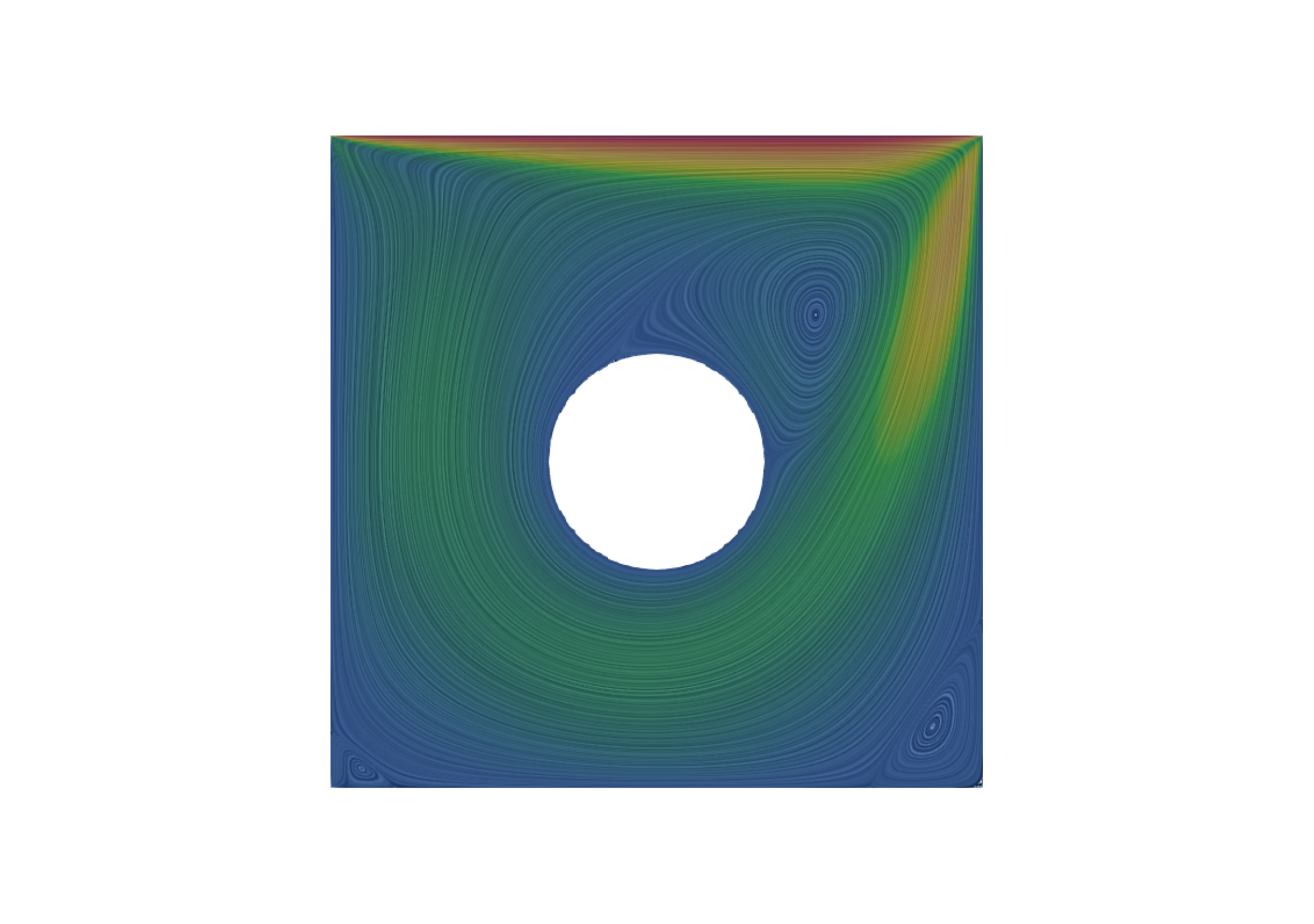}
        \caption{$\mathbf{Re=400}$}
        \label{fig:Re_400_LDC_LIC}
    \end{subfigure}
    \begin{subfigure}[b]{0.3\linewidth}
        \centering
        \includegraphics[width=\linewidth, trim=150 100 100 100, clip]{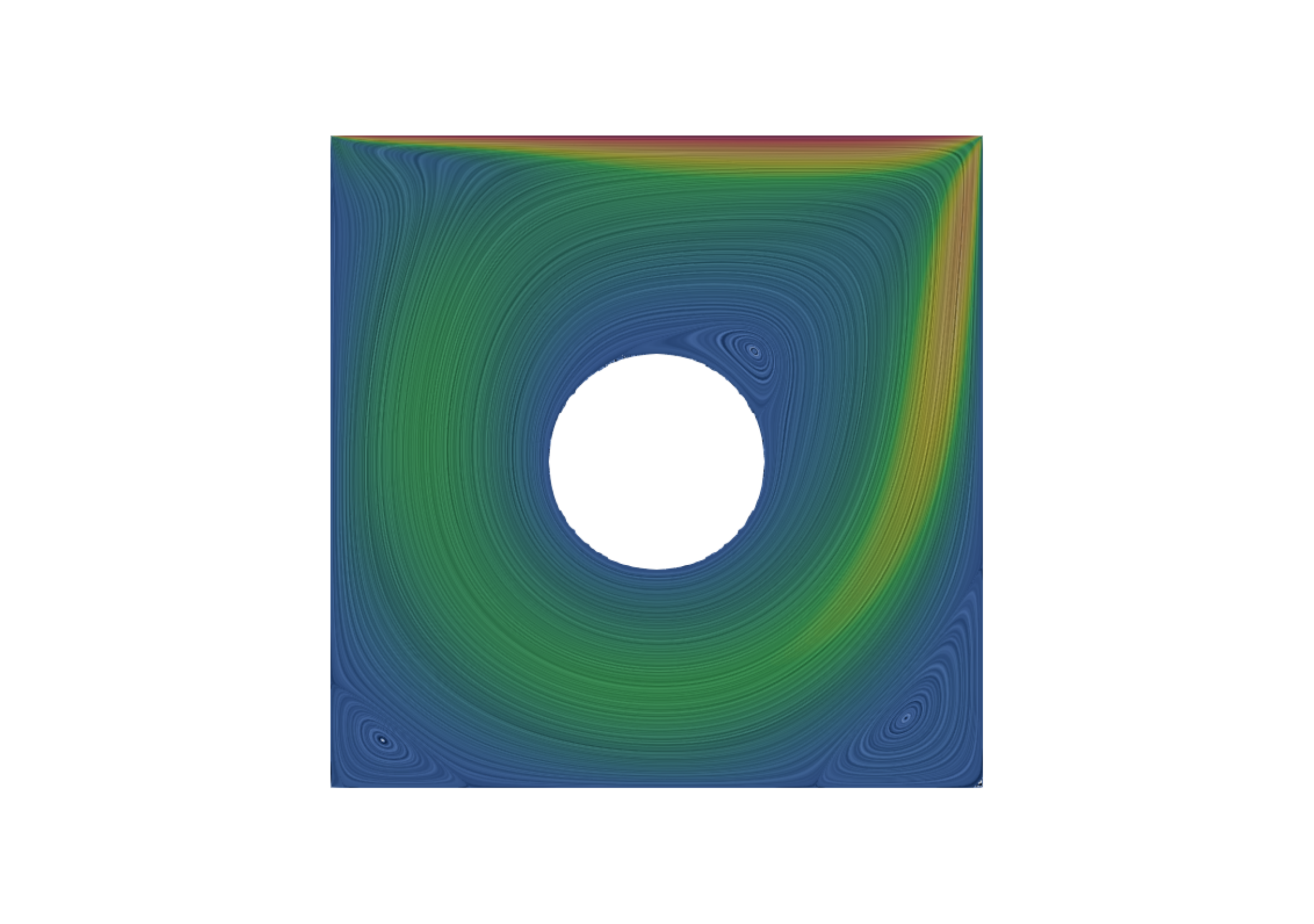}
        \caption{$\mathbf{Re=1000}$}
        \label{fig:Re_1000_LDC_LIC}
    \end{subfigure}
    \begin{subfigure}[b]{0.3\linewidth}
        \centering
        \includegraphics[width=\linewidth, trim=150 100 100 100, clip]{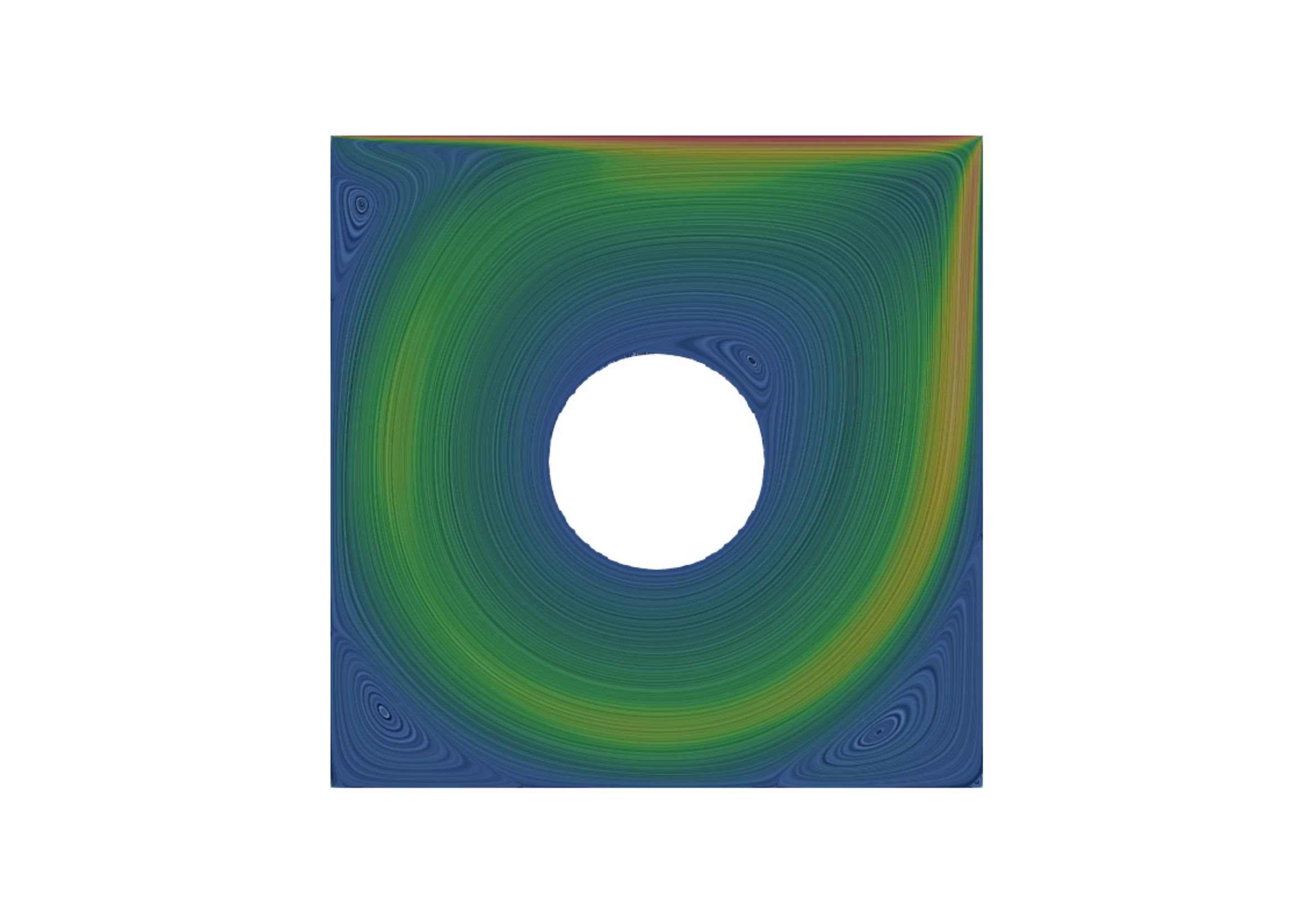}
        \caption{$\mathbf{Re=3000}$}
        \label{fig:Re_3000_LDC_LIC}
    \end{subfigure}
    \begin{subfigure}[b]{0.3\linewidth}
        \centering
        \includegraphics[width=\linewidth, trim=150 100 100 100, clip]{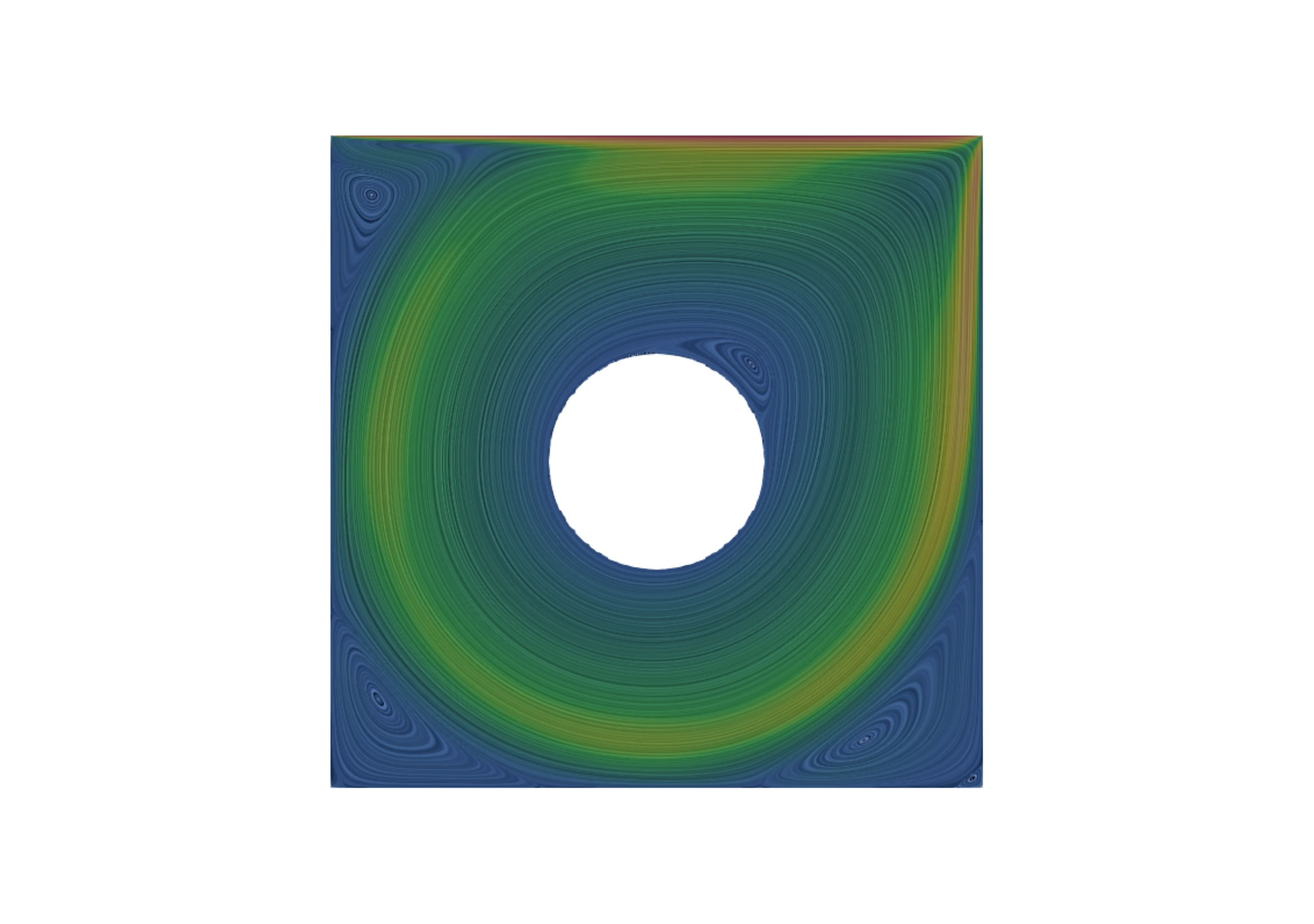}
        \caption{$\mathbf{Re=5000}$}
        \label{fig:Re_5000_LDC_LIC}
    \end{subfigure}
    \includegraphics[width=0.8\textwidth,trim=0.0in 9.0in 0.0in 0.1in, clip]{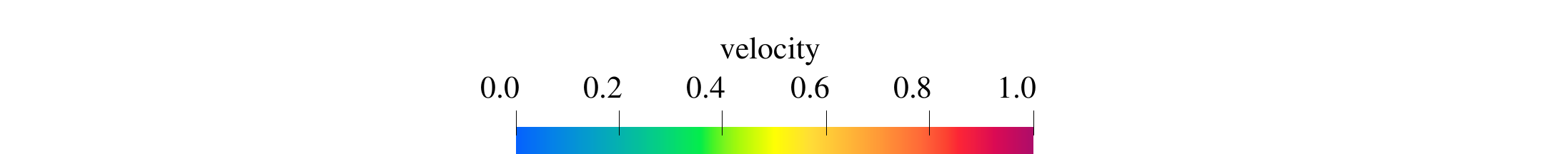}    
    \caption{Surface LIC for Different Reynolds Numbers for Lid-driven cavity case in 2D with a circular obstacle. The figure depicts the movement of the primary vortex in the top-right corner toward the circular obstacle as the Reynolds number increases.}
    \label{fig:Surface_LIC_Validation}
\end{figure}

The Lid-Driven Cavity (LDC) problem is frequently utilized to validate and assess the robustness of Computational Fluid Dynamics (CFD) frameworks. In this study, we follow the problem setup outlined by Huang et al.~\citep{huang2020simulation}. The boundary conditions for the LDC analysis are illustrated in \figref{fig:LDC_Circle_Schematic} (left). In our setup, the top lid moves in the x-direction, while the sidewalls of the 2D chamber and the circular obstacles adhere to no-slip boundary conditions. The circular obstacle has a diameter that is one-third of the chamber's length. This circular geometry is represented using an INR of a sphere with $z=0$, allowing for the classification of octants, surrogate boundary generation, and distance vector calculation through the implicit function $f_\theta(x, y, 0)$. We perform simulations with boundary-fitted quadrilateral mesh with a mesh size of 0.0014 as a comparative baseline. Then, we use the INR with a uniform background mesh size of 0.0039 (as the length of the chamber is taken as 2, this corresponds to the level of refinement 9 for the octree mesh). We validate the results by plotting the u-velocity along the vertical centerline and the v-velocity along the horizontal centerline, as shown on the right of \figref{fig:LDC_Circle_Schematic}. We perform comparisons across different Reynolds numbers (Re) of 100, 400, 1000, 3000, and 5000. The results of the comparison are shown in \figref{fig:CFD_Validation_LDC}, which indicates a close match between the boundary-fitted results and those obtained from our proposed method. Additionally, \figref{fig:Surface_LIC_Validation} visualizes the Line Integral Convolution (LIC) contours at steady state. LIC contours represent isocontours of a vector magnitude field, visualized in varying colors within a LIC image. The results depict the movement of the primary vortex in the top-right corner toward the circular obstacle as the Reynolds number increases, demonstrating the characteristic flow pattern changes.

In many engineering analysis cases that require significant design effort, a quick verification analysis is performed by taking a representative 2D slice of the complex 3D design. In our study presented above, the results for the 2D boundary-fitted mesh are validated by directly querying the INR of a sphere with $z=0$, $f_\theta(x,y,0)$. This ensures the INR-based representation can be used in similar regards for analysis.

\subsection{Lid driven cavity flow in 3D}
\label{section:ldc_3d}

\citet{yang2024simulating} deployed the Shifted Boundary Method for Navier-Stokes over polygonal meshes where distance vectors are obtained by exhaustive traversal across all the polygons. In this section, we compare the Navier-Stokes solution obtained from polygonal meshes based on \citep{yang2024simulating} and with INRs (representing the same polygonal mesh). The details of INRs construction are outlined in \appendixref{section:generating_INR_polygon}. All the geometries studied in \appendixref{section:Val_Implicit} are used for analysis, and they comprise geometries of varied complexities. \appendixref{section:Val_Implicit} details the analysis of the accuracy of the INR for compatibility with SBM.

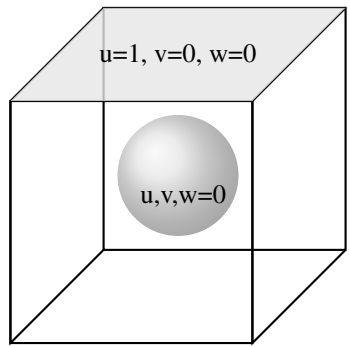
\begin{figure}[t!]
    \centering
    \begin{tikzpicture}[scale=0.8] 

        \def\boxwidth{4}
        \def\boxheight{4}
        \def\boxdepth{4}

        \draw[thick] (0,0,0) -- (\boxwidth,0,0) -- (\boxwidth,\boxdepth,0) -- (0,\boxdepth,0) -- cycle;
        \draw[thick] (0,0,0) -- (0,0,\boxheight) -- (0,\boxdepth,\boxheight) -- (0,\boxdepth,0);
        \draw[thick] (\boxwidth,0,0) -- (\boxwidth,0,\boxheight) -- (\boxwidth,\boxdepth,\boxheight) -- (\boxwidth,\boxdepth,0);
        \draw[thick] (0,0,\boxheight) -- (\boxwidth,0,\boxheight) -- (\boxwidth,\boxdepth,\boxheight) -- (0,\boxdepth,\boxheight) -- cycle;

        \fill[gray!20, opacity=0.8] (0,\boxheight,0) -- (\boxwidth,\boxheight,0) -- (\boxwidth,\boxdepth,\boxheight) -- (0,\boxdepth,\boxheight) -- cycle;

        \node[anchor=center] at (2, \boxheight, 2) {u=1, v=0, w=0}; 

        \shade[ball color=gray!40, opacity=0.5] (2,2,2) circle (1);
        \node[anchor=north west] at (1.225,2,2) {u,v,w=0};




    \end{tikzpicture}
    \caption{3D box domain with boundary conditions at the corners and a sphere centered inside. The boundary condition \( u = 1, v = 0, w = 0 \) is applied at the top face, and \( u = 0, v = 0, w = 0 \) is applied at all other boundaries and the boundary of the geometry.}
    \label{fig:3d_box_with_sphere_shaded}
\end{figure}

For the analysis, the geometries are placed within a cubic domain of $[-1.0,-1.0,-1.0]\times[1.0,1.0,1.0]$ with top lid sliding as presented in \figref{fig:3d_box_with_sphere_shaded}, all other faces and the geometry are at the no-slip boundary. The geometry's center of mass is located at the origin. This study is conducted with a uniform mesh of size $h = \frac{L}{2^7}$ everywhere and $h = \frac{L}{2^9}$ at the boundary octants for all cases. The analysis is performed with time-step, $\mathbf{dt}=1$ until a steady state is reached. 

\begin{figure}[t!]
    \centering
    \begin{subfigure}[b]{0.35\linewidth}
        \centering
        \includegraphics[width=\linewidth]{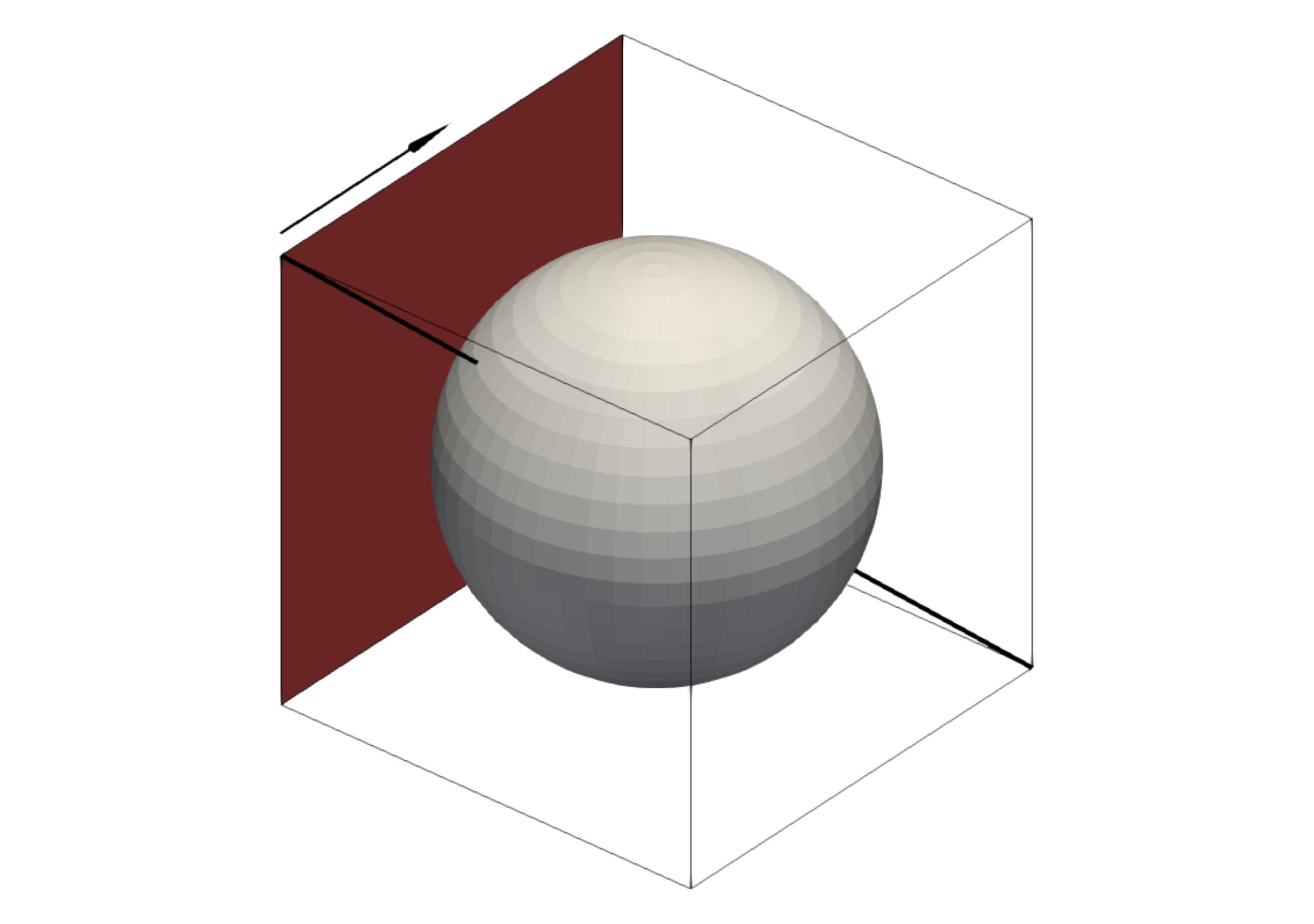}
        \caption{Sphere}
        \label{fig:sphere_ldc_3d_vis}
    \end{subfigure}
    \begin{subfigure}[b]{0.59\linewidth}
        \centering
        \raisebox{1em}{\createVelocityPlot{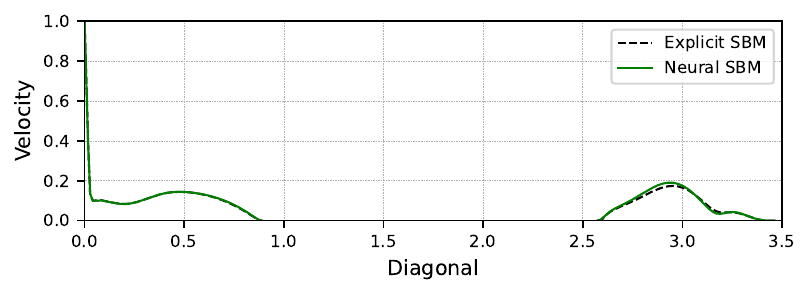}}
        \caption{}
        \label{fig:sphere_ldc_plot}
    \end{subfigure}
    \begin{subfigure}[b]{0.35\linewidth}
        \centering
        \includegraphics[width=\linewidth]{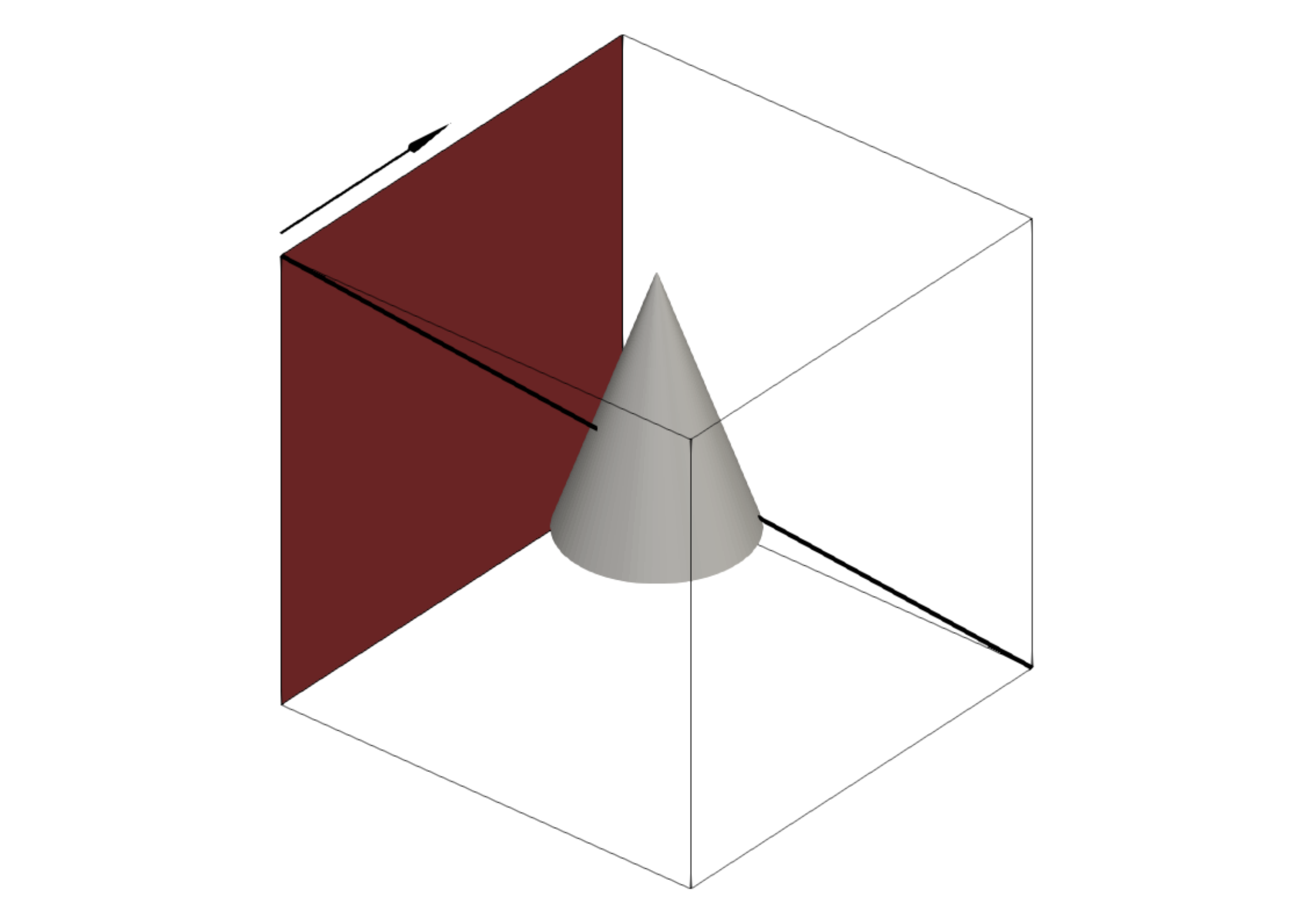}
        \caption{Cone}
        \label{fig:cone_ldc_3d_vis}
    \end{subfigure}
    \begin{subfigure}[b]{0.59\linewidth}
        \centering
        \raisebox{1em}{\createVelocityPlot{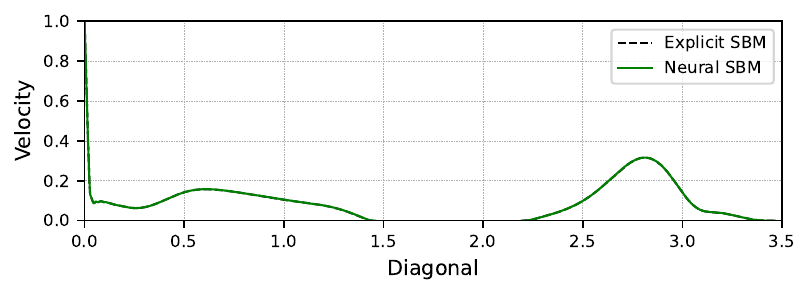}}
        \caption{}
        \label{fig:cone_ldc_plot}
    \end{subfigure}        
    \begin{subfigure}[b]{0.35\linewidth}
        \centering
        \includegraphics[width=\linewidth]{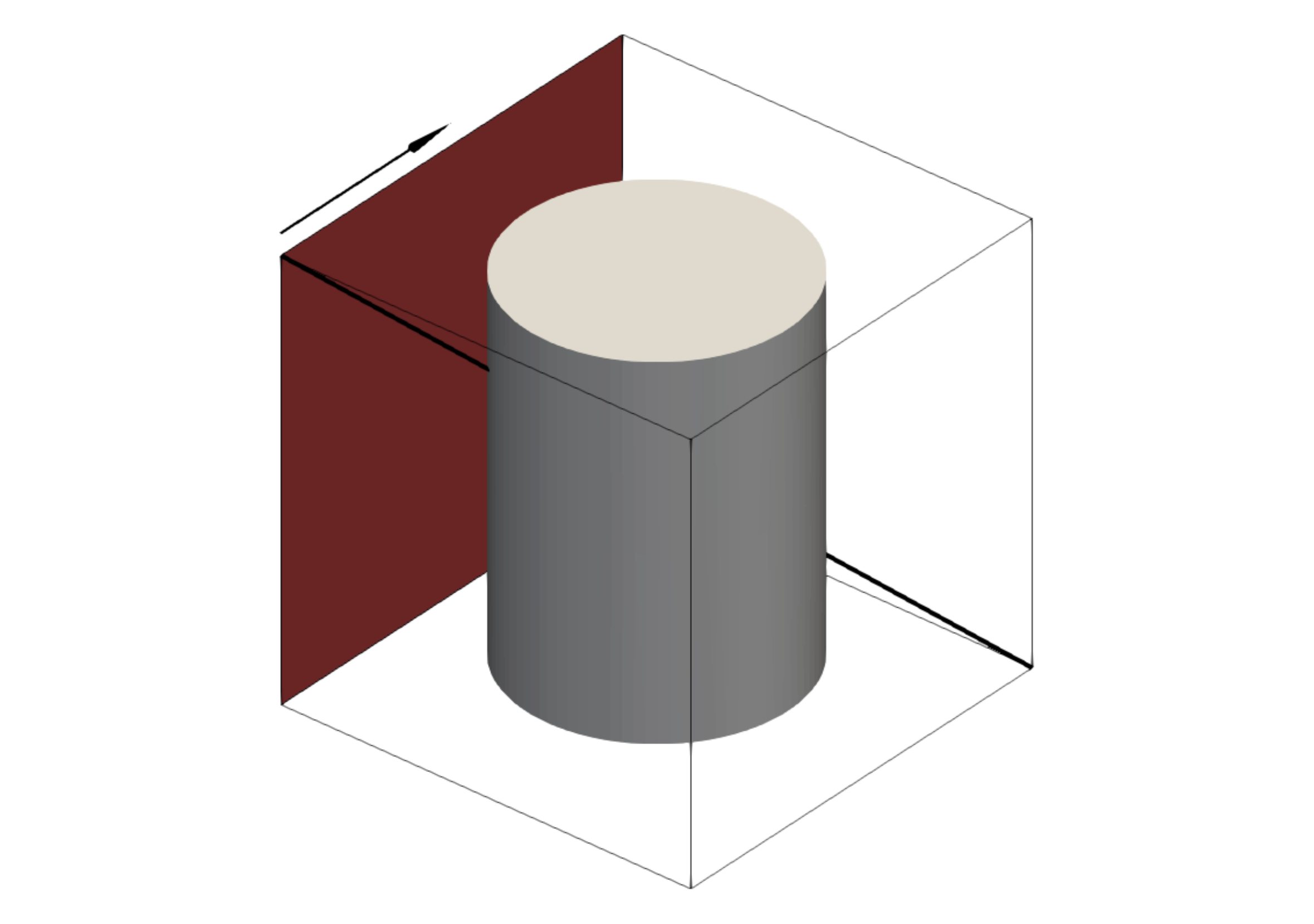}
        \caption{Cylinder}
        \label{fig:cylinder_ldc_3d_vis}
    \end{subfigure}
    \begin{subfigure}[b]{0.59\linewidth}
        \centering
        \raisebox{1em}{\createVelocityPlot{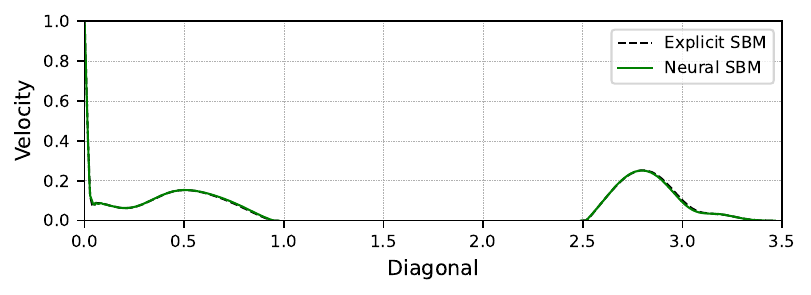}}
        \caption{}
        \label{fig:cylinder_ldc_plot}
    \end{subfigure}
    \begin{subfigure}[b]{0.35\linewidth}
        \centering
        \includegraphics[width=\linewidth]{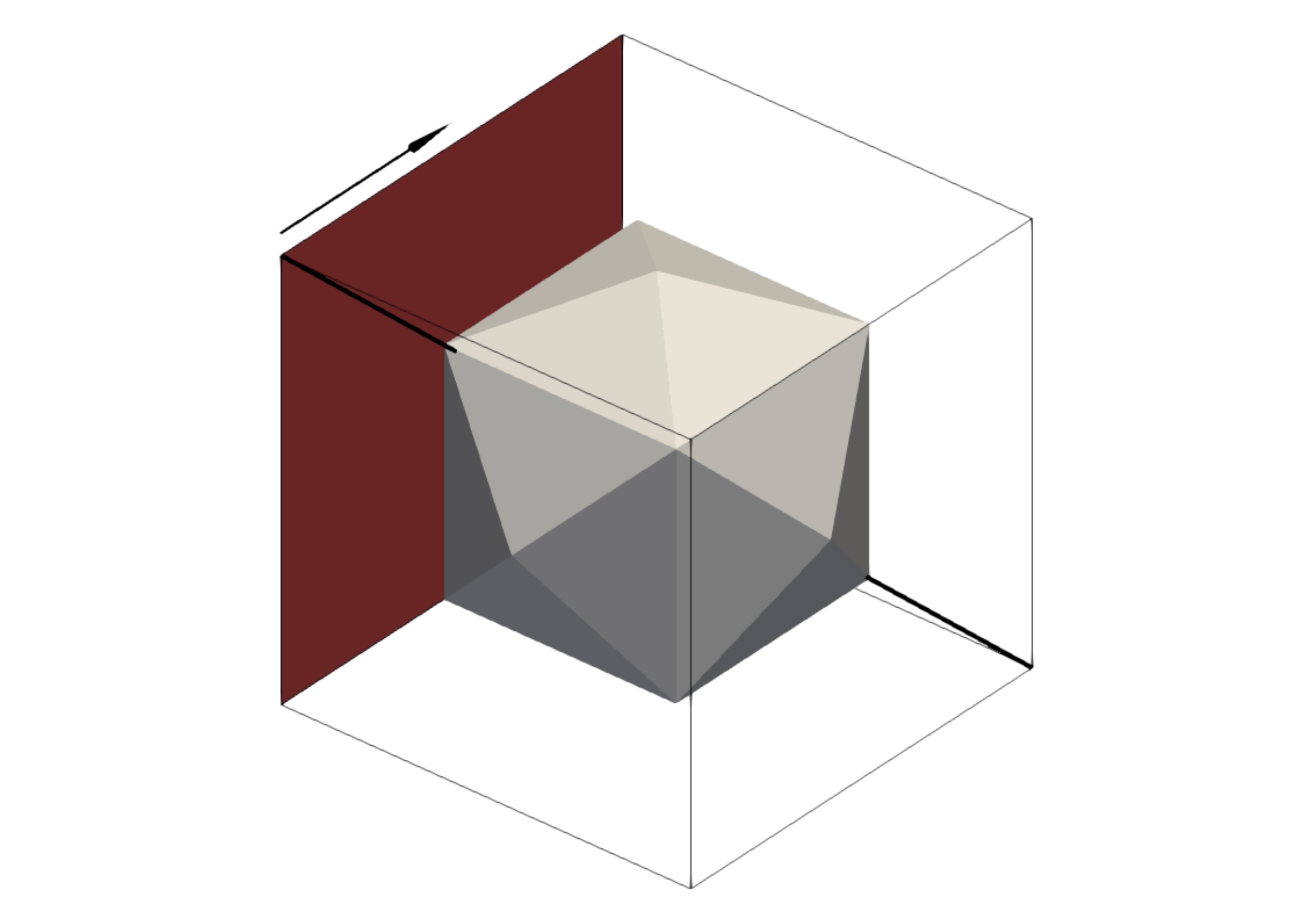}
        \caption{Tetrakis}
        \label{fig:tetrakis_ldc_3d_vis}
    \end{subfigure}
    \begin{subfigure}[b]{0.59\linewidth}
        \centering
        \raisebox{1em}{\createVelocityPlot{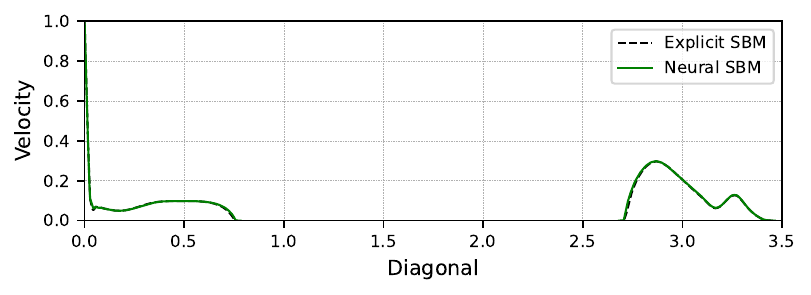}}
        \caption{}
        \label{fig:tetrakis_ldc_plot}
    \end{subfigure}
    \caption{Lid-driven cavity simulations for simple geometries. The geometries are placed inside the cube $[-1,-1,-1]\times [1,1,1]$ with center of mass at $[0,0,0]$. The shaded face is the sliding lid, and the arrow points in the direction of the boundary velocity. Comparison between INR (Neural SBM) and Polygonal Soup (Explicit SBM) is performed by plotting the velocity magnitude along the diagonal highlighted for Sphere ($Re=1800$), Cone ($Re=2000$), Cylinder ($Re=2400$), and Tetrakis ($Re=4000$).}
    \label{fig:3D_LDC_first_set}
\end{figure}

The comparison is made across different Reynolds numbers to present the robustness of the framework. The velocity magnitude is compared across a diagonal passing through the geometry. As shown in \figref{fig:sphere_ldc_3d_vis}, the red face is the sliding lid with the black arrow pointing towards the direction of the slide. The black line passing through the geometry is the diagonal over which the velocity magnitude is plotted. For ease of reference, we introduce the acronyms \emph{Neural SBM} for INR-based SBM analysis and \emph{Explicit SBM} for Polygonal Soup-based analysis. These naming conventions will be consistently used throughout the subsequent text.

\begin{figure}[t!]
    \centering
    \begin{subfigure}[b]{0.35\linewidth}
        \centering
        \includegraphics[width=\linewidth]{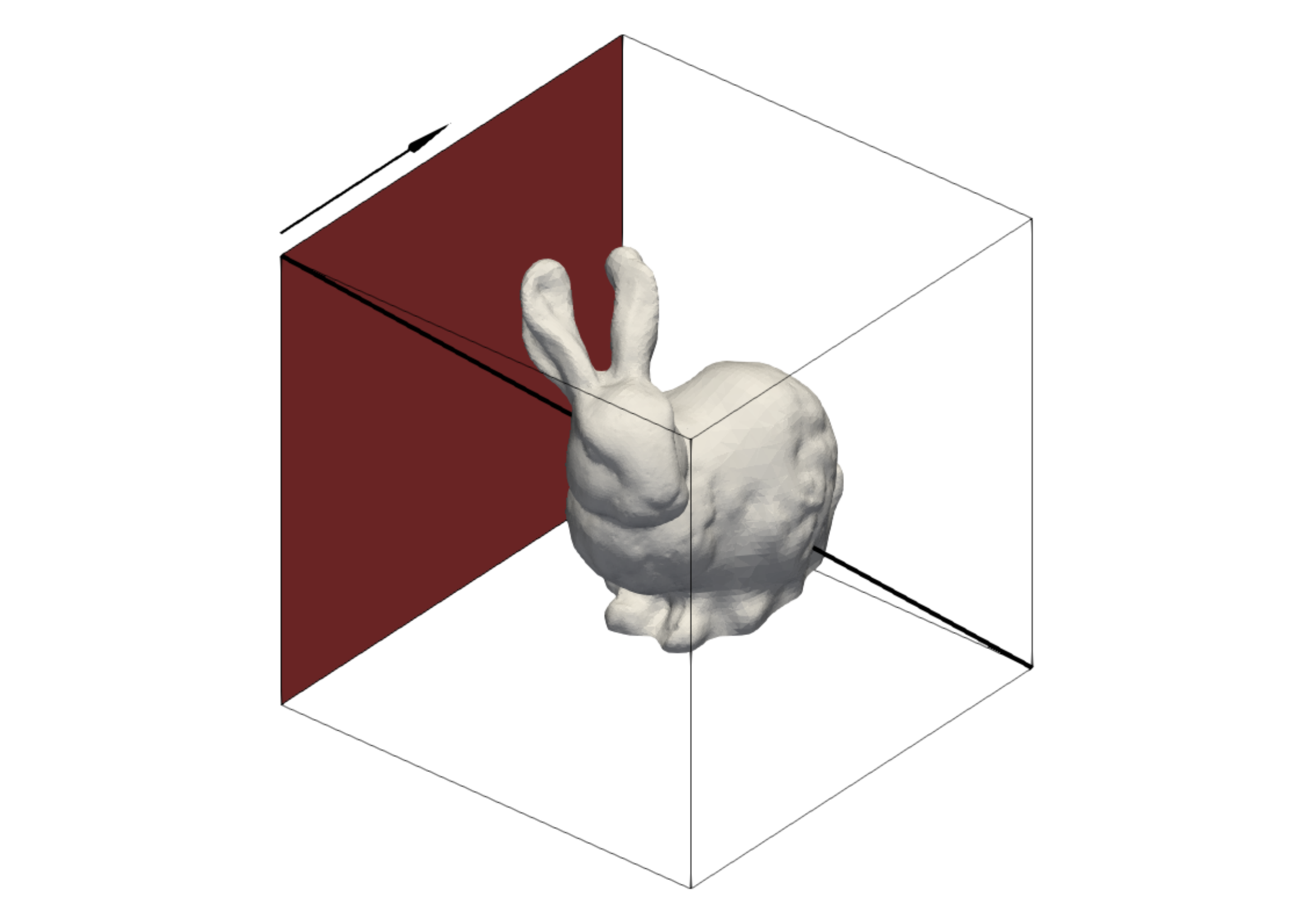}
        \caption{Sphere}
        \label{fig:bunny_ldc_3d_vis}
    \end{subfigure}
    \begin{subfigure}[b]{0.59\linewidth}
        \centering
        \raisebox{1em}{\createVelocityPlot{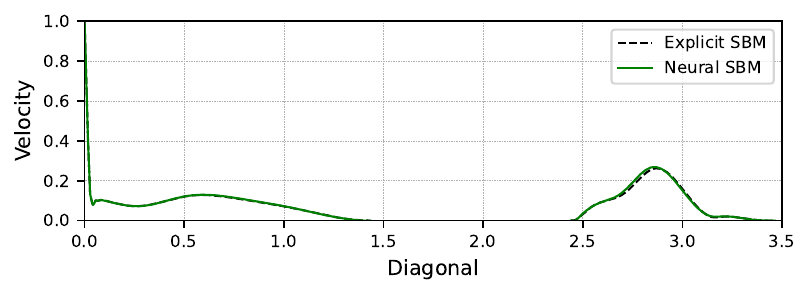}}
        \caption{}
        \label{fig:bunny_ldc_plot}
    \end{subfigure}
    \begin{subfigure}[b]{0.35\linewidth}
        \centering
        \includegraphics[width=\linewidth]{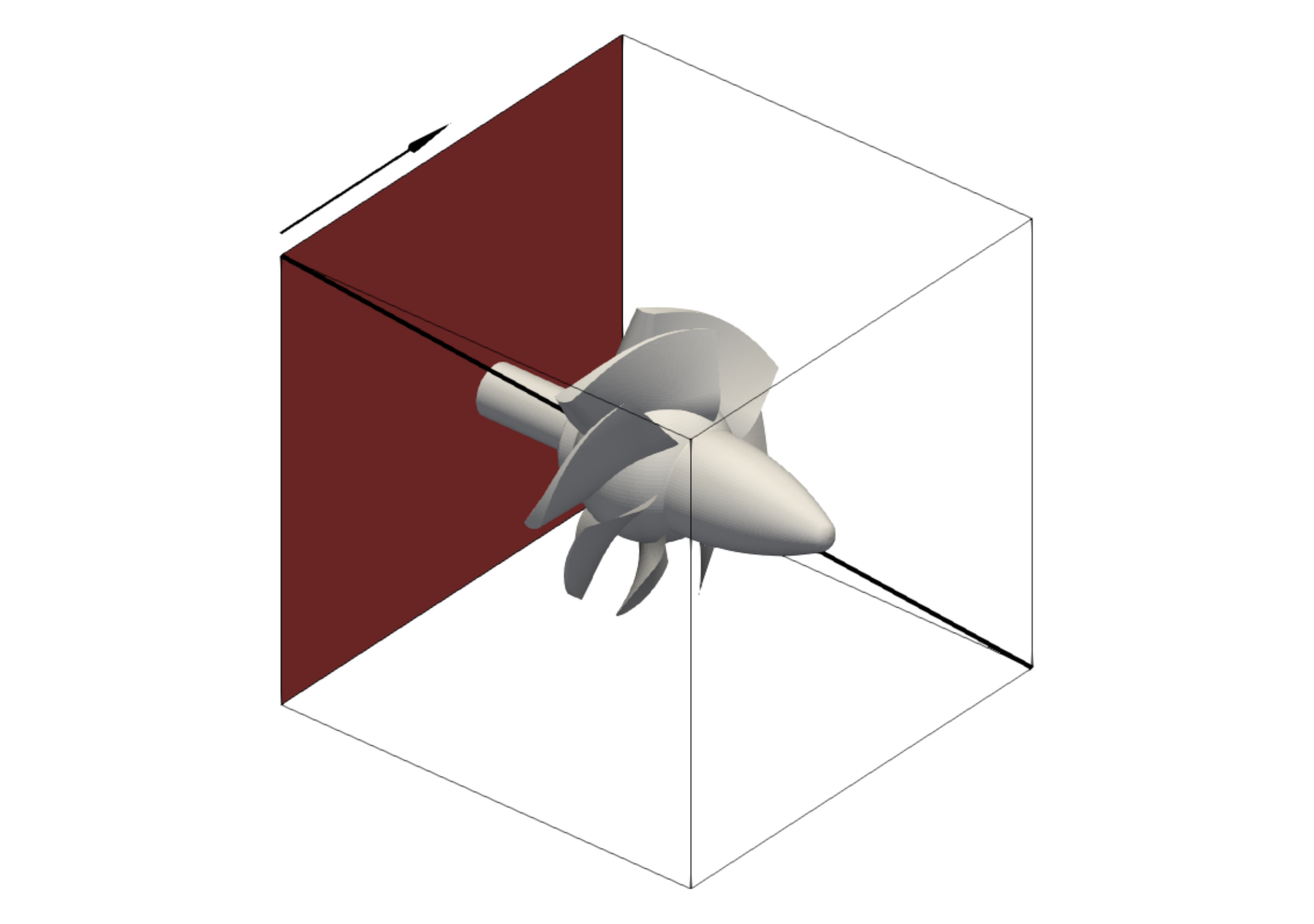}
        \caption{Sphere}
        \label{fig:turbine_ldc_3d_vis}
    \end{subfigure}
    \begin{subfigure}[b]{0.59\linewidth}
        \centering
        \raisebox{1em}{\createVelocityPlot{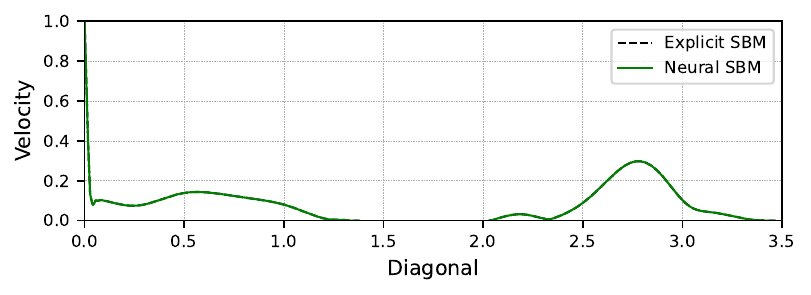}}
        \caption{}
        \label{fig:turbine_ldc_plot}
    \end{subfigure}
    \caption{Lid driven cavity simulations for complex geometries Bunny ($Re=1800$) and Turbine ($Re=1800$). The setup is the same as in \figref{fig:3D_LDC_first_set}.}
    \label{fig:3D_LDC_second_set}
\end{figure}

\figref{fig:3D_LDC_first_set} and \figref{fig:3D_LDC_second_set} compare the velocity magnitude along the diagonal of the cube.~\figref{fig:3D_LDC_first_set} contains simple geometries as presented in \secref{section:Val_Implicit}. The comparison is made for Sphere ($Re=1800$), Cone ($Re=2000$), and Cylinder ($Re=2400$) as shown in \figref{fig:3D_LDC_first_set}. A similar comparative analysis is performed for the Bunny ($Re=1800$), Tetrakis ($Re=4000$), and Turbine ($Re=1800$) geometries, as shown in \figref{fig:3D_LDC_second_set}. There is consistency in the velocity magnitude profiles between the INR and the Polygonal Soup, as presented in the figures depicted in solid-green for \emph{Neural SBM} and dotted-black for \emph{Explicit SBM}. This validation confirms that the INR can be used in place of the Polygonal Mesh for the purpose of the analysis performed with SBM.

\begin{figure}[t!]
    \centering
    \begin{subfigure}[b]{0.33\linewidth}
        \centering
        \includegraphics[width=\linewidth]{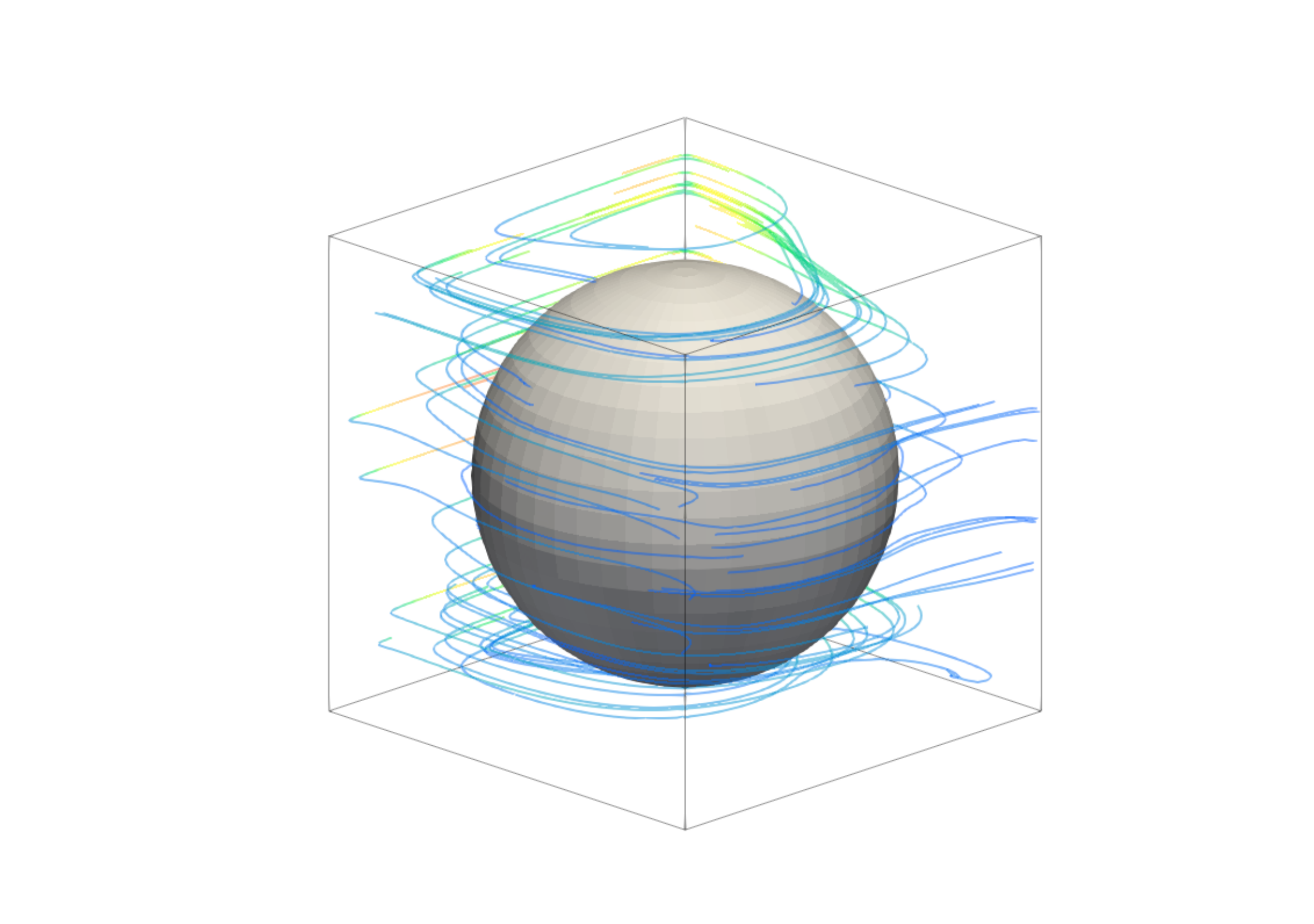}
        \caption{Sphere}
        \label{fig:sphere}
    \end{subfigure}
    \begin{subfigure}[b]{0.33\linewidth}
        \centering
        \includegraphics[width=\linewidth]{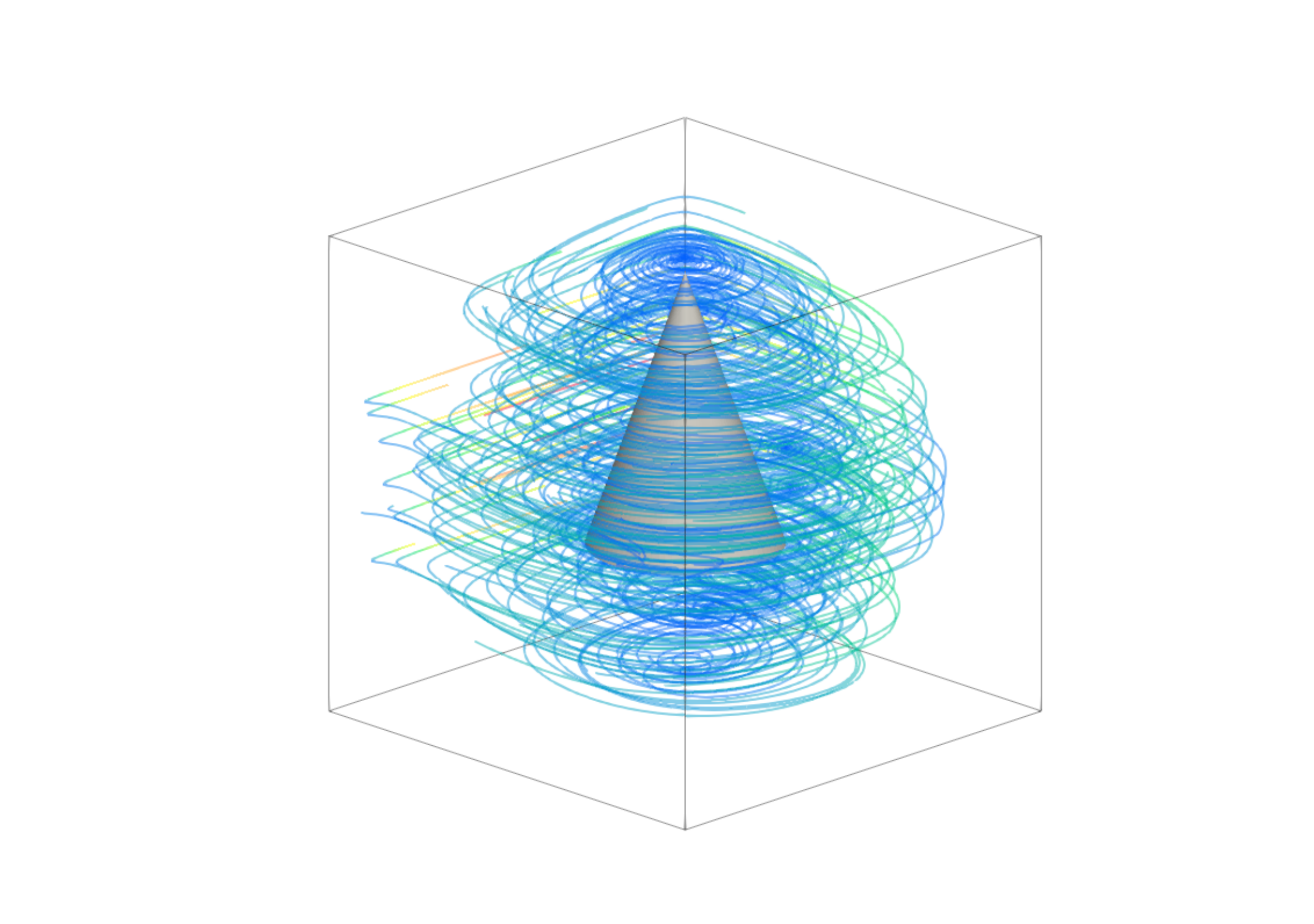}
        \caption{Cone}
        \label{fig:cone}
    \end{subfigure}
    \begin{subfigure}[b]{0.33\linewidth}
        \centering
        \includegraphics[width=\linewidth]{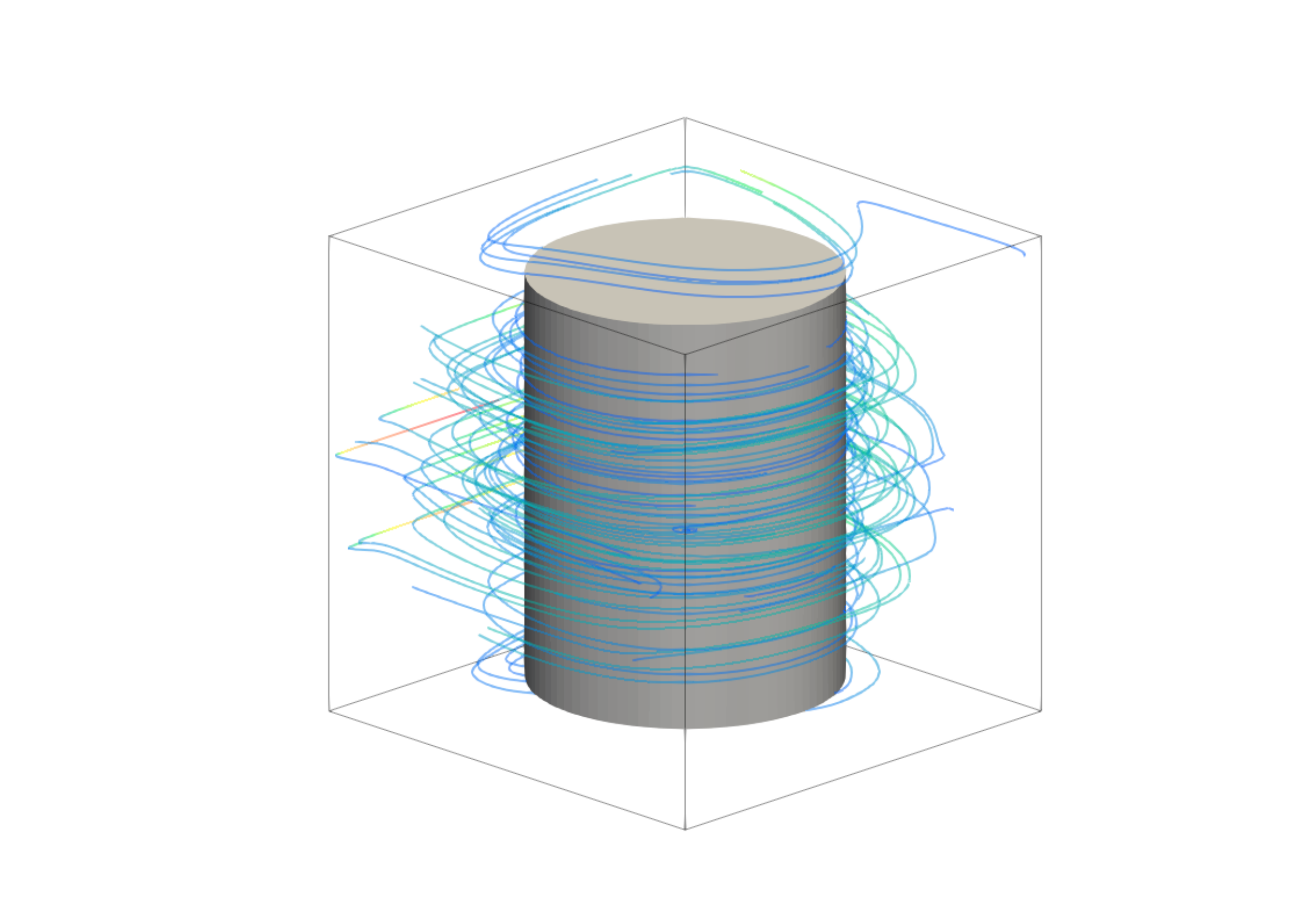}
        \caption{Cylinder}
        \label{fig:cylinder}
    \end{subfigure}
    \begin{subfigure}[b]{0.33\linewidth}
        \centering
        \includegraphics[width=\linewidth]{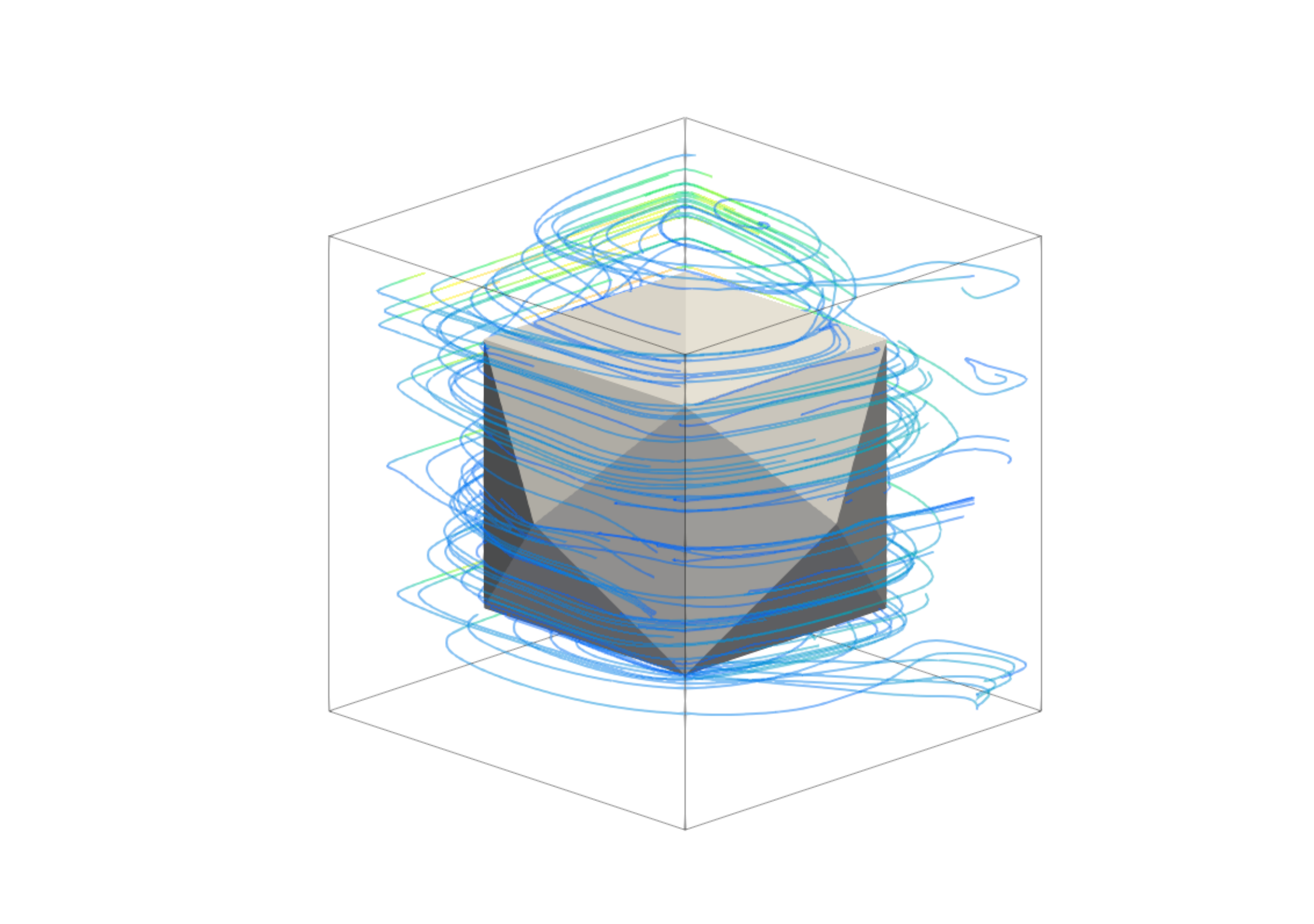}
        \caption{Tetrakis}
        \label{fig:tetrakis}
    \end{subfigure}
    \begin{subfigure}[b]{0.33\linewidth}
		\centering
		\includegraphics[width=\linewidth]{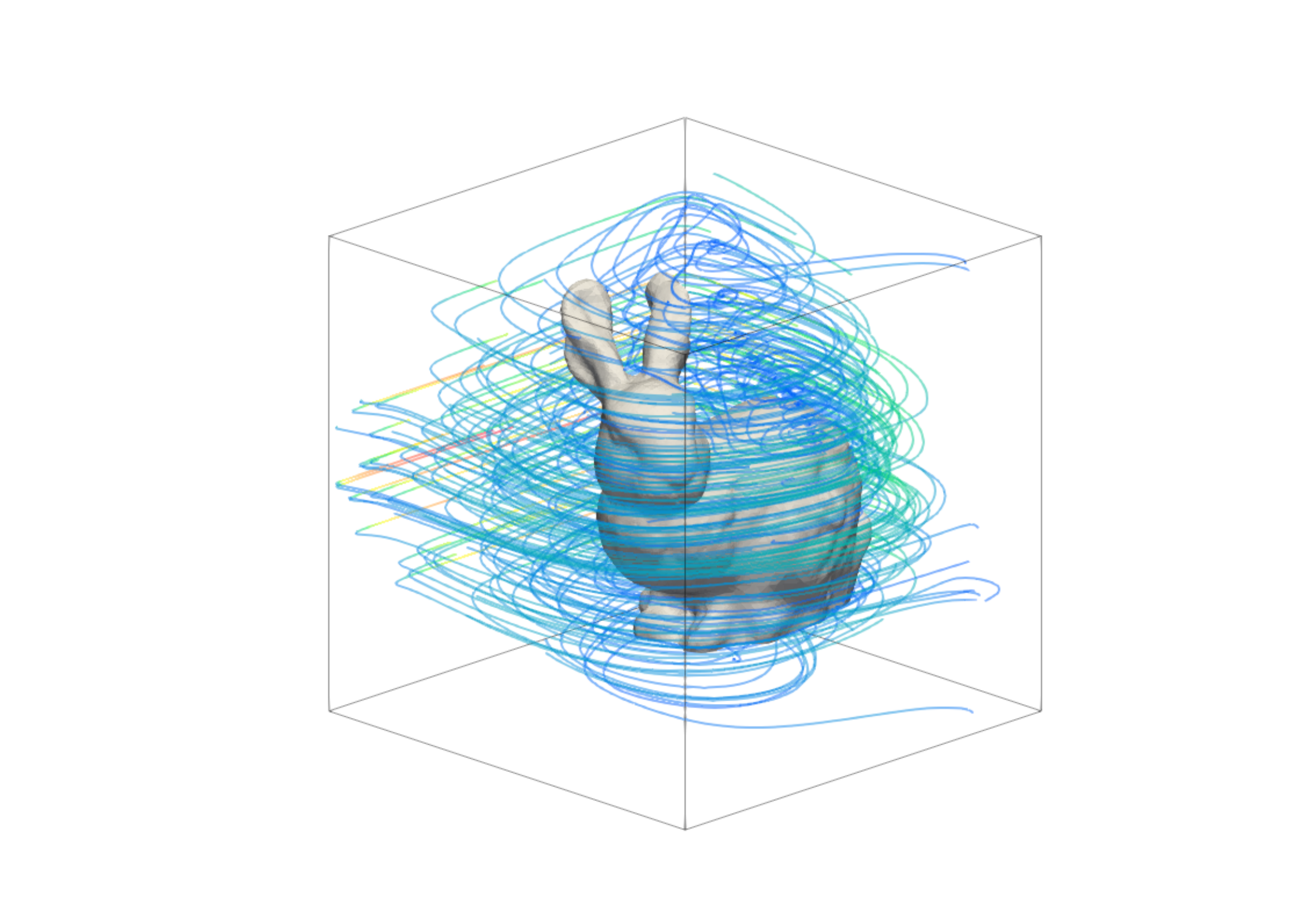}
		\caption{Bunny}
		\label{fig:bunny}
	\end{subfigure}
    \begin{subfigure}[b]{0.33\textwidth}
        \centering
        \includegraphics[width=\linewidth]{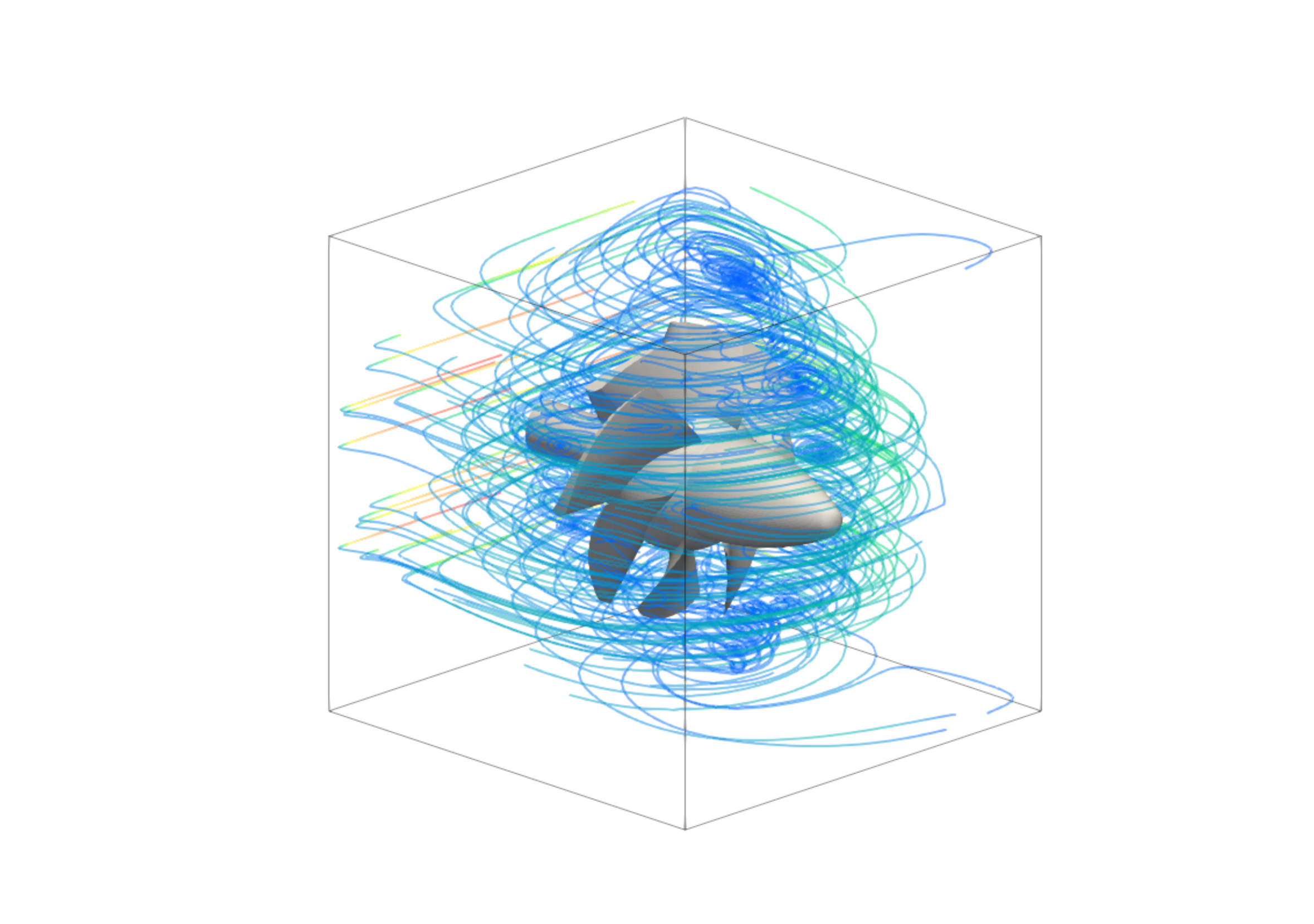}
        \caption{Turbine}
        \label{fig:turbine}
    \end{subfigure}
    \includegraphics[width=0.8\textwidth,trim=0.0in 9.0in 0.0in 0.1in, clip]{Figures_Compressed/LDC_Legend.pdf}    
    \caption{3D Streamline visualizations where the color of the streamline represents the magnitude of the velocity. The first row has Sphere ($Re=1800$), Cone ($Re=2000$), and Cylinder ($Re=2400$). The second row has Bunny ($Re=1800$), Tetrakis ($Re=4000$), and Turbine ($Re=1800$) }
    \label{fig:streamlines_3D_LDC}
\end{figure}

\figref{fig:streamlines_3D_LDC} demonstrates the velocity streamlines for all the cases with INR. The first row in the figure includes streamlines for Sphere($Re=1800$), Cone($Re=2000$), Cylinder($Re=2400$), and Tetrakis ($Re=4000$). Similarly, the second row in the figure includes streamlines for Bunny ($Re=1800$) and Turbine ($Re=1800$). The streamlines demonstrate intricate flow behavior in a Lid-driven Cavity setup with shapes of various complexities.

\clearpage

\subsection{Internal flow in a Gyroid}
\label{section:gyroid_flow}

\begin{figure}[b!]
    \centering
    \begin{subfigure}[b]{0.45\linewidth} 
        \centering
        \includegraphics[width=\linewidth]{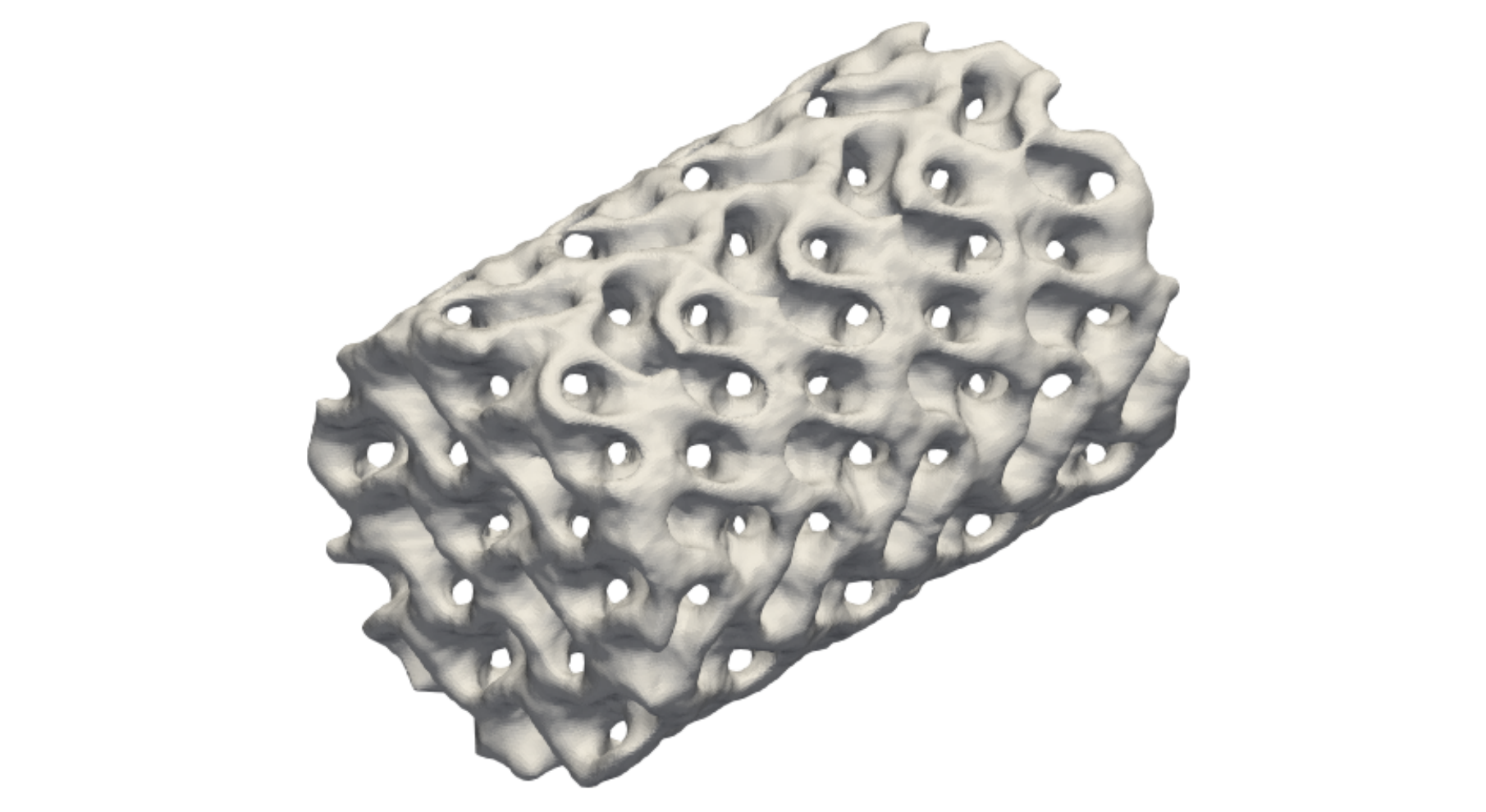}
        \caption{Gyroid Pipe reconstructed using marching cubes ($256^3$).}
        \label{fig:gyroid_reconstructed}
    \end{subfigure}\\
    \begin{subfigure}[b]{0.45\textwidth} 
        \centering
        \includegraphics[width=\linewidth]{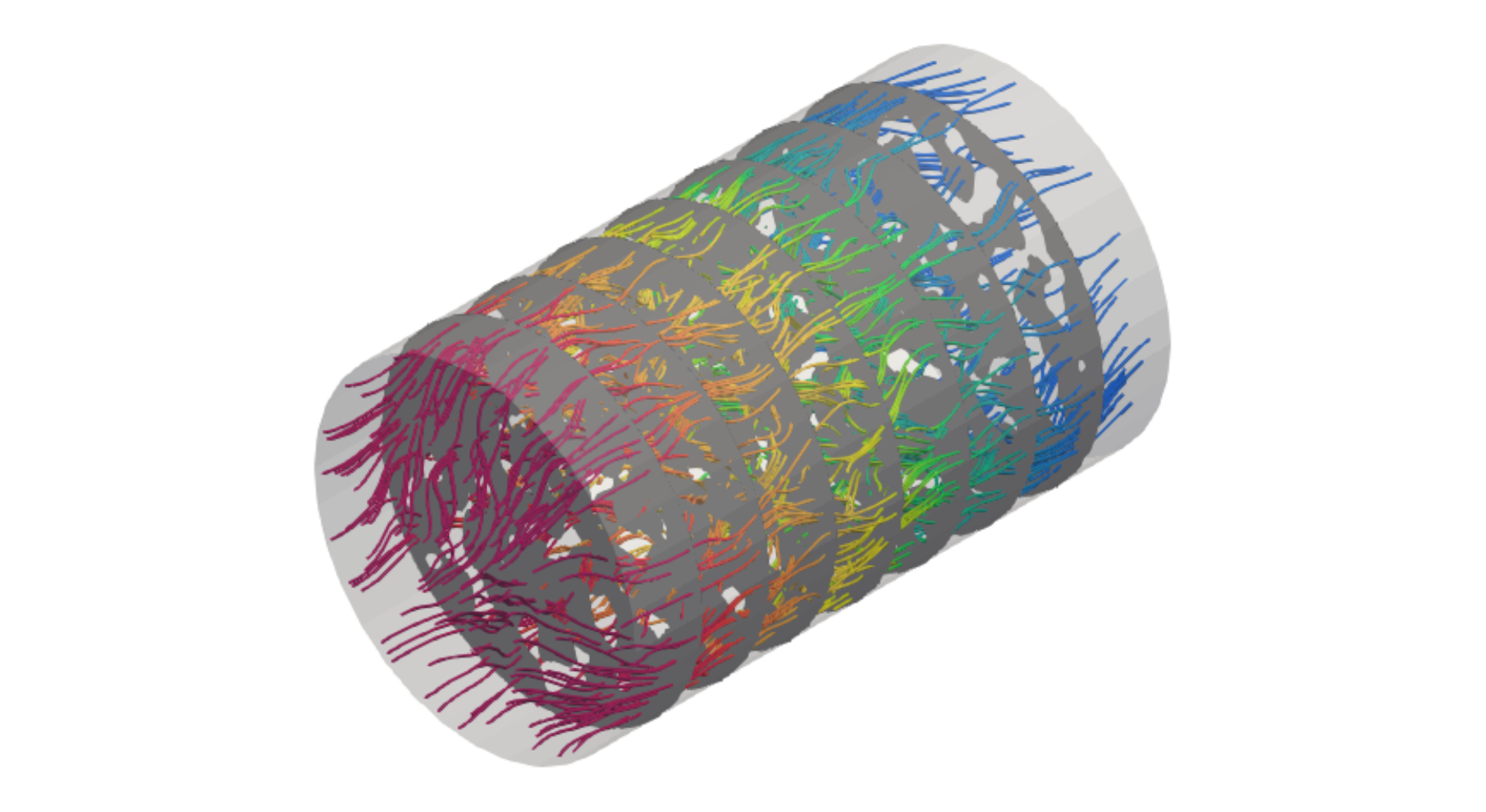}
        \caption{Gyroid flow showing geometry cross-sections.}
        \label{fig:gyroid_flow_alone}
    \end{subfigure}
    \hspace{0.05\linewidth}
    \begin{subfigure}[b]{0.45\textwidth} 
        \centering
        \includegraphics[width=\linewidth]{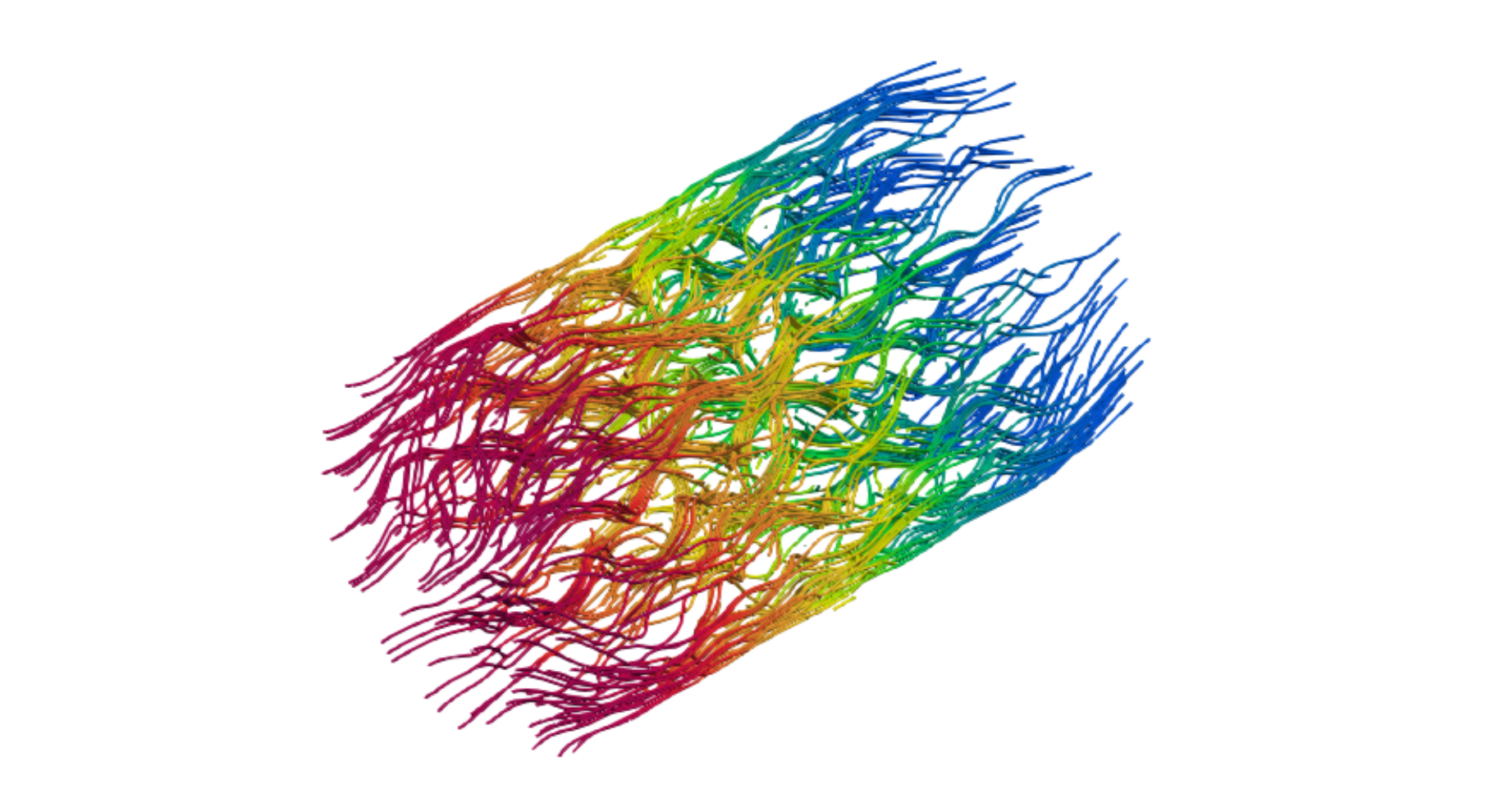}
        \caption{Streamlines of flow inside the Gyroid.}
        \label{fig:gyroid_flow_no_geometry}
    \end{subfigure}
    \begin{subfigure}[b]{1.0\textwidth} 
        \centering
        \includegraphics[width=\textwidth, trim={90 400 90 0}, clip]{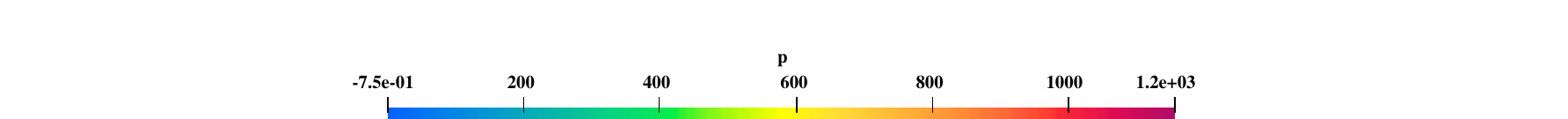}
    \end{subfigure}
    \caption{(a) Gyroid Pipe reconstructed using marching cubes ($256^3$). The figures visualize intricate geometry like the gyroid represented by INR. (b) Gyroid flow visualizations with slices of the domain. The flow is pressure-driven, and streamlines are colored according to the pressure magnitude. (c) Gyroid flow visualizations without slices. Streamlines are colored based on pressure.}
    \label{fig:gyroid_combined}
\end{figure}

Next, we perform a highly complex internal flow simulation using INR for a Gyroid shape~\citep{schoen1970infinite}. We illustrate the reconstruction of the Gyroid using INR in \figref{fig:gyroid_reconstructed}, showcasing the results obtained from our training algorithm detailed in \algoref{Algorithm 1 Implicit Network Training}.

For the flow analysis, we place the Gyroid inside a cylindrical domain with a length of 2 and a radius of 0.6. This arrangement allows us to thoroughly investigate fluid dynamics within this confined geometry. To accurately represent physical phenomena, we apply appropriate boundary conditions. In the cylinder's lateral surface and in the entire gyroid surface, we apply a no-slip boundary condition. At the left end of the cylinder, we apply a free-slip condition ($u_x=1$), allowing the fluid to move without resistance and facilitating unimpeded flow at that boundary. We create a pressure-driven flow scenario by applying zero pressure at the outlet.

The simulation is performed with a boundary refinement level of \( h = \nicefrac{1}{2^9} \), providing a high-resolution representation of the geometry. A Reynolds number of \( Re = 10 \) is selected, indicating laminar flow conditions that are appropriate for this level of detail. The time step is set to \( dt = 0.25 \), and the simulation is conducted until convergence is achieved. This careful setup ensures that the resulting flow patterns are stable and accurately reflect the underlying physics of the system.

The flow dynamics within the Gyroid structure are depicted in \figref{fig:gyroid_flow_alone}, which demonstrates the streamlines for the flow inside the Gyroid alongside slices of the flow domain. This visualization effectively captures the intricate flow behavior, highlighting how the unique geometry of the Gyroid influences the fluid motion. Additionally, \figref{fig:gyroid_flow_no_geometry} presents only the streamlines, providing a clear representation of the flow paths without the geometric context.



\subsection{Flow past bunny in a pipe}
\label{section:bunny_pipeflow}

The Stanford bunny is a very popular complex geometric object often used to evaluate the ability to handle complex objects \citep{turk1994zippered}. Here, we analyze flow past Stanford bunny represented by INRs in pipe flow boundary conditions. We apply a parabolic velocity profile at the inlet, which is typical for pipe flow scenarios. At the outlet of the domain, a zero pressure boundary condition is imposed to allow for the appropriate pressure-driven flow dynamics to develop.

The flow simulation is conducted at a Reynolds number of \( Re = 75 \), which represents a regime characterized by laminar flow conditions. The bunny's center of mass is positioned at the coordinates \([0, 0, 0]\) within a computational domain defined by the size \([-5, -4, -4] \times [10, 4, 4]\). This domain configuration provides ample space to analyze the fluid behavior as it interacts with the geometry of the bunny.

\begin{figure}[b!]
    \centering
    \includegraphics[width=0.7\linewidth]{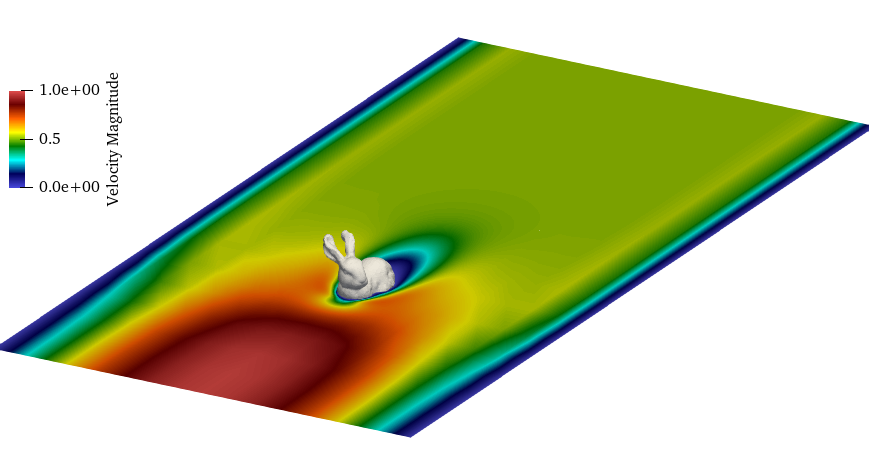}
    \caption{Figure demonstrates slice of velocity magnitude for bunny inside a pipe flow. The inlet has a parabolic velocity boundary condition, which is evident from no slip at the walls and high velocity at the center. The outlet is at pressure = 0.}
    \label{fig:bunny_slice}
\end{figure}

\begin{figure}[t!]
    \centering
    \includegraphics[width=0.8\linewidth]{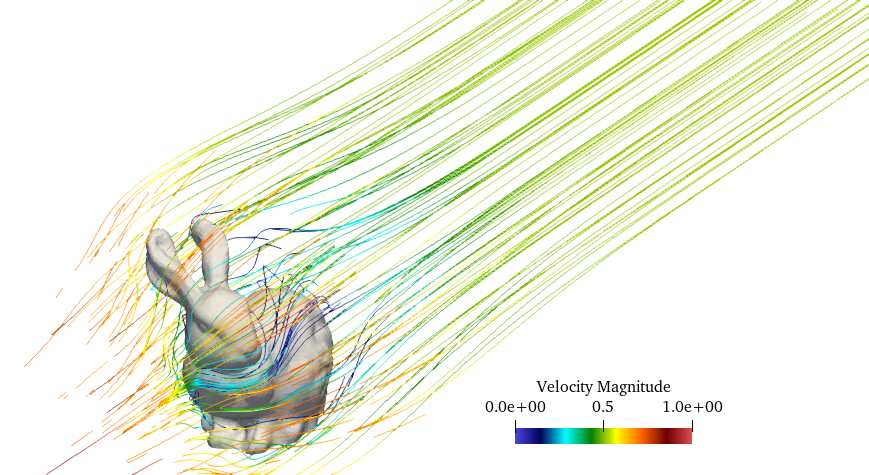}
    \caption{Streamlines for flow past bunny in the pipeflow. The streamlines are colored as per the velocity magnitude. The streamlines detail the intricate flow pattern around the bunny ears.}
    \label{fig:bunny_streamline}
\end{figure}

To accurately capture the flow features around the bunny, a boundary refinement strategy is employed. The region surrounding the bunny is refined at level 9, allowing for a detailed resolution of the intricate flow patterns that arise due to the geometry. Additionally, a spherical region centered at the origin with a radius of 1.2 is also refined at level 9. This refinement ensures that the flow characteristics near the bunny are captured with high fidelity. Furthermore, a cylindrical region extending from \([0, 0, 0]\) to \([8.0, 0.0, 0.0]\), with a radius of 1.2, is refined at level 8. This cylindrical region serves to model the flow along the pipe while maintaining sufficient resolution to observe how the bunny's shape influences the flow field. We perform the simulation with a time step of \( dt = 0.025 \) until a steady state.

\figref{fig:bunny_slice} displays a slice of the velocity magnitude within the flow domain, providing valuable insight into the flow characteristics as it interacts with the bunny shape. The visualization clearly indicates the presence of a stagnant region behind the bunny, where the velocity significantly decreases. This stagnation occurs as the fluid encounters the bunny's geometry, resulting in a local disruption of the flow field. The flow's inability to maintain its velocity in this area is indicative of the complex interactions between the fluid and the solid boundary, which can have significant implications for the overall flow dynamics, including potential impacts on pressure distribution and turbulence generation downstream.

In addition to the stagnant region, \figref{fig:bunny_streamline} presents the streamlines near the surface of the bunny, illustrating the intricate details of the flow patterns around the geometry. Notably, the streamlines around the bunny's ears reveal complex flow behavior characterized by sharp curvatures and variations in flow direction. These intricate details highlight how the bunny's shape influences the local flow field, causing the fluid to accelerate, decelerate, and change direction as it navigates the contours of the ears. The interplay between the streamlined flow and the geometric features of the bunny serves to underscore the capabilities of our framework in accurately capturing fluid dynamics in the presence of complex shapes.

\subsection{Generative AI Plane}
\label{sec:GenAI_plane}

The increasing popularity of generative AI approaches to create complex geometries provides a compelling case study opportunity. Here, we utilize a pre-trained diffusion-based model that performs diffusion on the neural network weights of INRs to generate a plane~\citep{erkocc2023hyperdiffusion}. We perform postprocessing steps to enhance the quality of the INRs, as detailed in \appendixref{section:preprocessinggenai}, and leverage our framework to perform flow analysis over the INR of the generated model.

\begin{figure}[t!]
    \centering
    \includegraphics[width=0.35\linewidth,clip,trim={3.0in 1.0in 3.0in 1.5in}]{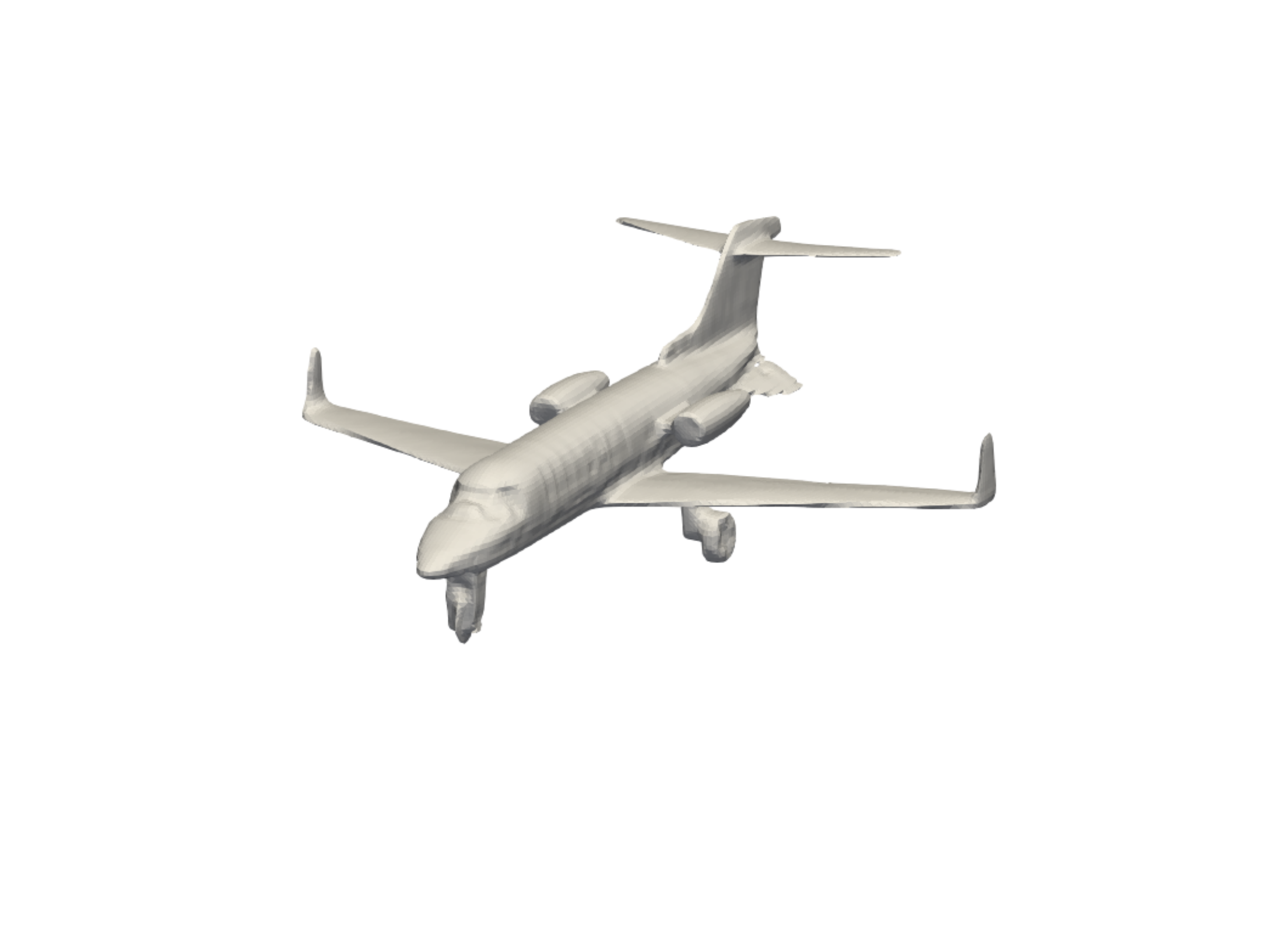}
    \hspace{0.03\linewidth}
    \begin{subfigure}{0.5\linewidth}
        \centering
        \includegraphics[width=\linewidth,clip,trim={3.0in 0.0in 0.0in 0.0in}]{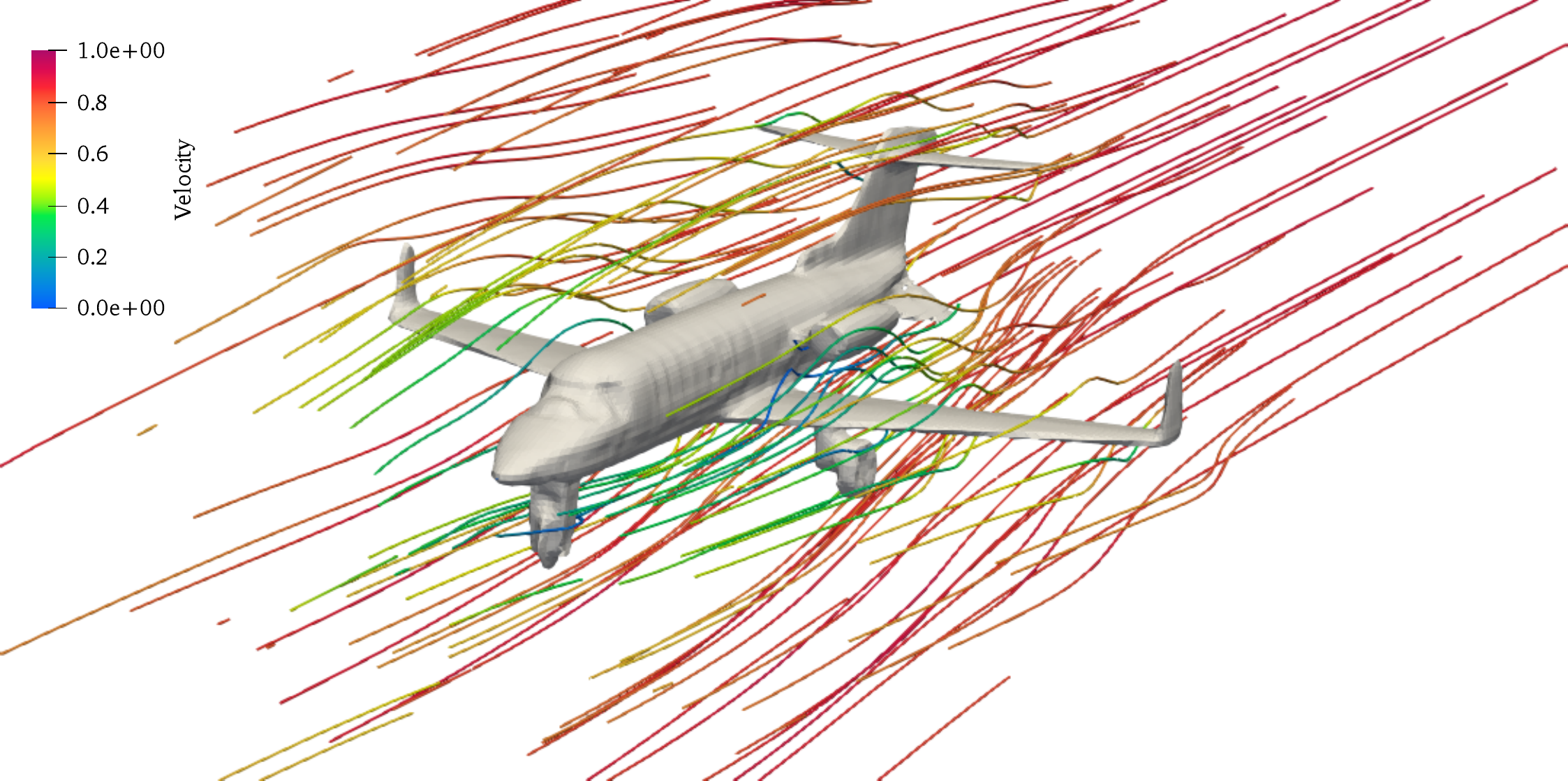}
        \includegraphics[width=1.0\linewidth,trim={0.0in 9.0in 0.0in 0.1in},clip]{Figures_Compressed/LDC_Legend.pdf}
    \end{subfigure}
    
    \caption{The figure demonstrates an unconditionally generated plane as a signed distance field based on \citet{erkocc2023hyperdiffusion}, with preprocessing to make it suitable for analysis. The visualization uses the Marching Cube method. The airplane is situated at the origin within the domain \([-4, -4, -5] \times [4, 4, 3]\). A parabolic boundary condition is applied with a maximum velocity of 1 at the inlet, and the flow is pressure-driven with a Reynolds number of 75.}
    \label{fig:plane}
\end{figure}

The computational domain is \([-4, -4, -5] \times [4, 4, 3]\), with the plane occupying a bounding box \([-0.62, -0.2, -0.65] \times [0.62, 0.23, 0.85]\) centered at the origin. To accurately capture the flow features around the region of interest, we employ a hierarchical boundary refinement strategy. We refine the geometry to level 11. A spherical region centered at \([0.0, 0.0, 0.0]\) with a radius of 0.85 is refined to level 10. Additionally, we refine a cylindrical region extending from \([0.0, 0.0, 0.0]\) to \([0.0, 0.0, -2.5]\) with a radius of 1.0 to level 9. Further downstream, another cylindrical region extending from \([0.0, 0.0, 0.0]\) to \([0.0, 0.0, -4.0]\) with a radius of 1.2 is refined to level 8, ensuring sufficient resolution to observe the downstream influence of the upstream geometry. The parabolic inlet-boundary condition is applied at $z=3$, and pressure outlet $\textit{p}=0$ is applied at $z=-5$. The flow is driven by a Reynolds number of 10000 with a timestep of $dt = 0.25$. \figref{fig:plane} visualizes the streamlines over the plane in colored with velocity magnitude.

\section{Discussion and Conclusions}
\label{sec:Conclusions}

In this work, we have demonstrated that Implicit Neural Representations (INRs) provide a suitable and scalable geometric representation for high-fidelity simulations using the Shifted Boundary Method (SBM). By directly querying the INR, we construct the incomplete octree mesh, surrogate boundary, and distance vector required for the SBM workflow, completely bypassing the need for explicit geometric representations. We have also introduced algorithmic adaptations that enable efficient INR-based simulation, as outlined in \algoref{Algorithm ImplicitOctreeGeneration}, \algoref{Algorithm: SurrogateBoundaryIdentificationUsingImplicitNetwork}, and \algoref{Algorithm: DistanceFunctionCalculationUsingImplicitNetworkOneGP}. These methods allow the direct integration of neural implicit geometries into Navier-Stokes simulations across a range of geometrical complexities, demonstrating both accuracy and robustness in fluid flow analysis.

Our framework is agnostic to the source of the INR, making it adaptable to various data-driven and AI-based geometric acquisition methods. While our validation involved INRs derived from polygonal meshes, our approach extends seamlessly to INRs obtained from:
\begin{itemize}
\item Real-world imaging and scanning methods (e.g., NeuS~\citep{wang2021neus}, IGR~\citep{gropp2020implicit}), allowing direct simulation on image-derived geometries for diagnostic and analysis purposes.
\item Generative AI-based 3D models (e.g., diffusion-based shape generation~\citep{chou2023diffusion,jiang2023sdf,erkocc2023hyperdiffusion}), enabling automated design exploration and evaluation of AI-generated objects.
\item Neural Radiance Fields (NeRFs)\citep{mildenhall2021nerf}, where recent work on truncated signed distance field learning\citep{10550893} could facilitate direct INR-based simulations on NeRF-derived geometries.
\end{itemize}
By eliminating the dependency on explicit mesh generation, this work represents a step toward a more flexible and efficient computational modeling workflow that integrates AI-driven shape representations with rigorous physics-based simulations.

\noindent\textbf{Challenges and Future Directions}: Despite the advantages of INRs, several open challenges remain. These can be broadly divided into challenges in INR representation, and challenges in INR-SBM strategies. These include (a) \textit{Capturing thin structures and complex topologies}: INR representations may struggle with geometries featuring fine details or sharp edges, an area requiring further research. (b) \textit{Handling sparse data}: INRs trained from sparse point clouds may lack sufficient 3D information, limiting their accuracy in representing complex objects. (c)
\textit{Enhancing generative model quality}: While 3D generative AI models are advancing, they still lag in producing high-resolution, physically consistent geometries suitable for detailed analysis, and (d) \textit{Shape editing limitations}: Unlike explicit meshes, modifying or fine-tuning INR-based shapes is non-trivial. Although recent efforts~\citep{yang2021geometry} have explored shape editing for neural fields, further innovations are needed to make INRs as interactive as traditional geometric models.

Addressing these challenges will shape the next generation of INR-driven computational analysis tools, expanding their role in engineering, design, and scientific computing. Our results provide a foundational step toward the seamless integration of AI-generated geometries into high-fidelity simulations, paving the way for faster, more flexible, and data-driven computational science workflows.

\section*{Acknowledgements}
This work is supported by the AI Research Institutes program supported by NSF and USDA-NIFA under AI Institute for Resilient Agriculture, Award No. 2021-67021-35329. We also acknowledge partial support through NSF awards CMMI-2053760 and DMREF-2323716.  We also acknowledge computational resources from the ISU HPC cluster Nova and TACC Frontera.

\clearpage
\newpage
\bibliographystyle{unsrtnat}
\bibliography{refs}

\newpage
\appendix

\begin{center}
{
{\usefont{OT1}{phv}{b}{n}\selectfont\Large{Supplementary Information}}
\vspace{0.5em}
}
\end{center}
\section{Generating INR from Polygonal Meshes}
\label{section:generating_INR_polygon}
This workflow outlines the process of converting a triangular polygonal mesh into a INR. To fulfill the requirements for simulating using SBM, the training process leverages a carefully designed loss function and a strategic sampling approach. The loss function is structured to penalize errors in distance vector magnitude and direction, particularly in regions within the narrowband $\delta$, while also promoting the correct classification of grid points. The sampling strategy prioritizes regions within the narrowband, ensuring that the model learns to represent critical geometric features accurately. This approach ensures that the INR remains compatible with the SBM, resulting in a highly accurate representation that adapts to complex boundary geometries. We discuss below some of the critical aspects of the NN training process for generating the INR.

\begin{enumerate}
    \item \textbf{Geometry Re-scaling}: Define a cubic domain $\boldsymbol{\Omega} = [-1, -1, -1] \times [1, 1, 1]$. All the geometries are rescaled such that the volume occupied by $\Omega^-$ and $\Omega^+$ are in close tolerance for training purposes.
    \item \textbf{Hybrid Sampling}: We use a hybrid sampling method, where points are sampled uniformly on the surface \( \Gamma \), in the narrow band defined by a specified width \( \delta \),and as well as uniformly in cube $\Omega$. \newline
\( P_S \) be the set of points sampled uniformly on the true boundary \( \Gamma \):
\[
P_S = \{ \mathbf{x} \in \Gamma \mid \exists \, \mathbf{u} \in [0, 1]^2, \, \mathbf{x} = \text{Sample}(\mathbf{u}) \}
\]
\( P_{NB} \) be the set of points sampled uniformly within the narrow band around the true boundary \( \Gamma \):
\[
P_{NB} = \{ \mathbf{x} \in \mathbb{R}^n \mid \text{dist}(\mathbf{x}, S) \leq \delta \text{ and } \exists \, \mathbf{u} \in [0, 1]^m,\mathbf{x} = \text{Sample}(\mathbf{u}) \}
\]
\( P_{U} \) be the set of points sampled uniformly in the overall sampling space \( \Omega \):
\[
P_{U} = \{ \mathbf{x} \in \Omega \mid \exists \, \mathbf{u} \in [0, 1]^n, \, \mathbf{x} = \text{Sample}(\mathbf{u}) \}
\]
The total sampled points \( P \) is then defined as the union of these three sampling techniques:
\[
P = P_S \cup P_{NB} \cup P_{U}
\]
The number of points taken from each set can be controlled to balance the sampling strategy based on the requirements of the problem. The ablation study of impact of varying, $\boldsymbol{n (P_{S})}$,$\boldsymbol{n (P_{U})}$, and $\boldsymbol{n (P_{NB})}$ is presented in \appendixref{appendix:Neural Implicit}.

\item  
\textbf{Loss Function}: The loss function is defined in \eqnref{equation:loss_function}, which takes the location of point $\mathbf{x}$, prediction $f_{\theta}(x)$, true distance $\mathbf{s}$, and true normal $\hat{n}$. The loss function uses clamped loss, which ensures that the distance near the narrow band, given by the width \(\delta\), is given more priority. This approach helps to stabilize the training process by focusing on the values of \(\mathbf{s}\) that are within a certain proximity to the true boundary $\mathbf{\Gamma}$. Similarly, the eikonal constraint and the property of INR as described in \eqref{eikonal} which is termed as Geometric Regularization Loss in \citet{gropp2020implicit} is applied wherever $|\mathbf{s}|<\omega$, where $\omega$ is a region close to the true boundary $\mathbf{\Gamma}$.
\begin{equation}
\begin{aligned}
    L(f_\theta(\mathbf{x}), \mathbf{x}, \mathbf{s}, \hat{\mathbf{n}}) = & \int_{\Omega} \left( \text{clamp}(\mathbf{s}, \delta) - \text{clamp}(f_\theta(\mathbf{x}), \delta) \right)^2 \, d\mathbf{\Omega} \\
    & + \begin{cases}
        \lambda_g \int_{\Omega} \left( \left\| \nabla_{\mathbf{x}} f_\theta(\mathbf{x}) \right\| - 1 \right)^2 \, d\mathbf{\Omega} + \tau \int_{\Omega} \left( \frac{\nabla_{\mathbf{x}} f_\theta(\mathbf{x})}{\left\| \nabla_{\mathbf{x}} f_\theta(\mathbf{x}) \right\|} \cdot \hat{\mathbf{n}}(\mathbf{x}) - 1 \right)^2 \, d\mathbf{\Omega} & \text{if } |\mathbf{s}| < \omega \\
        0 & \text{otherwise}
    \end{cases}
\end{aligned}
\label{equation:loss_function}
\end{equation}
Here, $\lambda_g$ and $\tau$ are the Lagrange multipliers (hyperparameters) for the eikonal constraint and the normal similarity constraint, respectively. \appendixref{appendix:Neural Implicit}presents the ablation study of the above loss function as well as a comparison of the given loss function with other plausible loss functions.
\item \textbf{Network Architecture}: The foundational architecture used in INR is Multi-layer perceptron. The work here compares a Fully Connected Network (eight hidden layers with 512 neurons each) with an Implicit Net of the same size with one skip-in layer. The Implicit Net, which was proposed in \citet{park2019deepsdf} as an Auto-Decoder network and used by~\citep{wang2021neus,sitzmann2020implicit,gropp2020implicit} performs well in comparison to Fully Connected Network as presented in \cref{Appendix}{appendix:Neural Implicit}.
\item \textbf{Evaluation Metric}:
\label{section:Evaluation_Metric}
Evaluating the network for the particular task at hand is very crucial. The network should perform well in near boundary regions, but how close to the boundary is a hyper-parameter, which would depend on the computational discretization size we want for our computational analysis. In this analysis, we obtain a 3D grid of size $1024^3$ in the bounding box and select the points near the boundary region given by a threshold($\delta$). Then, Normalized Mean Squared Error is obtained to analyze the performance of the network as given by \eqnref{nmse}.
\begin{equation}
    \text{NMSE}_\delta = \frac{\frac{1}{N}\sum_{i=1}^{N} (s_i - f_\theta(x_i))^2}{\Delta}
    \label{nmse}
\end{equation}
where \( |y_i| < \delta \), and \( \Delta \) is the characteristic dimension.
\end{enumerate}

\algoref{Algorithm 1 Implicit Network Training} presents the full training pipeline where a polygonal soup is taken as input and converted into respective INR. Hybrid Sampling and the loss function are described to train the architecture as explained to obtain appropriate INR of a particular shape. In this work, $target(p)$ and $normal(\hat{n})$ are obtained using libgl package~\citep{libigl}. The INRs are evaluated on Normalized Mean Square error as described in \secref{section:Evaluation_Metric}. \cref{Appendix}{appendix:Neural Implicit} has more detailed results on the different experiments that were performed.

\begin{algorithm}[h!]
  \footnotesize
    \caption{\textsc{ImplicitNetworkTraining:} Obtain Implicit Network from a Polygonal Soup}
    \label{Algorithm 1 Implicit Network Training}
    \begin{algorithmic}[1]
\Require Polygonal soup $P$, Bounding box $B$, Number of samples $N$, Implicit Neural Network $f_\theta$, Loss function $L(f_\theta, s)$
\Ensure Trained neural network $f_\theta$
\State Initialize a list $\mathcal{P}_U$ for sampled points within the bounding box

\For{$i = 1$ to $N$}
    \State $p_i \gets \text{sample point uniformly from } B$
    \State $\mathcal{P}_U \gets \mathcal{P}_U \cup p_i$ \Comment{Store uniform samples in bounding box}
\EndFor

\State Initialize a list $\mathcal{P}_S$ for sampled surface points
\For{each polygon $\triangle \in P$}
    \State Sample barycentric coordinates $(u, v, w)$ where $u + v + w = 1$ and $u, v, w \geq 0$
    \State Project point $q$ onto the surface of polygon $\triangle$ using barycentric coordinates
    \State $\mathcal{P}_S \gets \mathcal{P}_S \cup q$ \Comment{Store surface samples}
\EndFor

\State Define a narrow band around the surface with distance threshold $\delta$
\For{each point $q \in \mathcal{P}_S$}
    \State Sample points $q_{\text{band}}$ around $q$ within distance $\delta$
    \State $\mathcal{P}_{\text{NB}} \gets \mathcal{P}_{\text{NB}} \cup q_{\text{band}}$ \Comment{Store points in narrow band}
\EndFor

\State Combine all sampled points: $\mathcal{P} = \mathcal{P}_S \cup \mathcal{P}_U \cup \mathcal{P}_{\text{NB}}$
\For{each point $p \in \mathcal{P}$}
    \State Pass $p$ to the implicit neural network $f_\theta(p)$
    \State Compute the loss $L(f_\theta, \text{target}(p))$ \Comment{Calculate loss for each point}
\EndFor

\While{not converged}
    \State Update parameters $\theta \gets \theta - \eta \nabla_\theta L(f_\theta, \text{targets})$ \Comment{Optimize network parameters}
\EndWhile

\State \Return Trained implicit network $f_\theta$
\end{algorithmic}
\end{algorithm}

\clearpage

\section{Ablation Studies on INR Generation for Polygonal Meshes}
\label{appendix:Neural Implicit}

The sphere is one of the most computationally analyzed geometry. The experiments, until specifically specified, take 100K uniformly sampled points, 25K points in narrowband with width $\delta = 0.001$, and 25K points in the surface.

\begin{figure}[h]
    \centering
    \includegraphics[width=0.2\linewidth]{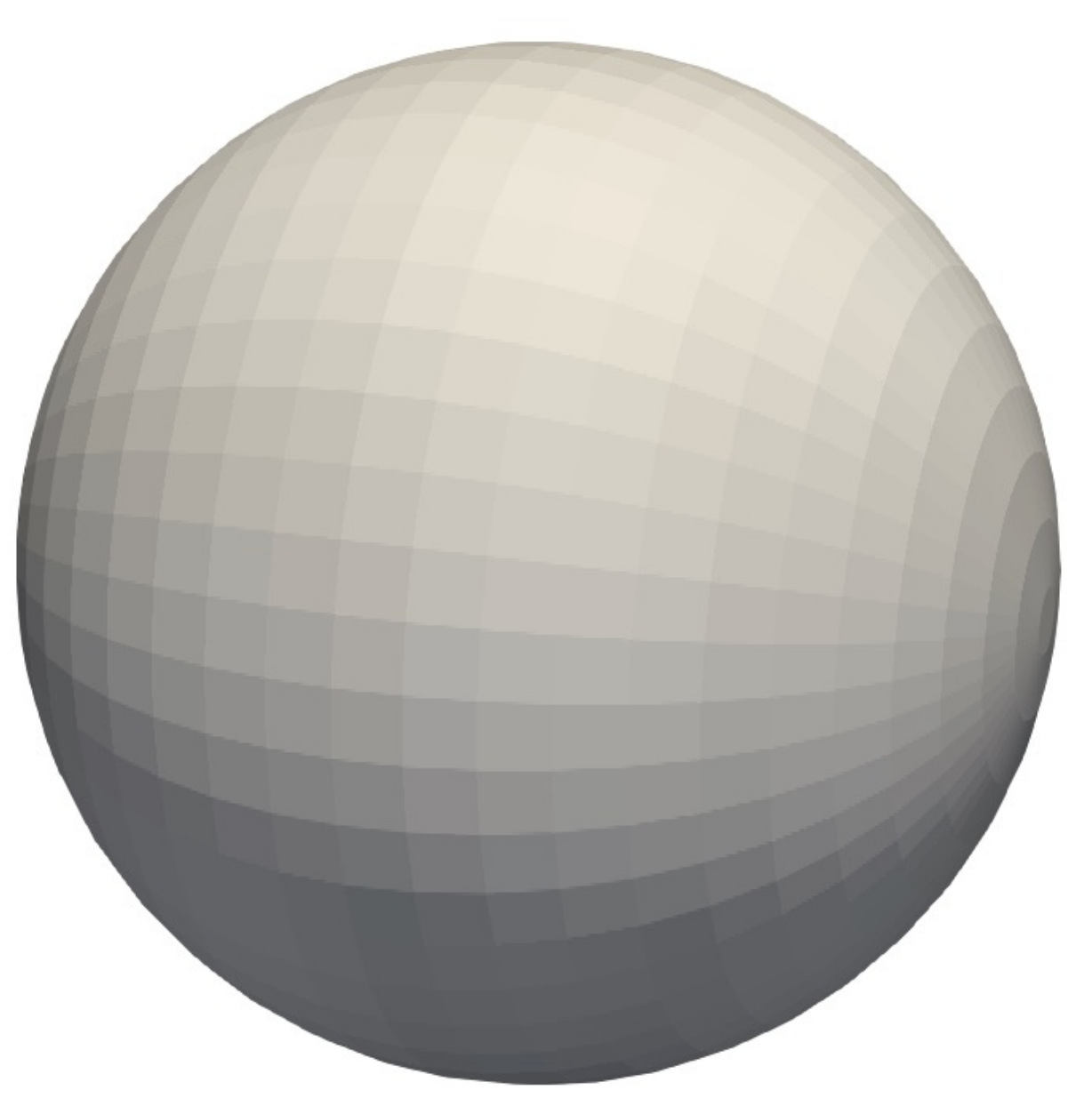}
    \caption{Ico-Sphere}
    \label{fig:ico_sphere}
\end{figure}

In this study, we perform following comparisons:
\begin{enumerate}
    \item Comparison across different loss functions with Fully Connected Network.
    \item Comparison of the Implicit Net with different loss functions along the modified Implicit Loss as given in \eqnref{equation:loss_function}.
    \item Ablation study of loss function in \eqnref{equation:loss_function}.
    \item Ablation study of the sampling strategy.
\end{enumerate}

\begin{figure}[ht]
    \centering
    \includegraphics[trim=0cm 6cm 0cm 8cm, clip, width=1\linewidth]{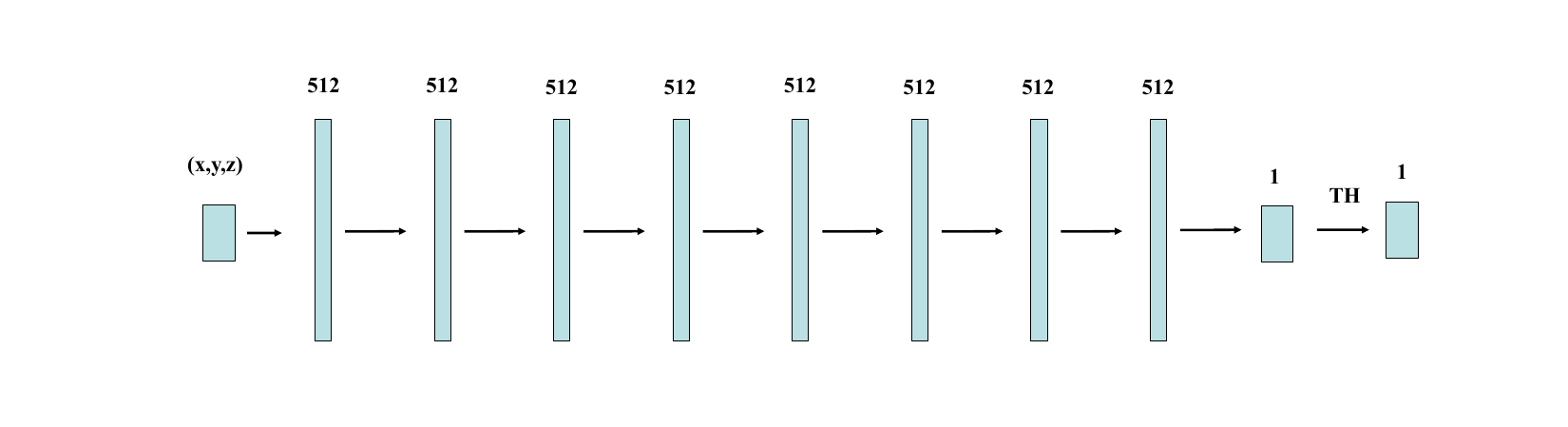}
              \caption{Fully Connected Network for Implicit Representation}
              \label{fig:fcn}
\end{figure}

\begin{figure}[htbp]
    \centering
    \begin{subfigure}[b]{0.22\textwidth}
        \includegraphics[width=\textwidth]{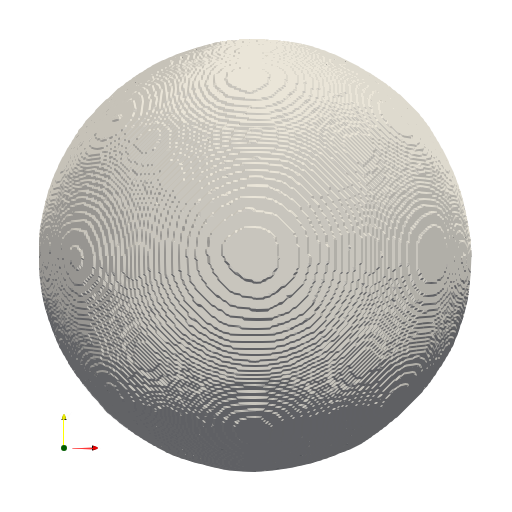}
        \caption{L1 Clamped Loss}
        \label{fig:pdf3}
    \end{subfigure}
    \hfill
    \begin{subfigure}[b]{0.22\textwidth}
        \includegraphics[width=\textwidth]{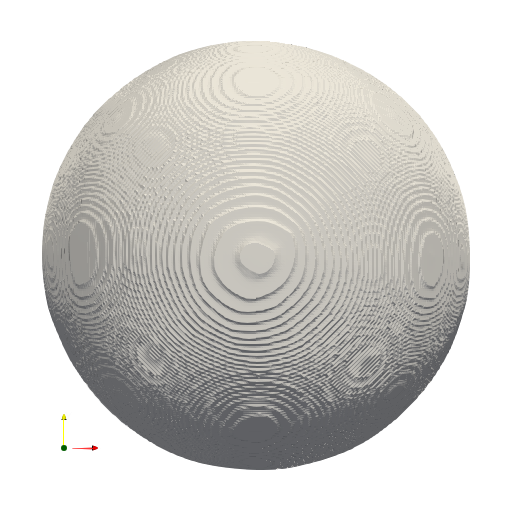}
        \caption{L2 Clamped Loss}
        \label{fig:pdf2}
    \end{subfigure}
    \hfill
    \begin{subfigure}[b]{0.22\textwidth}
        \includegraphics[width=\textwidth]{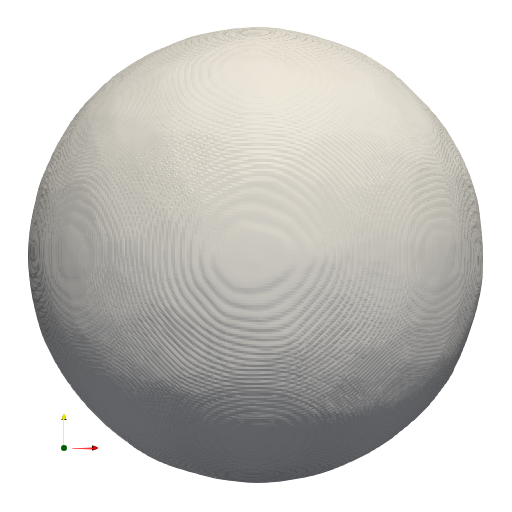}
        \caption{L2 Smoothened Loss}
        \label{fig:pdf1}
    \end{subfigure}
    \caption{Comparison of Different Loss Functions with Fully Connected Network}
    \label{fig:reconstruction_fcn}
\end{figure}

We take the fully connected network as shown in  \figref{fig:fcn} and compare the NMSE with $\delta=0.1$ and similarly perform comparison across different plausible loss functions as presented in \tabref{tab:loss_functions}. For a very high threshold $\delta=0.1$ as well, the errors are significantly higher. \figref{fig:reconstruction_fcn} demonstrates reconstruction with staircase effect for the combination of the fully connected network and the presented loss functions. 

\begin{table}[htbp]
    \centering
        \caption{Comparison of Different Loss Functions with Fully Connected Network}
    \begin{adjustbox}{max width=\textwidth}
        \begin{tabular}{|c|l|c|c|}
            \hline
            \textbf{SN} & \textbf{Loss Function} & \textbf{Expression} & $NMSE_{0.1}$ \\ \hline
            a & $L_1$ Clamped Loss~\citep{park2019deepsdf} & $L_{1,Clamped}(\mathbf{s}, f_\theta(\mathbf{x})) = \frac{1}{N}\sum_{i=1}^{N} \left| \text{clamp}(\mathbf{s}_i, \delta) - \text{clamp}(f_\theta(\mathbf{x}), \delta) \right|$ & 0.0566 \\ \hline
            b & $L_2$ Clamped Loss & $L_{2,Clamped}(\mathbf{s}, f_\theta(\mathbf{x})) = \frac{1}{N}\sum_{i=1}^{N} \left( \text{clamp}(\mathbf{s}_i, \delta) - \text{clamp}(f_\theta(\mathbf{x}), \delta) \right)^2$ & 0.0407 \\ \hline
            c & $L_2$ Smooth Loss & $L_{2,Smooth}(\mathbf{s}, f_\theta(\mathbf{x})) = \frac{1}{N}\sum_{i=1}^{N} (1+\alpha ^{\left | \mathbf{s}_i \right|})(\mathbf{s}_i - f_\theta(\mathbf{x}))^2$ & 0.01895 \\ \hline
        \end{tabular}
    \end{adjustbox}

    \label{tab:loss_functions}
\end{table}
\begin{figure}[t!]
    \centering
    \includegraphics[trim=0cm 6cm 0cm 6cm, clip, width=1\linewidth]{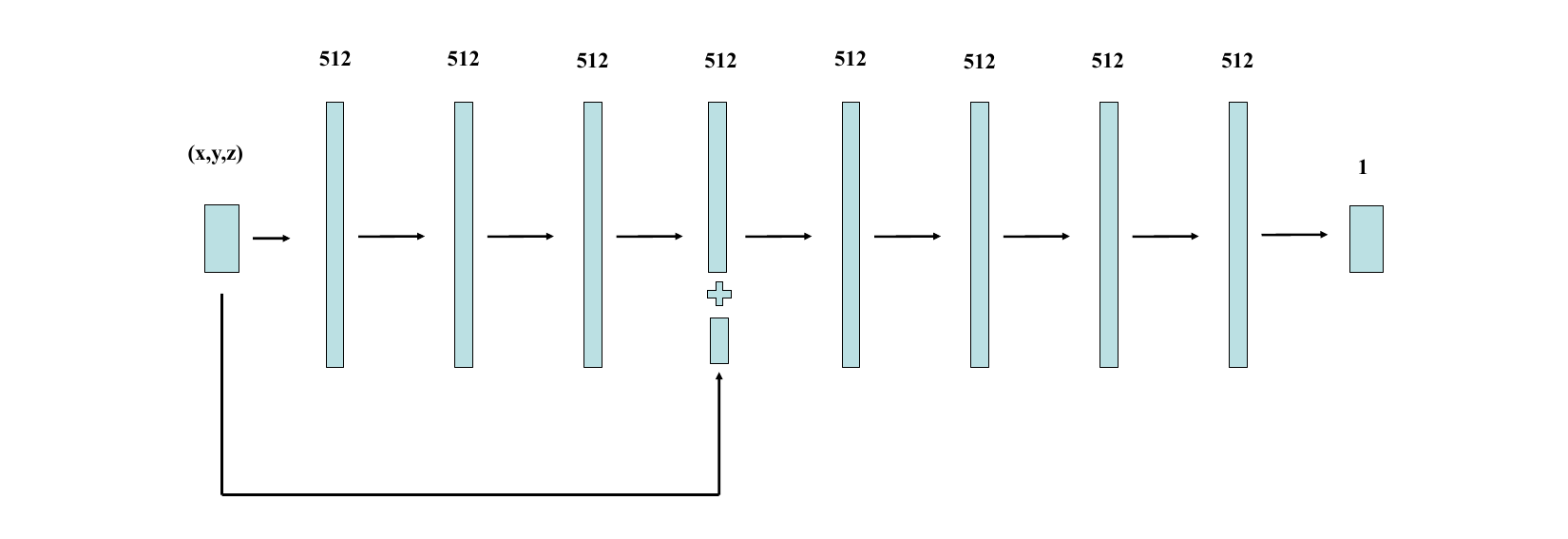}
    \caption{Implicit Geometric Network with skip in}
    \label{fig:igrnet}
\end{figure}

\textbf{\figref{fig:igrnet}} gives the architecture of the network with 8 hidden layers of 512 units each with one skip-in connection. This network was proposed by \citet{park2019deepsdf} and has been thoroughly used in the field of INR. \tabref{tab:implicitnet_loss_functions} makes comparison between loss functions used and defined in \tabref{tab:loss_functions} along with loss function from \eqnref{equation:loss_function}. \eqnref{equation:loss_function} is hybrid of better performing $L_2 Clamped Loss$ from \tabref{tab:loss_functions} and geometric regularization as proposed in \citet{gropp2020implicit}. The comparison is made with a very low threshold, $\delta = 10^{-10}$. The errors are significantly lower in comparison to \tabref{tab:loss_functions} despite of difference in the threshold. Similarly, the model with loss function as \eqnref{equation:loss_function} has the least value of the error. This inspires us to perform the ablation study to analyze the effect of removing different terms from the loss function, as shown in \tabref{tab:ablation_of_loss}. The Eikonal loss and Normal loss are defined below which are components of the original loss function.
\[\text{Eikonal Loss:} \quad \lambda_g \int_{\Omega} \left( \left\| \nabla_{\mathbf{x}} f_\theta(\mathbf{x}) \right\| - 1 \right)^2 \, d\Omega\]
\[\text{Normal Loss:} \quad \tau \int_{\Omega} \left( \frac{\nabla_{\mathbf{x}} f_\theta(\mathbf{x})}{\left\| \nabla_{\mathbf{x}} f_\theta(\mathbf{x}) \right\|} \cdot \hat{\mathbf{n}}(\mathbf{x}) - 1 \right)^2 \, d\Omega\]

\begin{table}[t!]
    \centering
        \caption{Comparison of Different Loss Functions with Implicit Net}
        \label{tab:implicitnet_loss_functions}
        \begin{tabular}{|c|c|c|}
            \hline
            \textbf{SN} & \textbf{Loss Function} & $\mathbf{NMSE_{2^{-10}}}$ \\ \hline
            a & $L_1$ Clamped Loss~\citep{park2019deepsdf} & $2.17 \times 10^{-7}$ \\ \hline
            b & $L_2$ Clamped Loss & $1.96 \times 10^{-7}$ \\ \hline
            c & $L_2$ Smooth Loss & $2.08 \times 10^{-7}$ \\ \hline
            d & \textbf{IGR Loss (\eqnref{equation:loss_function})} & $\mathbf{1.54 \times 10^{-7}}$ \\ \hline
        \end{tabular}

\end{table}
\begin{table}[t!]
    \centering
        \caption{Ablation study of IGR Loss Functions based on \( NMSE_{2^{-10}} \)}
        \begin{tabular}{|c|c|}
            \hline
            \textbf{Loss Function} & $\mathbf{NMSE_{2^{-10}}}$ \\ \hline
            \textbf{IGR Loss (\eqnref{equation:loss_function})} & $\mathbf{1.54 \times 10^{-7}}$ \\ \hline
            w/o Eikonal loss & $1.97 \times 10^{-7}$ \\ \hline
             w/o Eikonal and Normal loss& $1.96 \times 10^{-7}$ \\ \hline

        \end{tabular}

    \label{tab:ablation_of_loss}
\end{table}
\begin{table}[t!]
    \caption{Comparison of sampling strategy \(NMSE_{2^{-10}}\)}
    \centering
        \begin{tabular}{|c|c|c|c|}
            \hline
            \textbf{Approach} & Total Points & (Uniform, Surface,Narrow Band) & $\mathbf{NMSE_{2^{-10}}}$ \\ 
            \hline
            \textbf{Only Uniform} & 150K & 150K, 0, 0 & $2.0 \times 10^{-7}$ \\
            \hline
            \textbf{Only Narrow Band} & 150K & 0, 0, 150K & $2.25 \times 10^{-7}$ \\
            \hline
            \textbf{Only Surface} & 150K & 0, 150K, 0 & $2.4 \times 10^{-7}$ \\
            \hline
            \textbf{Hybrid} & 150K & 90K, 28K, 32K & $\mathbf{3.81 \times 10^{-8}}$ \\
            \hline
        \end{tabular}

    \label{tab:hybrid_sampling}
\end{table}

It can be clearly seen from \ref{tab:implicitnet_loss_functions} that the loss function proposed in \eqnref{equation:loss_function} performs well in comparison to other intuitive loss functions. 
All of the above experiments were performed with consideration to the same sampling set. In \tabref{tab:hybrid_sampling}, the loss function and the network architecture are fixed, and a comparison is made between different sampling strategies with an equal number of points but sampled differently The number of points in hybrid sampling is obtained through Bayesian Optimization which has the least value of the Normalized Mean Square Error. All the INR is based on the hybrid sampling technique presented here.

\section{Accuracy of INRs for SBM}
\label{section:Val_Implicit}
\begin{figure}[t!]
    \centering
    \begin{subfigure}[t]{0.45\textwidth}
        \centering
        \includegraphics[width=0.45\textwidth]{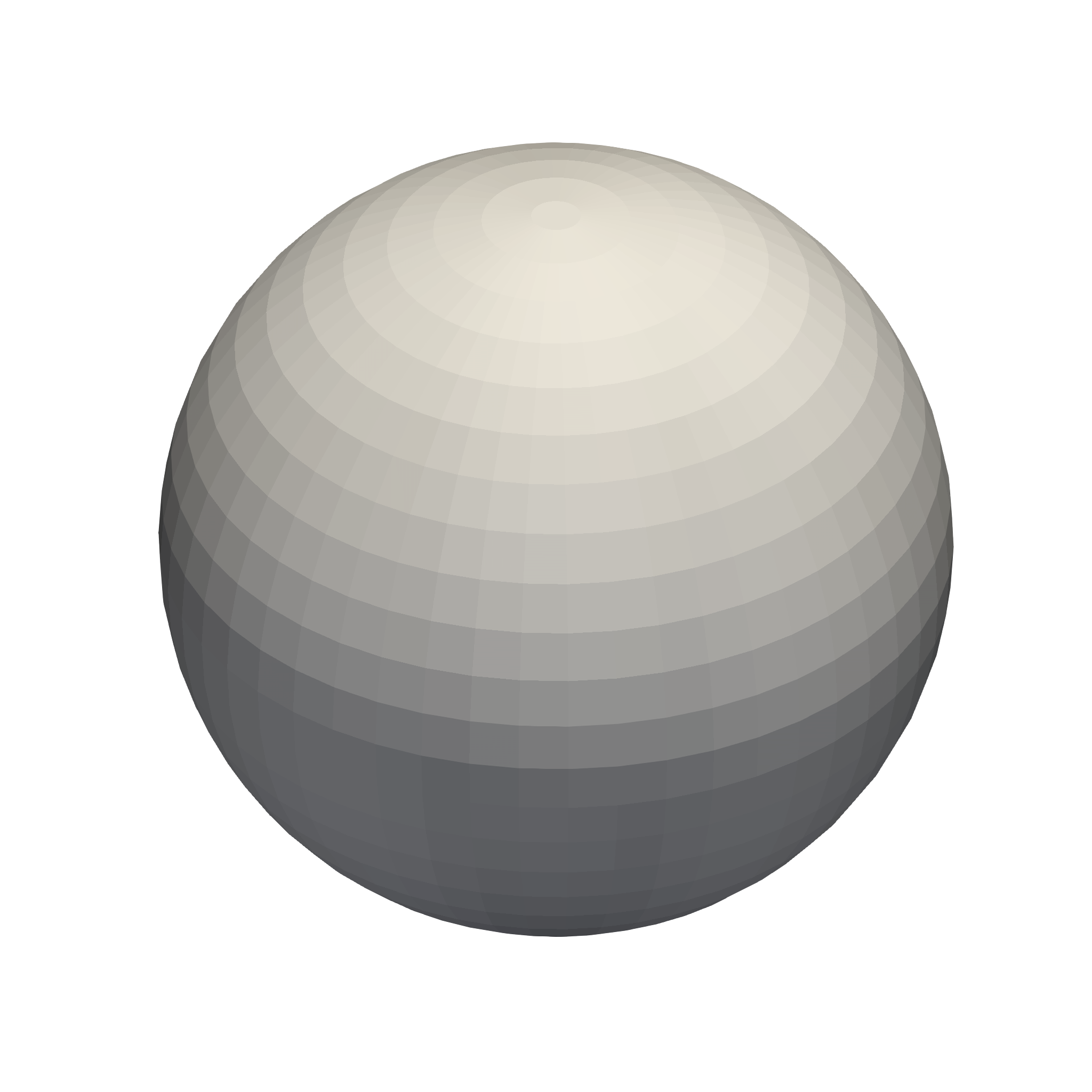}
        \includegraphics[width=0.45\textwidth]{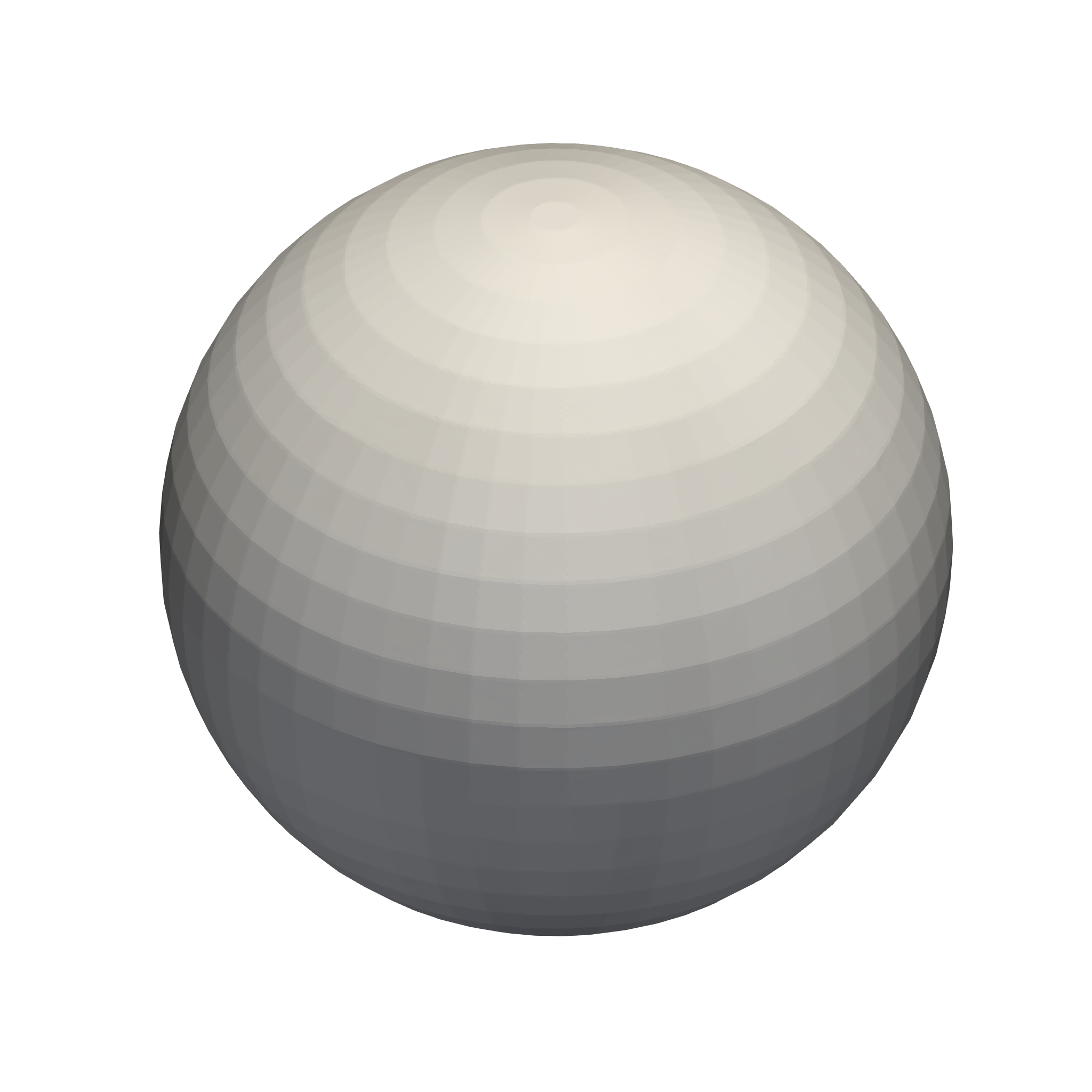}
        \caption{Sphere}
    \end{subfigure}
    \begin{subfigure}[t]{0.45\textwidth}
        \centering
        \includegraphics[width=0.45\textwidth]{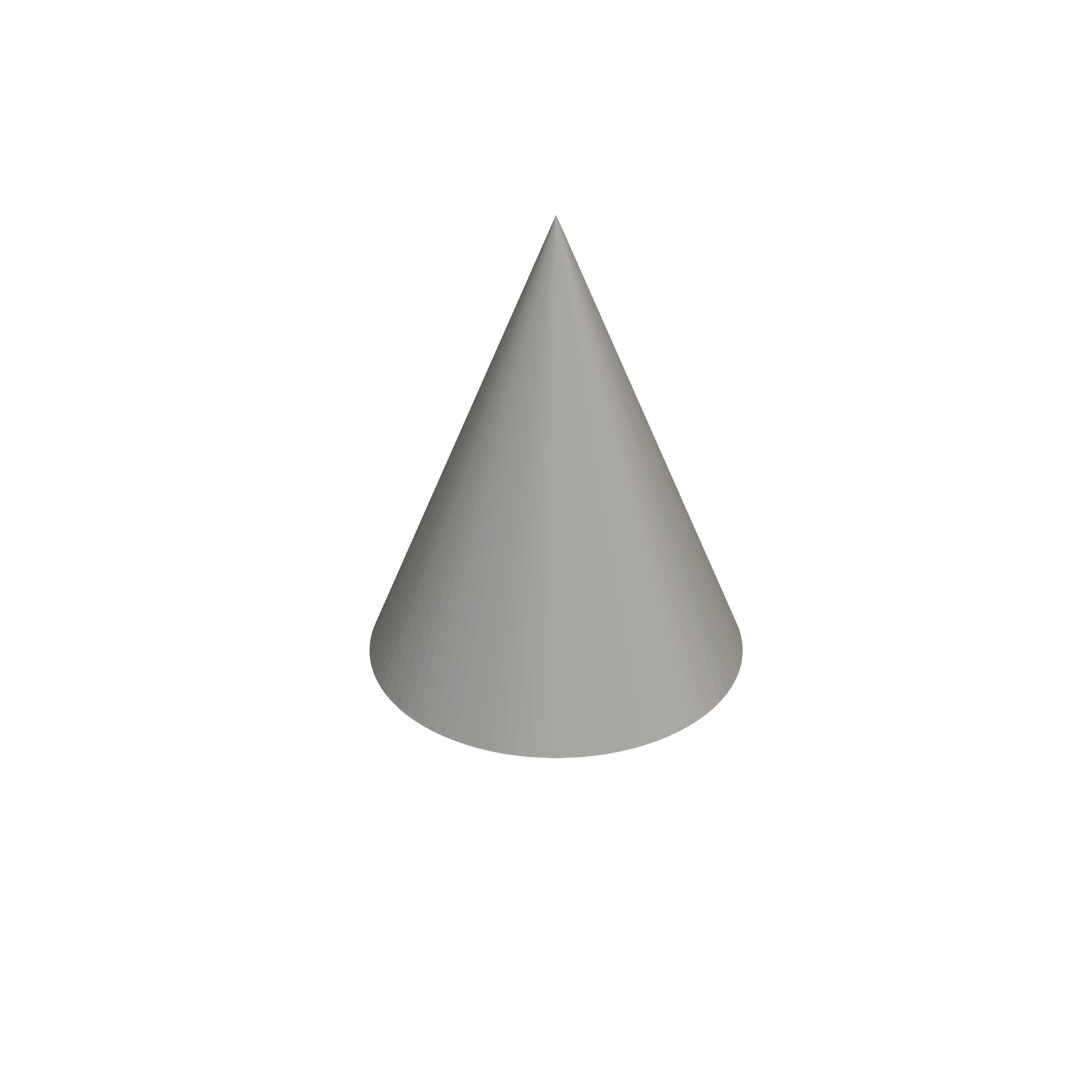}
        \includegraphics[width=0.45\textwidth]{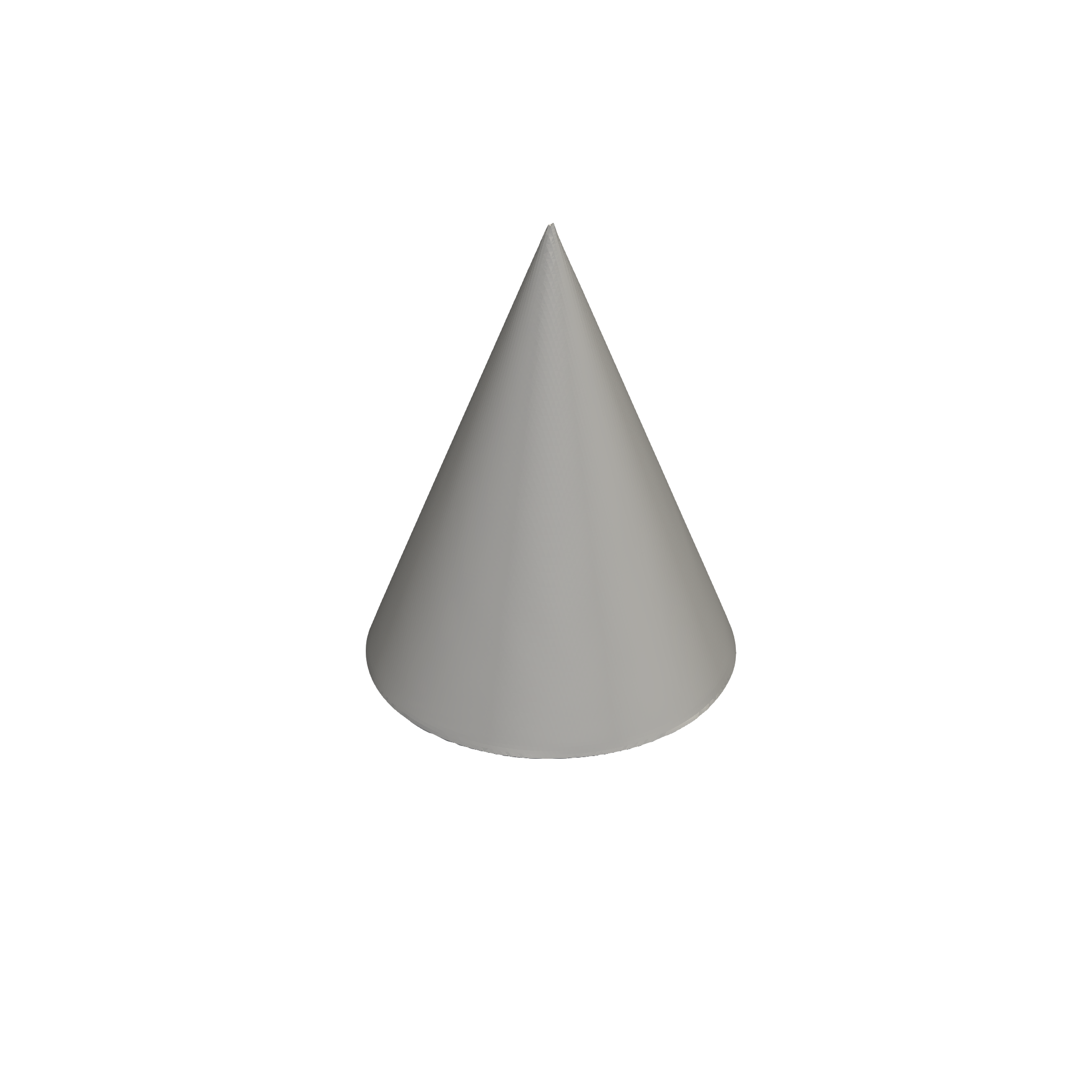}
        \caption{Cone}
    \end{subfigure}        
    \begin{subfigure}[t]{0.45\textwidth}
        \centering
        \includegraphics[width=0.45\textwidth]{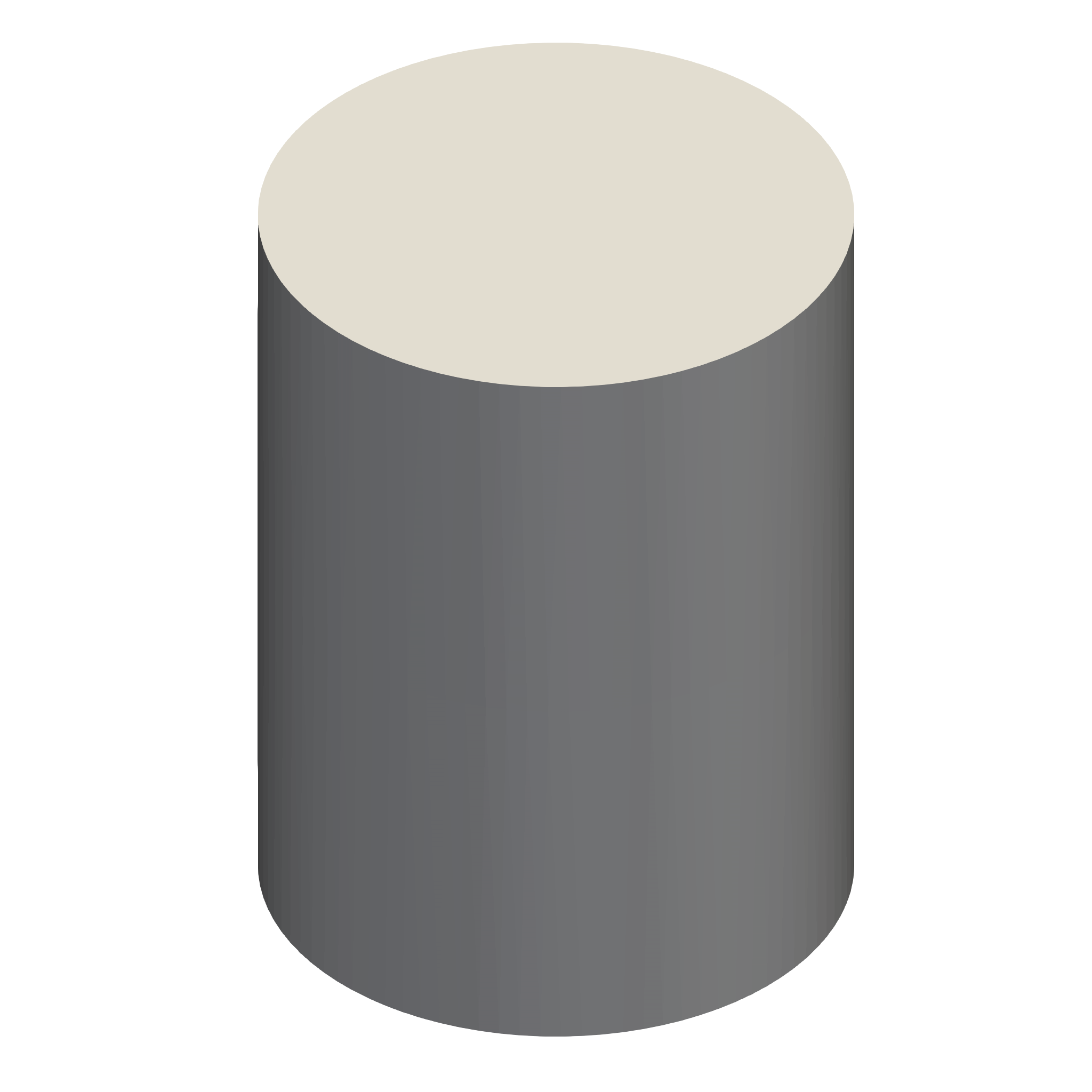}
        \includegraphics[width=0.45\textwidth]{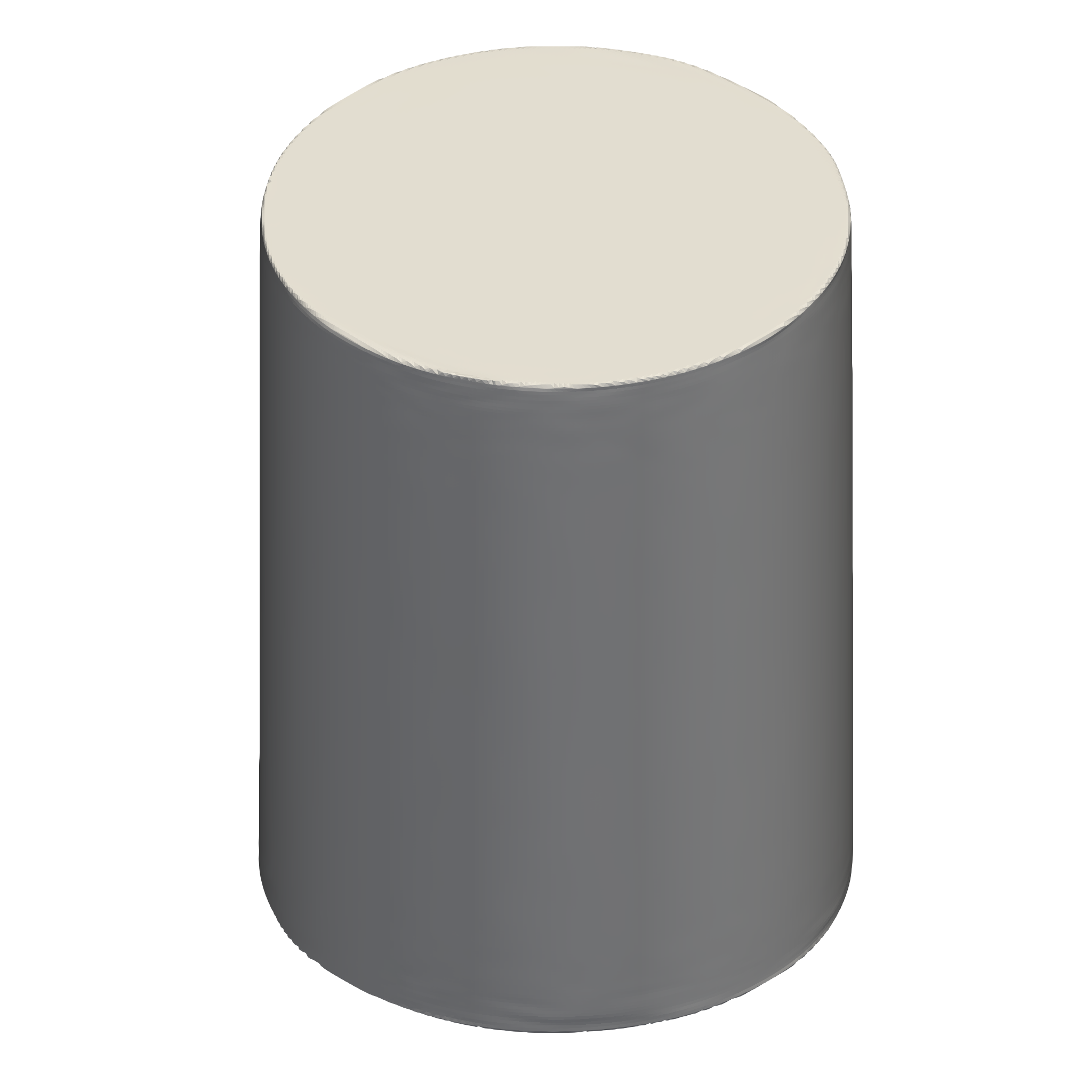}
        \caption{Cylinder}
    \end{subfigure}
    \begin{subfigure}[t]{0.45\textwidth}
        \centering
        \includegraphics[width=0.45\textwidth]{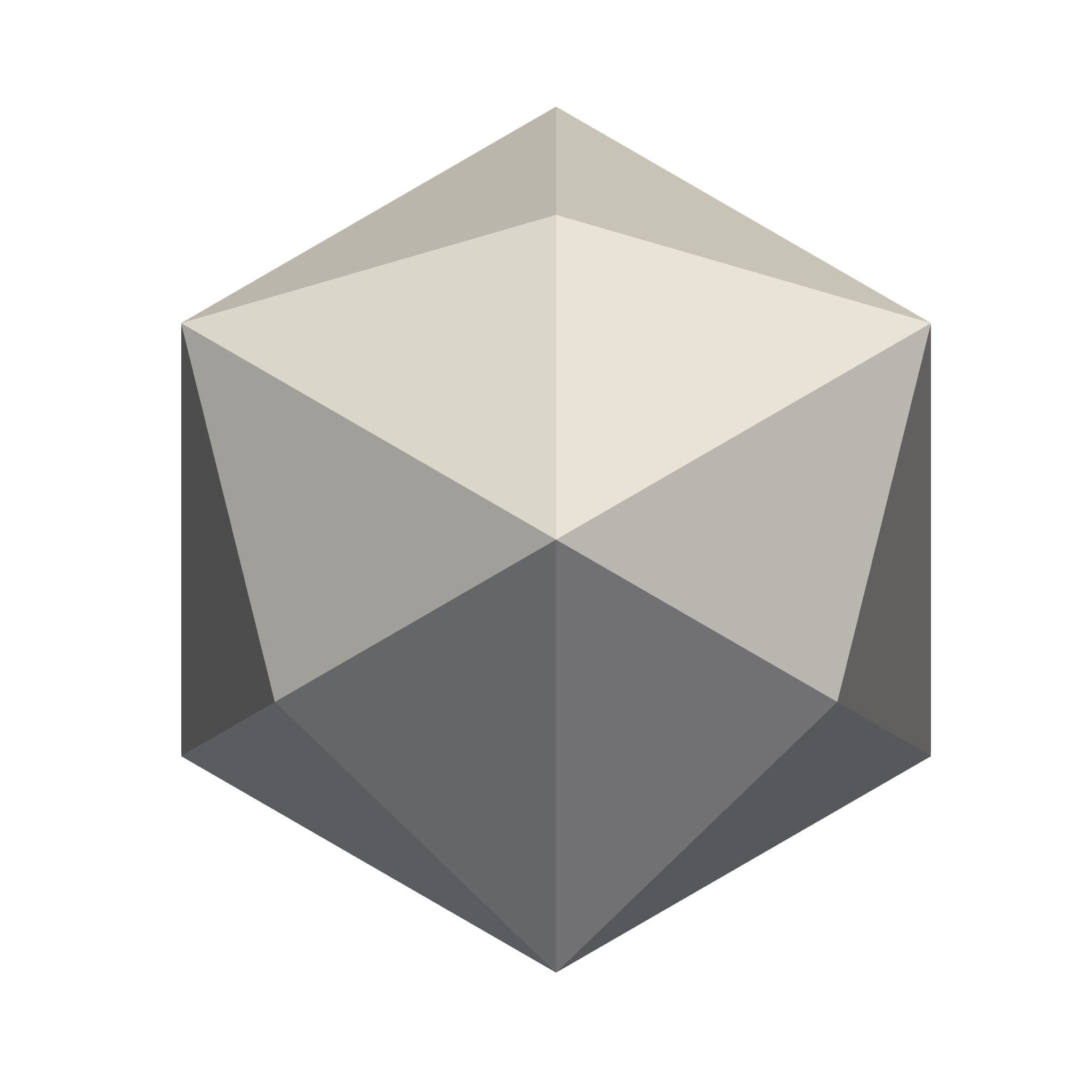}
        \includegraphics[width=0.45\textwidth]{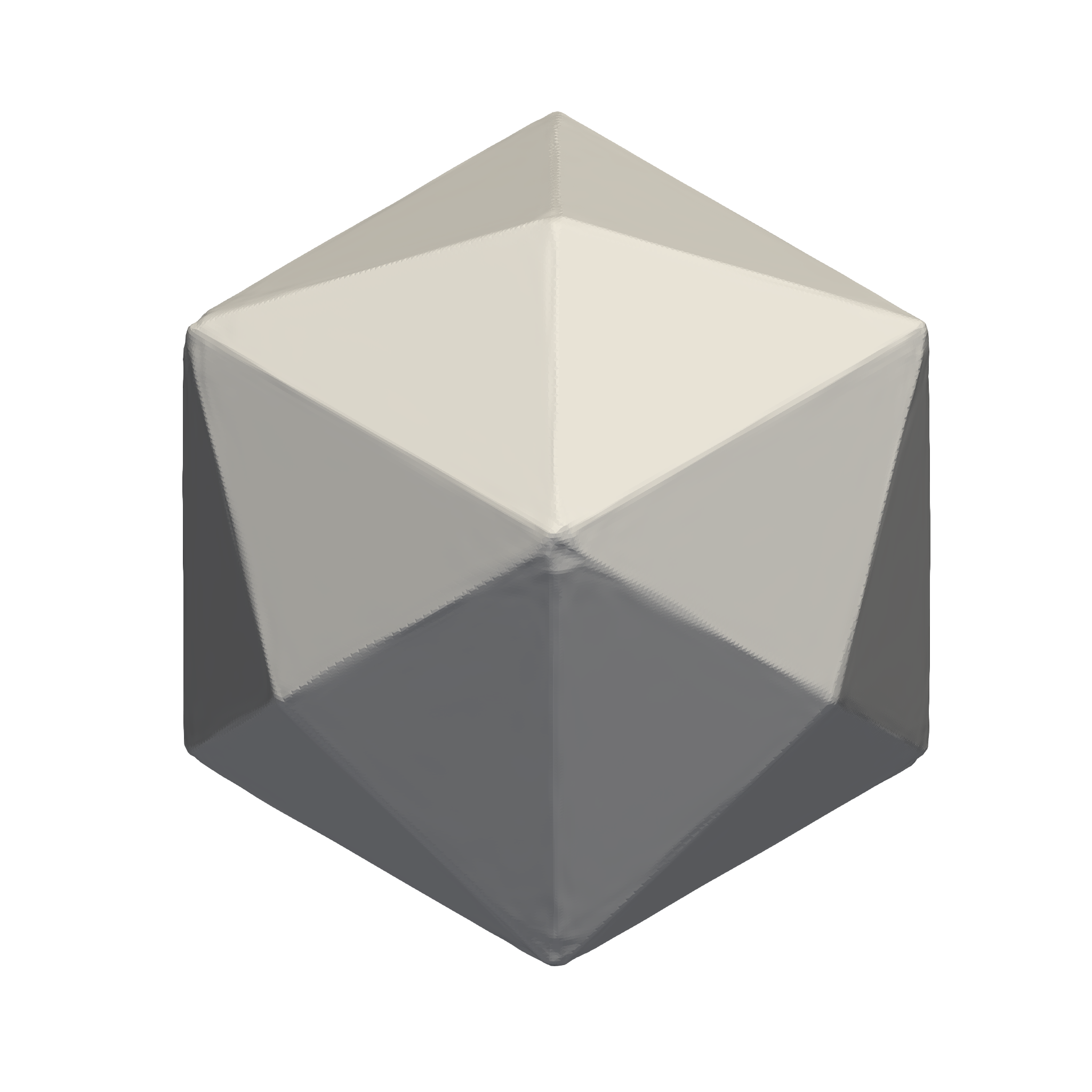}
        \caption{Tetrakis}
    \end{subfigure}
    \begin{subfigure}[t]{0.45\textwidth}
        \centering
        \includegraphics[width=0.45\textwidth]{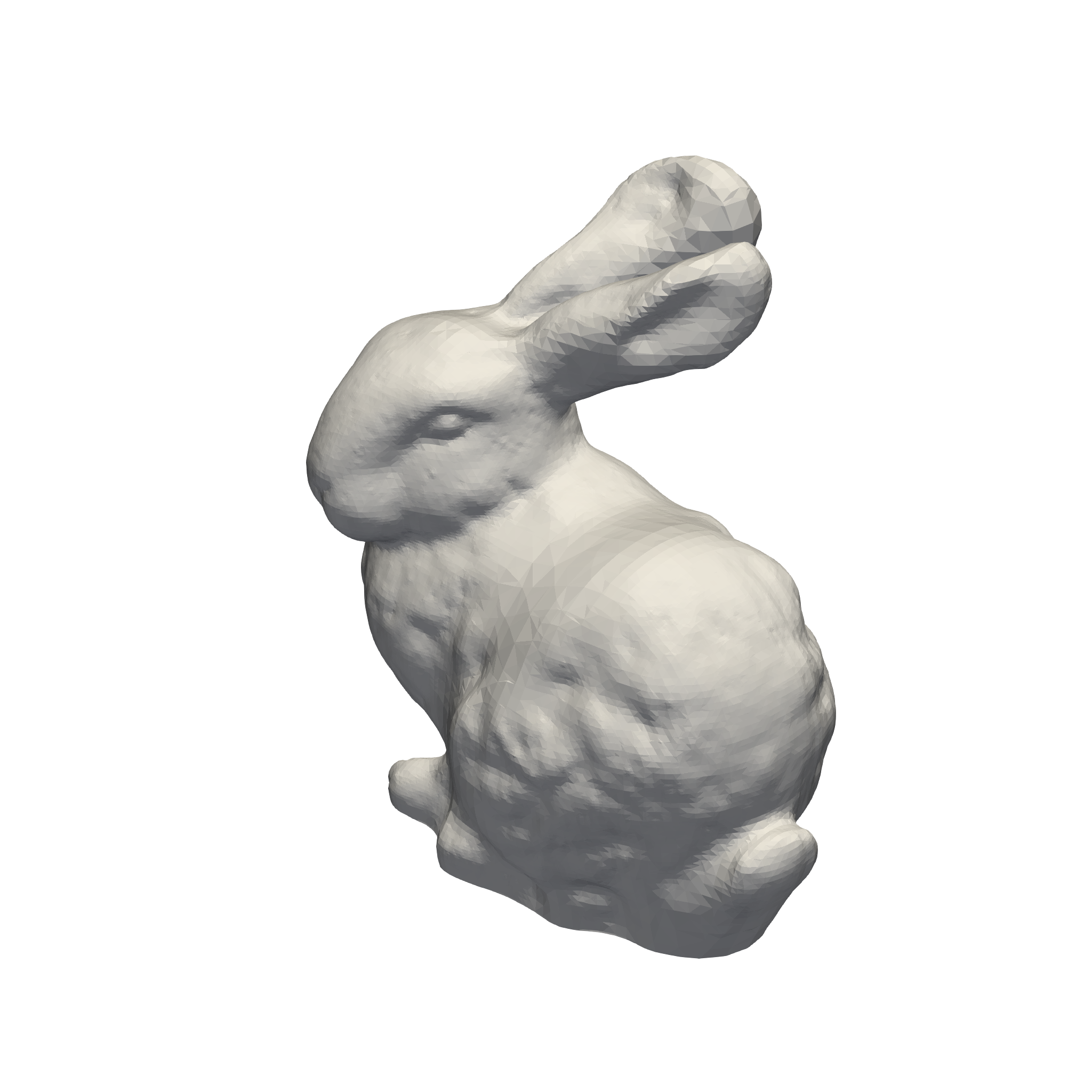}
        \includegraphics[width=0.45\textwidth]{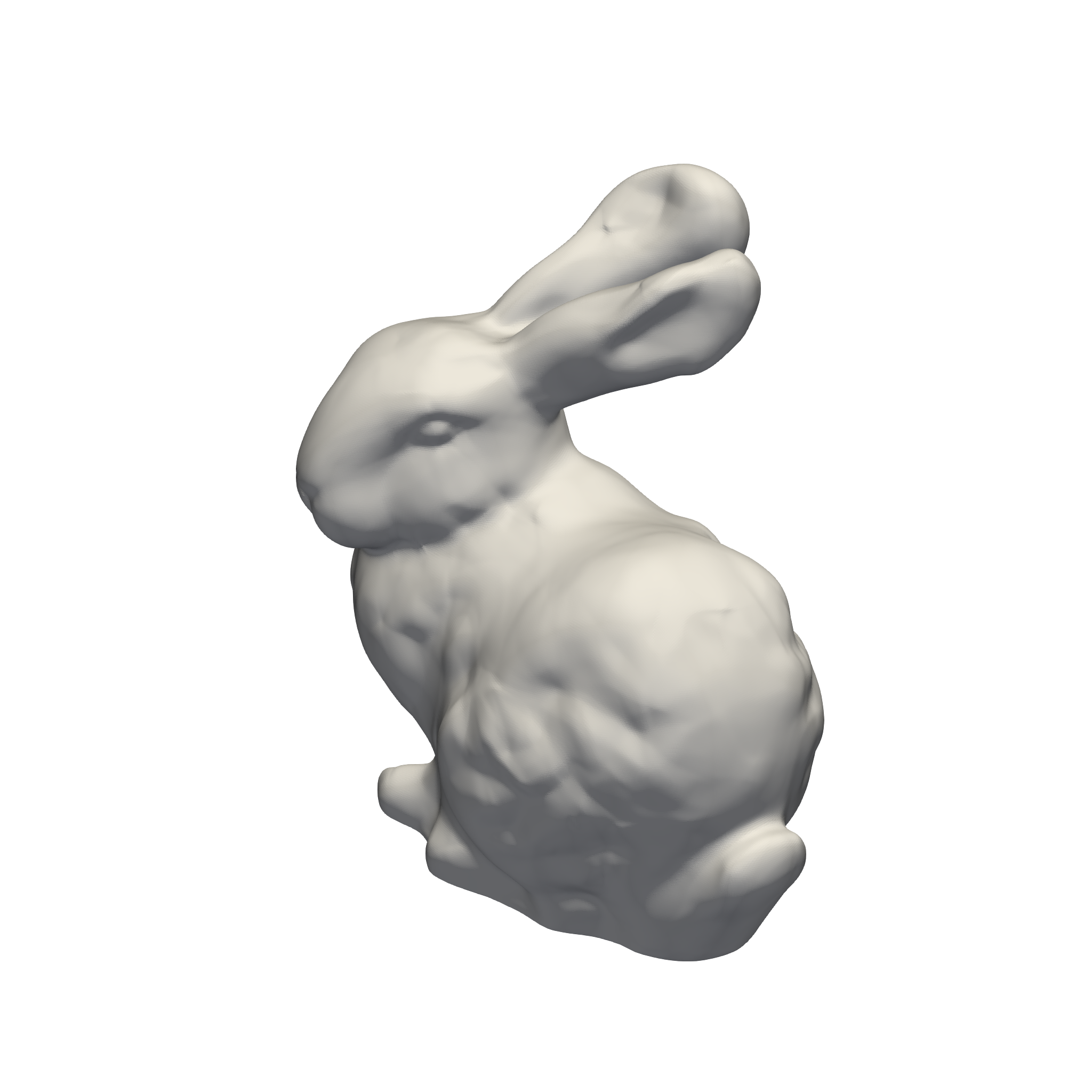}
        \caption{Bunny}
    \end{subfigure}
    \begin{subfigure}[t]{0.45\textwidth}
        \centering
        \includegraphics[width=0.45\textwidth]{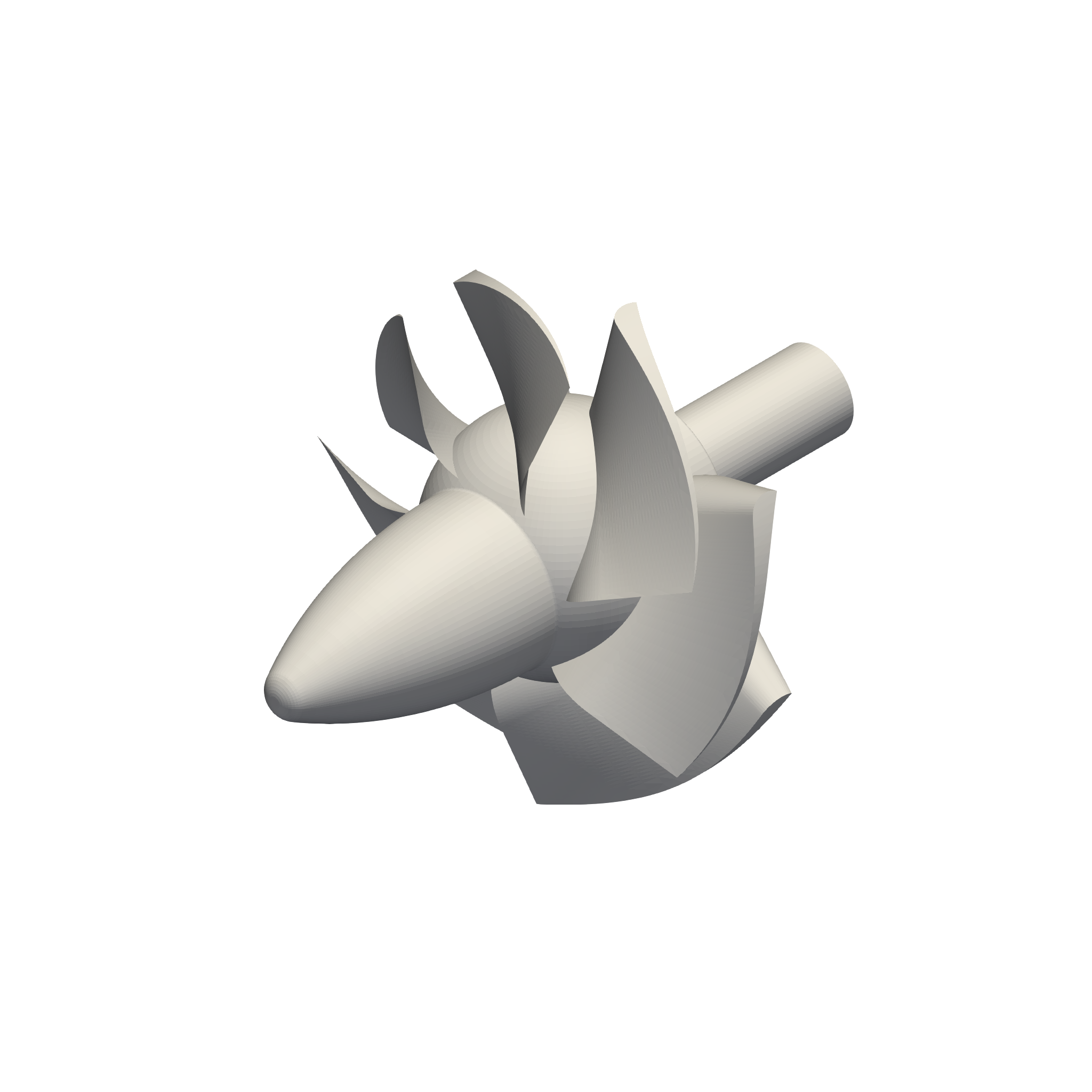}
        \includegraphics[width=0.45\textwidth]{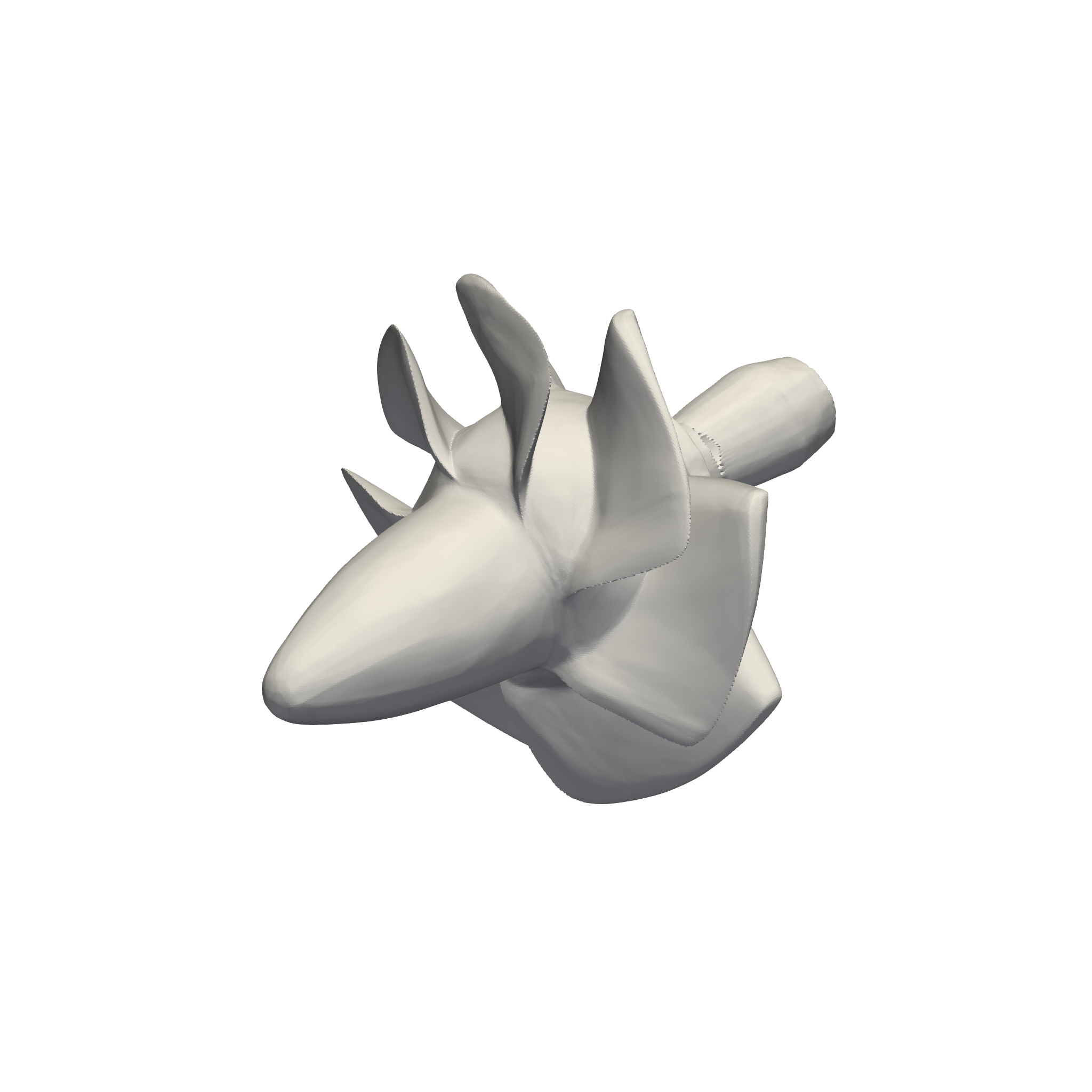}
        \caption{Turbine}
    \end{subfigure}
    \caption{Original(left) and reconstruction(right) using marching cubes for visualization. The marching cubes is performed in $256^3$. Visually the geometries look similar except in places of sharp corners or high curvature changes. The discrepancy is the cumulative effect of marching cubes's resolution and difficulty of learning sharp curvatures by INR.} 
    \label{fig:original_reconstructed}
\end{figure}

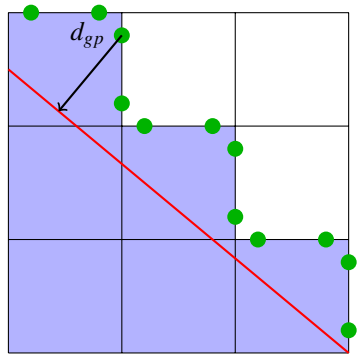
\begin{figure}[htbp]
    \centering
    \begin{tikzpicture}[scale=1.5]
        \fill[blue!30] (0,0) rectangle (1,3)
                       (1,0) rectangle (2,2)
                       (2,0) rectangle (3,1);
        
        \draw[thin] (0,0) grid (3,3);
        
        \draw[thick, red] (0,2.5) -- (3,0);
        
        \foreach \x/\y in {0.2/3,0.8/3,1/2.8,1/2.2,1.2/2,1.8/2,2/1.2,2/1.8,2.2/1,2.8/1,3/0.2,3/0.8}{
            \fill[green!70!black] (\x, \y) circle (2pt);
        }
        
        \draw[->, thick] (1,2.8) -- (0.443,2.131);
        \node at (0.7, 2.8) {$d_{gp}$};
        
    \end{tikzpicture}
    \caption{Distance Vector, $\mathbf{d_{GP}}$, corresponding to the gauss points (marked in green) at the surrogate boundary. Distance Vector points in the shortest distance to the true boundary (line in red).}
    \label{fig:evaluating the distance vector}
\end{figure}

INRs of different shapes as classified as complex and simple as presented in  \tabref{complexity_comparison} are obtained using \algoref{Algorithm 1 Implicit Network Training}. \figref{fig:original_reconstructed} is the visualization of the polygonal soup and the corresponding reconstruction obtained from the INR obtained by performing marching cube in $\mathbf{256^3}$~\citep{cline19873d}. The end goal of this work is to use the Implicit Neural Representations for Shifted Boundary Method Analysis, which requires a correct distance vector for the Gauss points in the surrogate boundary as presented in  \figref{fig:evaluating the distance vector}.

\begin{figure}[t!]
    \centering
    \begin{subfigure}[t]{0.2\linewidth}
        \centering
        \includegraphics[width=\linewidth, trim=1.5in 1.5in 1.5in 1.5in, clip]{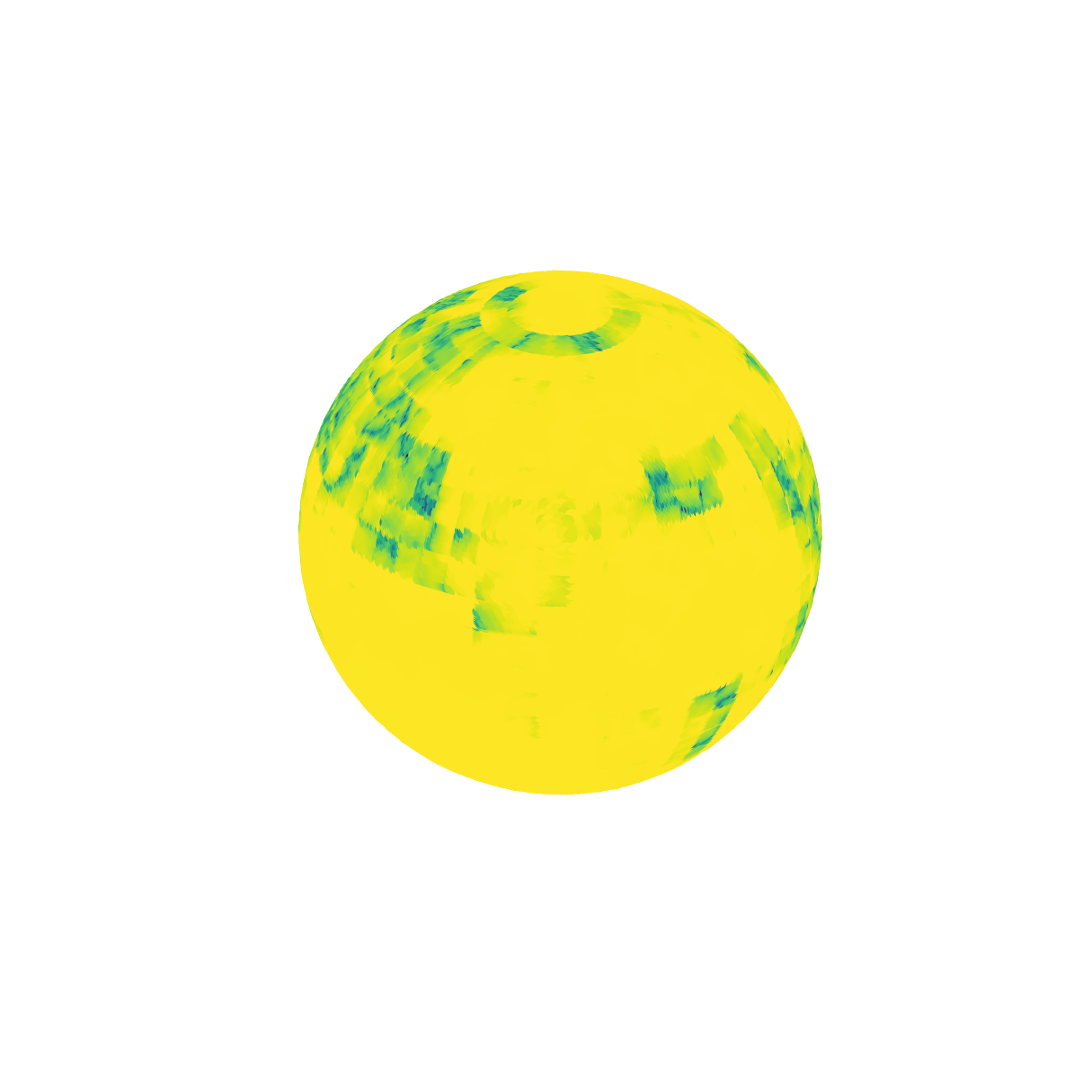}
        \caption{Sphere}
    \end{subfigure}
    \begin{subfigure}[t]{0.2\linewidth}
        \centering
        \includegraphics[width=\linewidth, trim=1.5in 1.5in 1.5in 1.5in, clip]{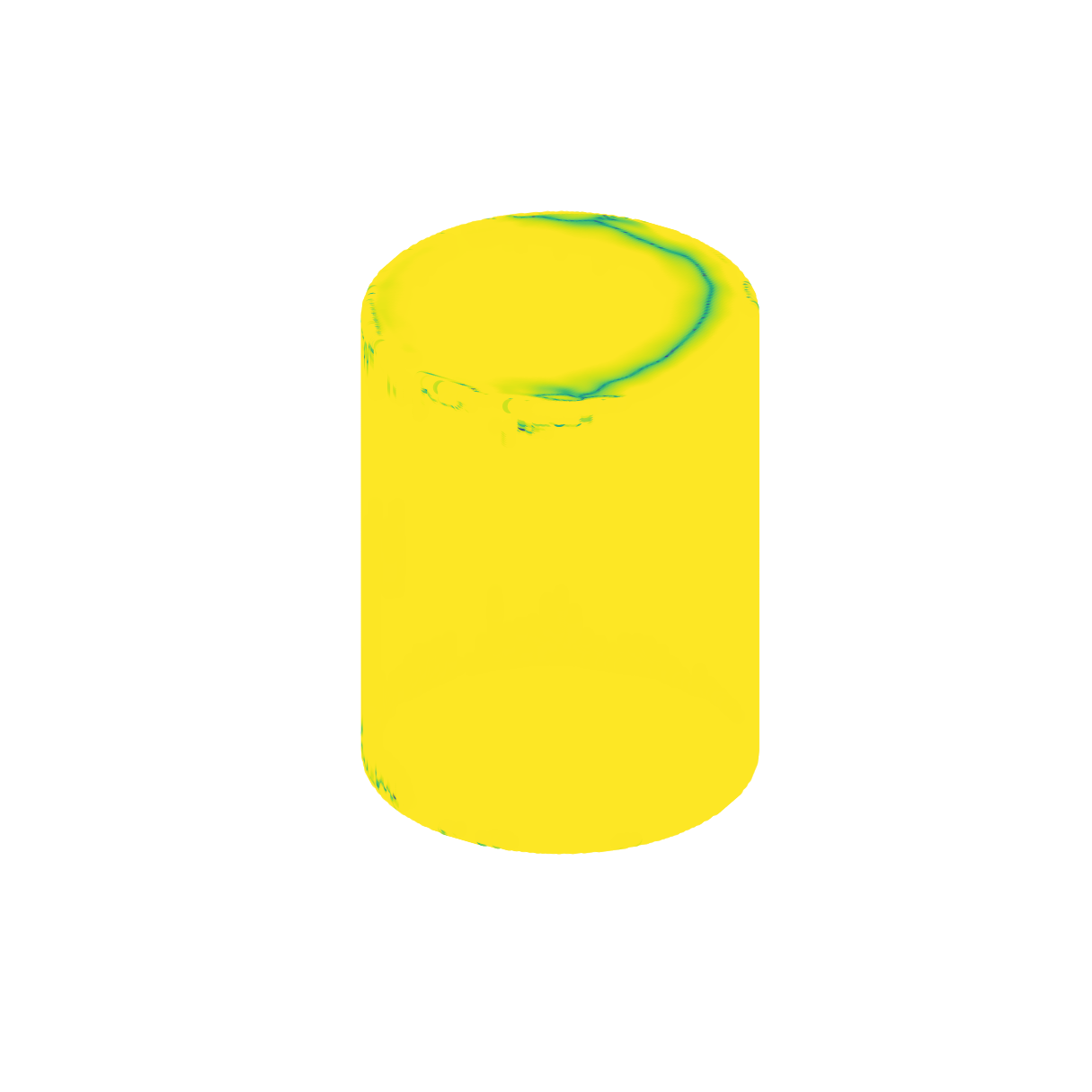}
        \caption{Cylinder}
    \end{subfigure}
    \begin{subfigure}[t]{0.2\linewidth}
        \centering
        \includegraphics[width=\linewidth, trim=2.3in 2.8in 2.3in 1.0in, clip]{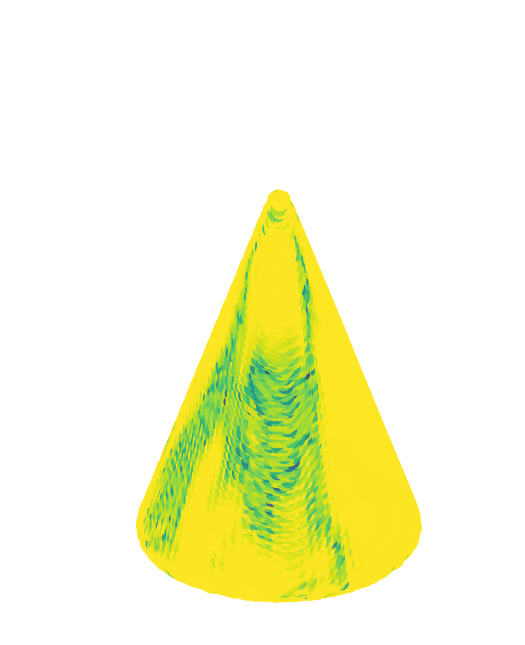}
        \caption{Cone}
    \end{subfigure}
    \begin{subfigure}[t]{0.2\linewidth}
        \centering
        \includegraphics[width=\linewidth, trim=1.5in 2.0in 1.5in 1.5in, clip]{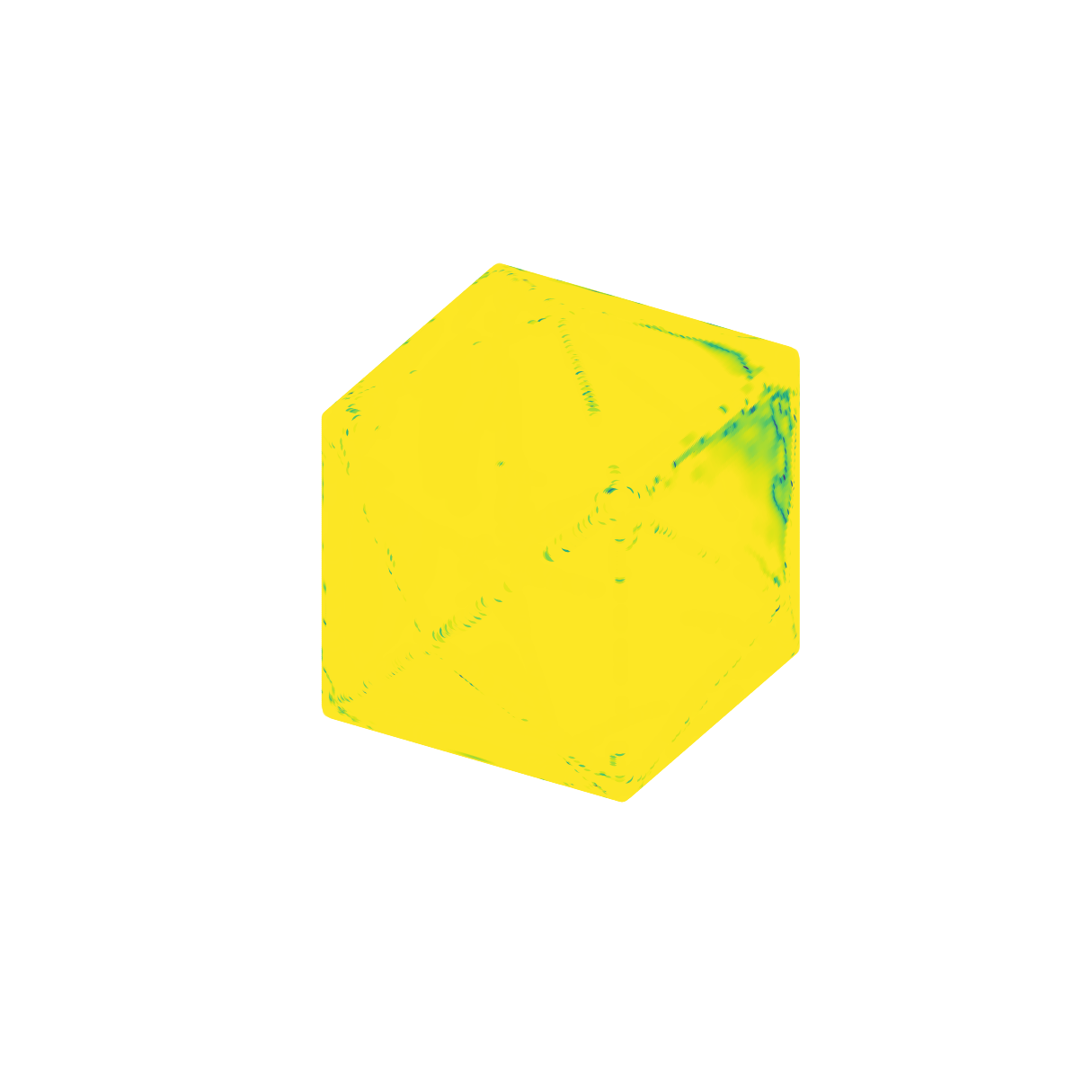}
        \caption{Tetrakis}
    \end{subfigure}
    \begin{subfigure}[t]{0.2\linewidth}
        \centering
        \includegraphics[width=\linewidth, trim=1.5in 2.0in 1.5in 1.5in, clip]{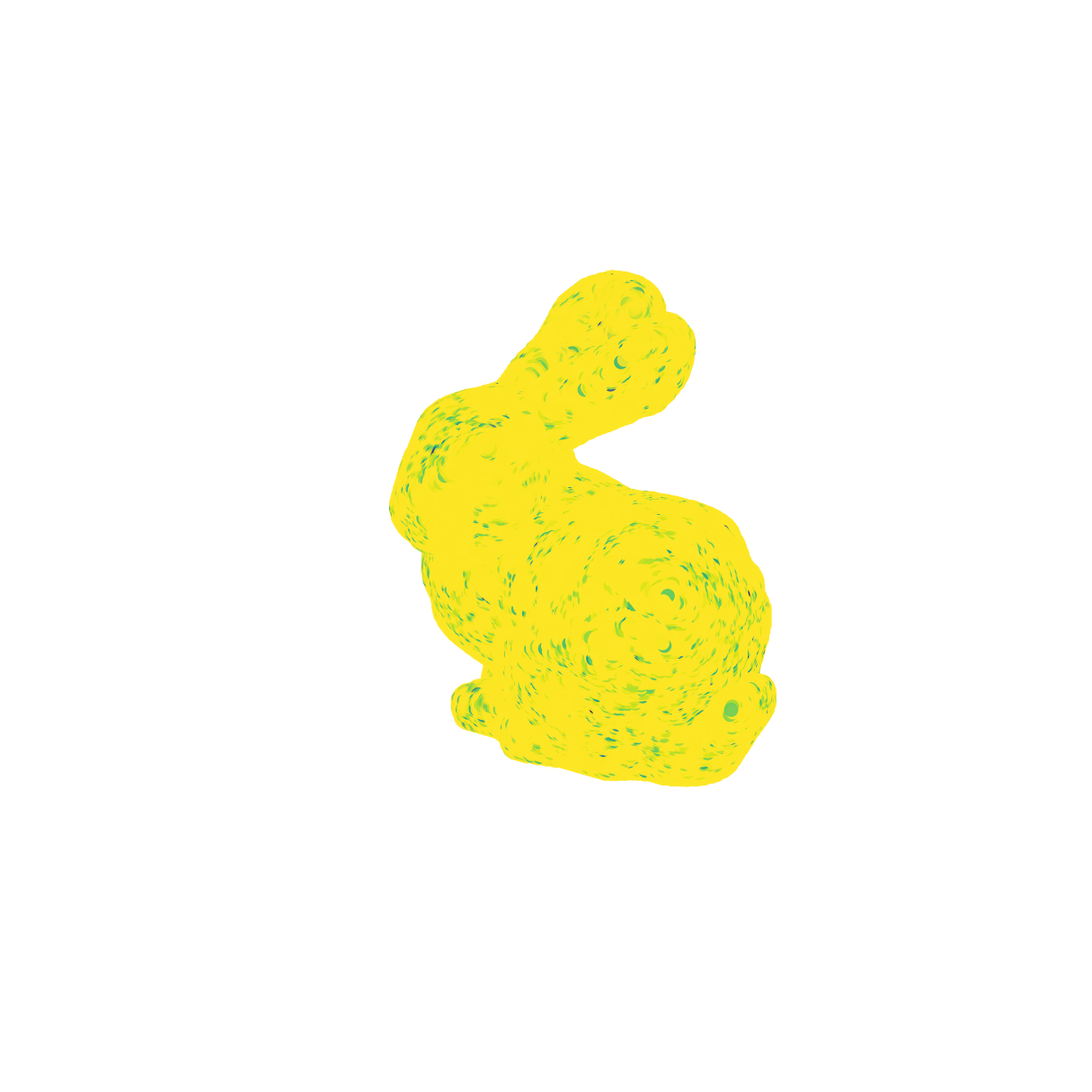}
        \caption{Bunny}
    \end{subfigure}
    \begin{subfigure}[t]{0.2\linewidth}
        \centering
        \includegraphics[width=\linewidth, trim=2.0in 2.3in 2.0in 2.0in, clip]{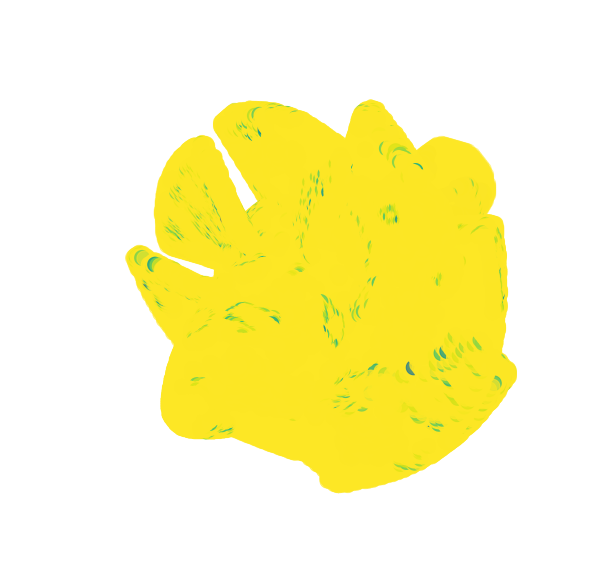}
        \caption{Turbine}
    \end{subfigure}
    \begin{subfigure}[t]{0.4\textwidth}
        \centering
        \includegraphics[width=\textwidth]{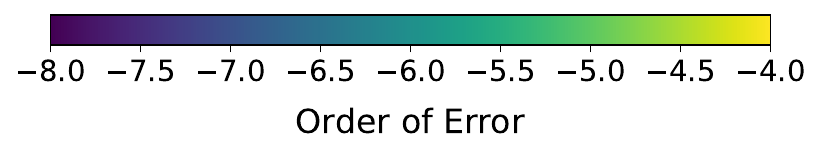}
    \end{subfigure}
    \caption{Plot of $log_{10}(|f_\theta(x_{gp})-s(x_{gp})|)$ for refinement $h=\nicefrac{\Delta}{2^8}$. The error mostly is in order of $10^{-4}$ for all the geometries. The plot shows the spatial variation of error in the magnitude of the distance vectors.}
    \label{fig:comparison_gp_distv}
\end{figure}

The analysis of the accuracy of representation is performed by obtaining Gauss points (two per axis) with the surrogate boundary for the level of refinement $h=\frac{\Delta}{2^8}$, where  $\Delta$ is the characteristic length of the bounding box for the INRs. For error computation, the signed distance value is obtained from the implicit representation, $f_\theta(x_{gp})$, and the actually signed distance $s(x_{gp})$ is obtained from libgl library. \figref{fig:comparison_gp_distv} visualizes $log_{10}(|f_\theta(x_{gp})-s(x_{gp}|)$ across different geometries. The accuracy of the direction of distance vector $\mathbf{d_{gp}}$ is computed by obtaining cosine similarly between the distance vector obtained from \algoref{Algorithm: DistanceFunctionCalculationUsingImplicitNetworkOneGP}, $d_{gp}^{f_\theta}$ and the true distance-vector $d_{gp}^{true}$. \figref{fig:comparison_gp_distv} visualizes  $log_{10}(1 - <d_{gp}^{true}\cdot d_{gp}^{f_\theta})>)$ across different geometries. Analysis of the two figures reveals that regions exhibiting significant changes in curvature are particularly prone to error. This is consistent with areas where large deviations in the magnitude of the distance function are observed. In these regions, a corresponding increase in the error of the cosine similarity metric is also evident. This behavior is expected due to the underlying eikonal constraint, which governs the relationship between the gradient of the distance function and the surface geometry. The eikonal equation imposes that the gradient of the distance function maintains a unit norm, and deviations from this condition in regions of high curvature can lead to both larger distance magnitude errors and higher discrepancies in the cosine similarity. Therefore, the correlation between these errors is a natural consequence of the mathematical properties imposed by the eikonal constraint.

\tabref{tab:complexity_objects} presents a comparison of Normalised Mean Square as defined earlier. The maximum mean squared error is of the order $10^{-6} << h$. Similarly mean cosine similarity of the distance vector with (Standard Deviation) is presented for each geometry. Essentially, the INR gives fairly accurate distance vectors for both complex and simple geometries to proceed further with analysis.

\begin{table}
    \centering
    \caption{Comparison of Normalized Mean Squared error for the signed distance and Cosine Similarity of Distance Vectors for gauss points at the surrogate boundary.}
    \label{tab:complexity_objects}
    \begin{tabular}{|c|c|c|c|}
        \hline
        Complexity & Object & NMSE$_{GP}$ & Mean  Cosine Similarity of Distance Vector$_{GP}$ (S.D) \\
        \hline
        Simple Shape & Sphere & $3.75 \times 10^{-8}$ & 1(0.00044) \\
        & Cone & $2 \times 10^{-7}$ & 0.996(0.045) \\
        
        & Cylinder & $5.6 \times 10^{-7}$ & 0.999(0.025) \\
        \hline
        Complex Shape & Bunny & $9.75 \times 10^{-7}$ & 0.995(0.014) \\
        & Tetrakis & $7 \times 10^{-7}$ & 0.997(0.015)  \\
        & Turbine & $3.84 \times 10^{-6}$ & 0.98(0.13) \\
        \hline
    \end{tabular}
    \label{complexity_comparison}
\end{table}

\begin{figure}[t!]
    \centering
    \begin{subfigure}[t]{0.2\linewidth}
        \centering
        \includegraphics[width=\linewidth, trim=1.5in 1.5in 1.5in 1.5in, clip]{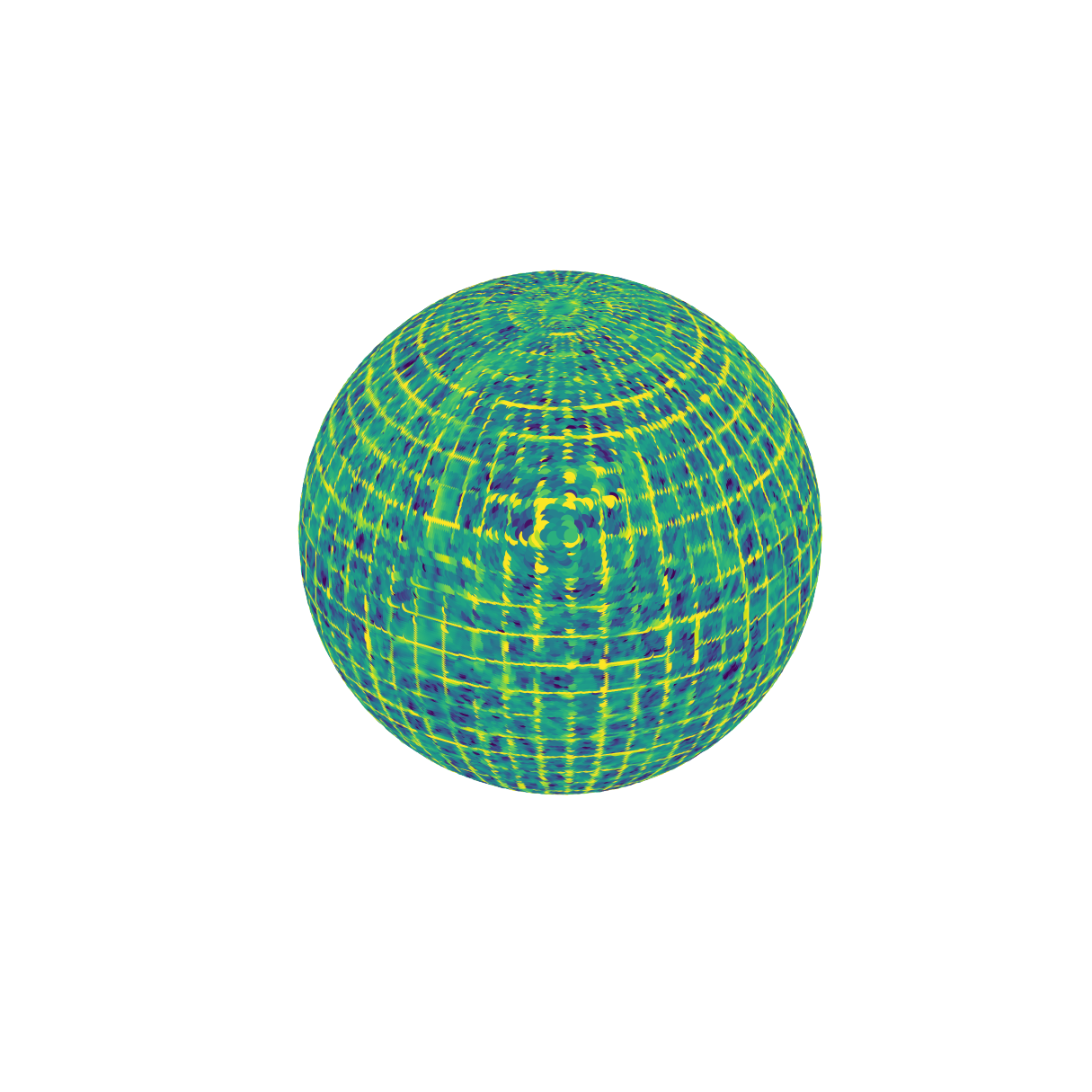}
        \caption{Sphere}
    \end{subfigure}
    \begin{subfigure}[t]{0.2\linewidth}
        \centering
        \includegraphics[width=\linewidth, trim=1.5in 1.5in 1.5in 1.5in, clip]{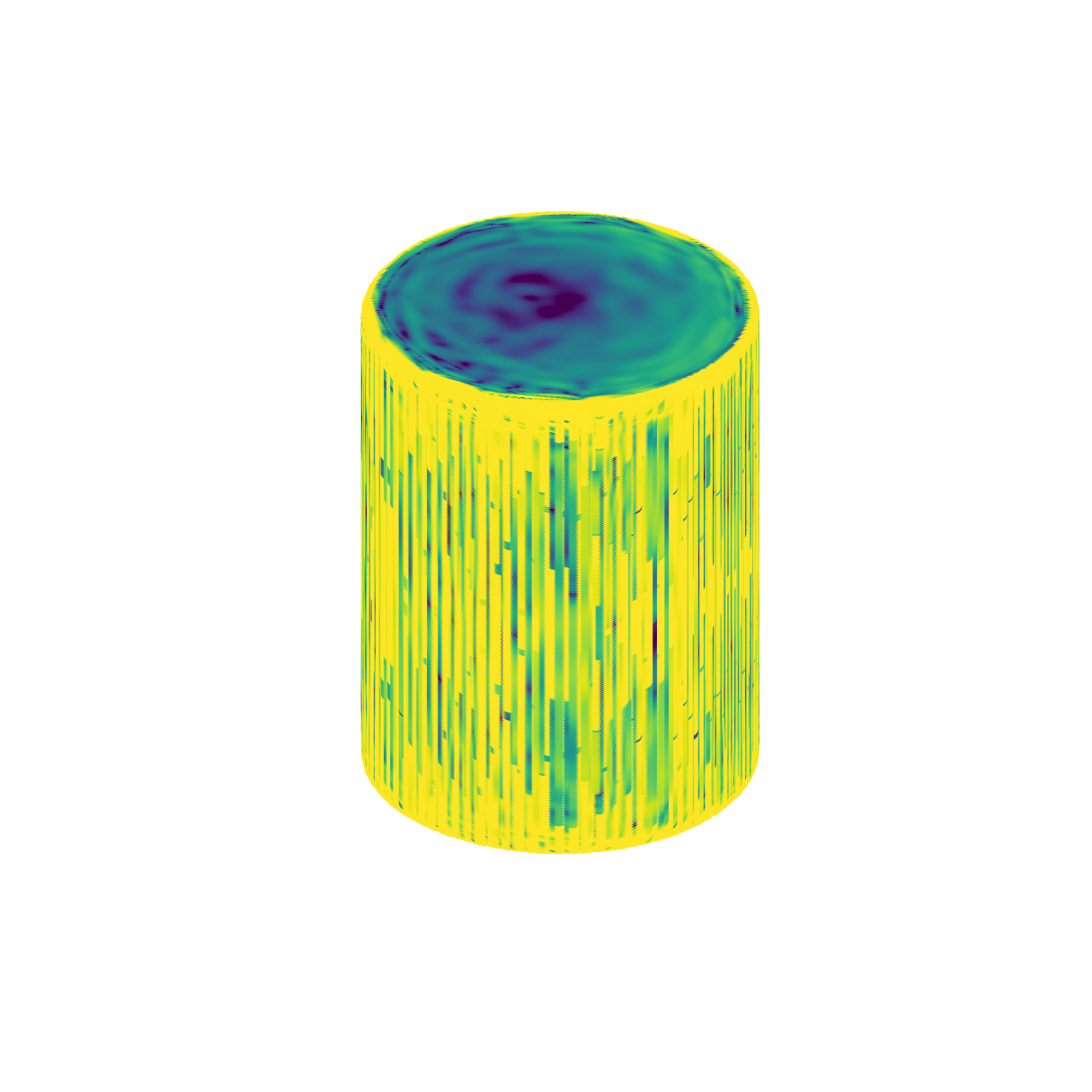}
        \caption{Cylinder}
    \end{subfigure}
    \begin{subfigure}[t]{0.2\linewidth}
        \centering
        \includegraphics[width=\linewidth, trim=2.3in 2.8in 2.3in 1.0in, clip]{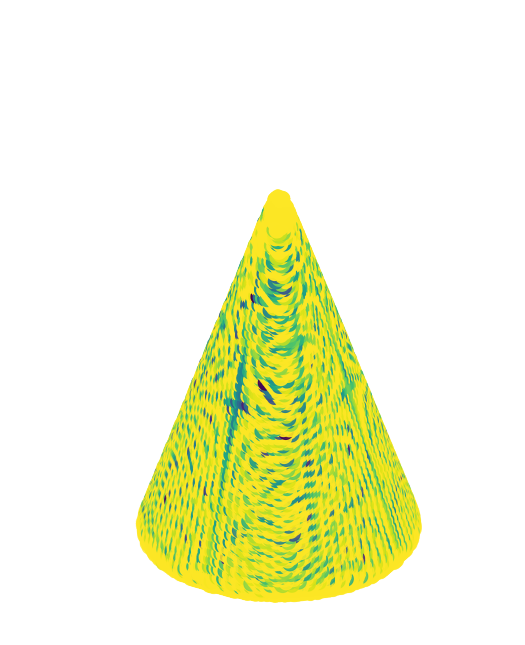}
        \caption{Cone}
    \end{subfigure}
    \begin{subfigure}[t]{0.2\linewidth}
        \centering
        \includegraphics[width=\linewidth, trim=1.5in 2.0in 1.5in 1.5in, clip]{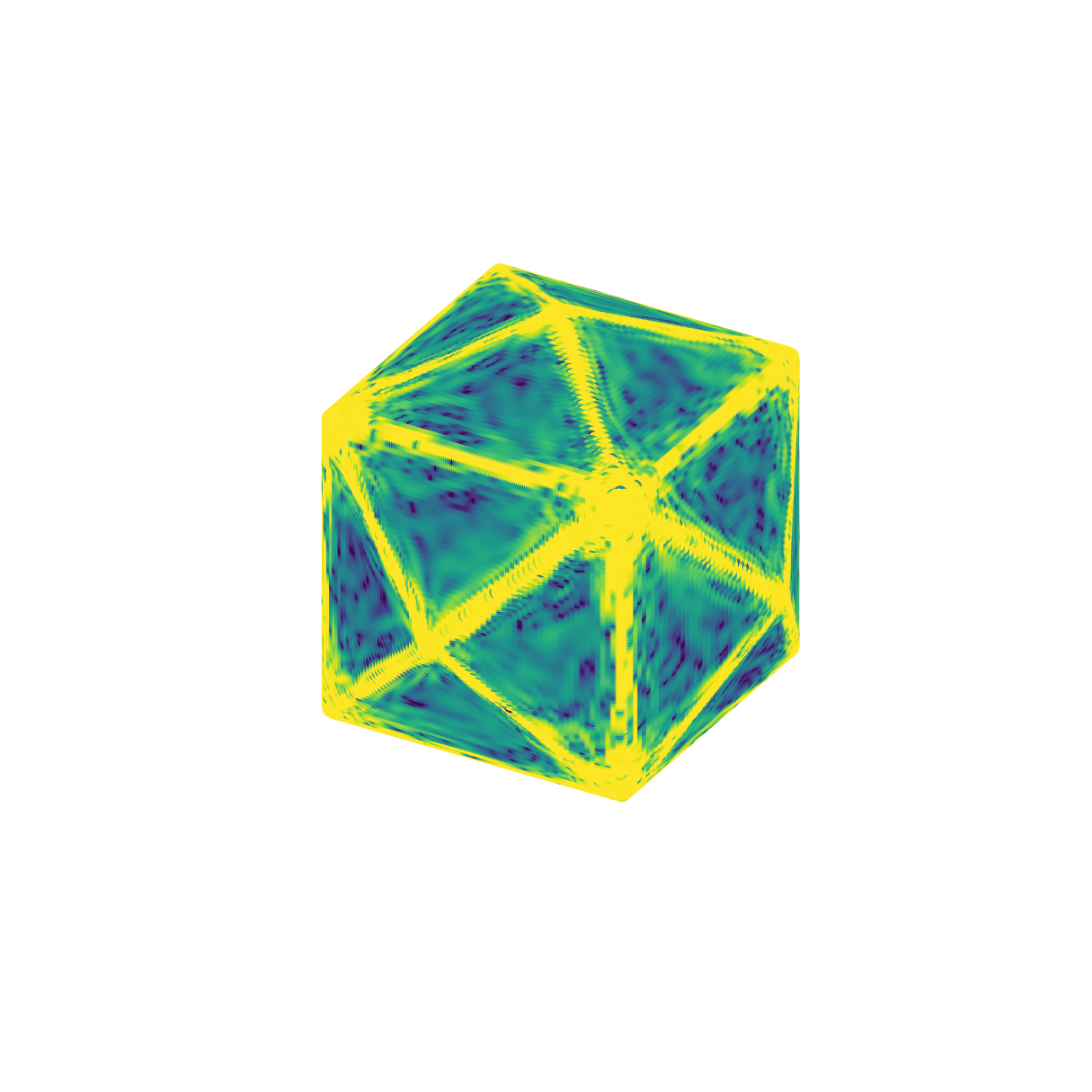}
        \caption{Tetrakis}
    \end{subfigure}
    \begin{subfigure}[t]{0.2\linewidth}
        \centering
        \includegraphics[width=\linewidth, trim=1.5in 2.0in 1.5in 1.5in, clip]{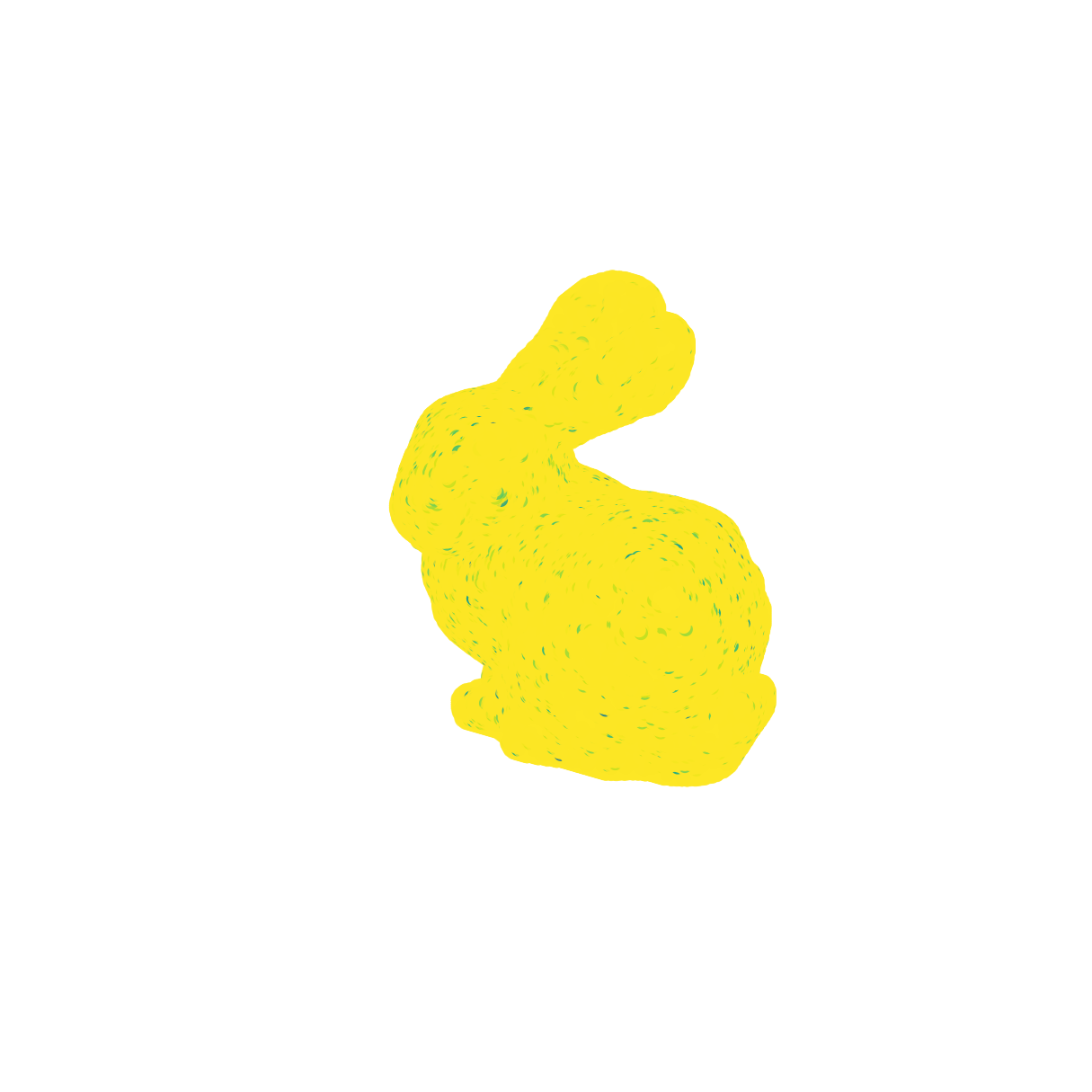}
        \caption{Bunny}
    \end{subfigure}
    \begin{subfigure}[t]{0.2\linewidth}
        \centering
        \includegraphics[width=\linewidth, trim=2.0in 2.3in 2.0in 2.0in, clip]{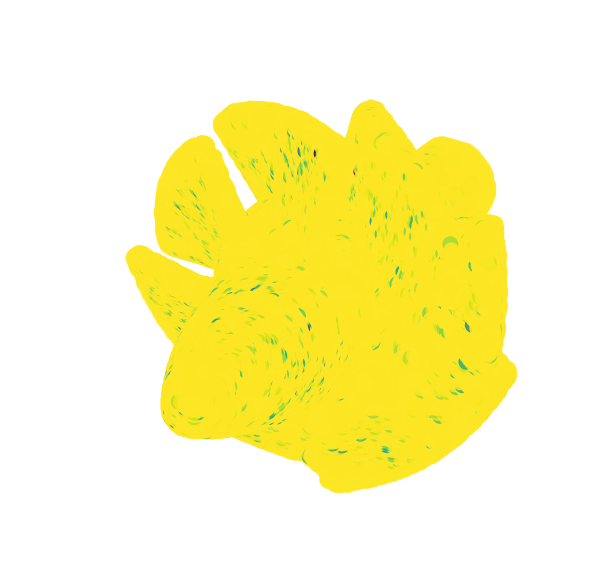}
        \caption{Turbine}
    \end{subfigure}
    \begin{subfigure}[t]{0.4\linewidth}
        \centering
        \includegraphics[width=\linewidth]{Figures_Compressed/geom_colorbar.pdf}
    \end{subfigure}
    \caption{Plot of $log_{10}(1 - \langle d_{gp}^{\text{true}} \cdot d_{gp}^{f_\theta} \rangle)$ for refinement $h=\nicefrac{\Delta}{2^8}$. The error shows the magnitude of the misalignment of the distance vector. As is more evident in tetrakis, the error is very high in the edges where there is a sharp change in curvature. Overall the error is still less than $10^{-4}$.}
    \label{fig:comparison_gp_cs}
\end{figure}

\clearpage
\section{Accuracy of Implicit Neural Representation to Explicit Representation Conversion}
\label{appendix:implicit_explicit_conversion}
Marching Cube is one of the most popular ways to convert INRs to a triangular soup, i.e., Explicit Representation. A triangular soup is completely defined by the vertices of the triangles. If the triangles were completely at the zero-level set of INR, the query of vertices through INRs would give a relatively small value (almost zero). We take the INR of three shapes (airplane, bunny, turbine), and compute the root-mean-squared error by performing Marching Cube at a resolution of 256 cubes. When INR is the only geometric representation available, the implicit to explicit conversion has an overhead of error in geometric representation.
\begin{table}[t!]
\centering
\caption{RMSE of Signed Distance Value of the vertices of triangular soup obtained by performing marching cube at a resolution of 256 cube.}
\begin{tabular}{lc}
\toprule
\textbf{Shape} & \textbf{RMSE} \\
\midrule
Airplane & 0.0391 \\
Bunny    & 0.0128 \\
Turbine  & 0.0060 \\
\bottomrule
\end{tabular}

\label{tab:rmse_shapes}
\end{table}
\section{Identification of Surrogate Boundary for Shifted Boundary Method}
\label{section:surrogate}

Once the elements are marked as TrueIntercepted, FalseIntercepted, Interior, or Exterior, nodal information is also required to obtain the surrogate boundary.
First, the NeighborsFalseIntercepted (neighbor of false intercepted element) elements are identified through the ghostwrite and ghost-read processes by setting the nodal values of FalseIntercepted elements to 1. After marking the NeighborsFalseIntercepted elements, the surrogate boundary is optimally constructed based on two scenarios as described in \algoref{alg:boundary}:
\begin{enumerate} 
    \item If an element is marked as Intercepted, the face(s) of the element with all nodes marked as Exterior is added to the surrogate boundary. 
    \item If an element is marked as NeighborsFalseIntercepted, we examine the faces of the element. Any face where all nodes are marked as FalseIntercepted or Exterior is included in the surrogate boundary. 
\end{enumerate} 
For more details about the algorithm interested reader are referred to \citet{yang2024optimal}.

\begin{algorithm}[h!]
\footnotesize
\caption{\textsc{GetBoundary:} Get surrogate boundary}
\label{alg:boundary}
\begin{algorithmic}[1]
\Require Octree $\mathcal{O}$, Element marker $\mathcal{M}$
\Ensure Surrogate boundary $\Tilde{\Gamma}$, Updated marker $\mathcal{M}$
\item[]
\For{each element $\in \mathcal{O}$}
    \State FaceBits $\leftarrow$ [false] \Comment{Initialize FaceBits as an array of false values}
    \For{each face $\in$ element} \Comment{Loop over the faces of the element}
        \If{$\mathcal{M}$[element] == Intercepted and \texttt{allof}(nodes $\in$ face == In)} \Comment{Condition for Intercepted element}
            \State FaceBits[face] $\leftarrow$ True
        \ElsIf{$\mathcal{M}$[element] == Neighbors} \Comment{Condition for Neighbor elements}
            \State BoundaryFace $\leftarrow$ True
            \For{each node $\in$ face}
                \If{(node == In) or (node == FalseInterceptedNode)} \Comment{Node is either In or FalseIntercepted}
                    \State \textbf{continue}
                \Else
                    \State BoundaryFace $\leftarrow$ False
                    \State \textbf{break}
                \EndIf
            \EndFor
            \If{BoundaryFace}
                \State FaceBits[face] $\leftarrow$ True
            \EndIf
        \EndIf
    \EndFor
    \If{\texttt{cycle}(FaceBits)} \Comment{Check if opposite faces of the element form part of $\Tilde{\Gamma}$}
        \State $\mathcal{M}$[element] $\leftarrow$ FalseIntercepted
    \Else
        \For{each face $\in$ element} \Comment{Add faces to the surrogate boundary $\Tilde{\Gamma}$}
            \If{FaceBits[face] == True}
                \State $\Tilde{\Gamma} \leftarrow \Tilde{\Gamma}.\texttt{push\_back}$(face)
            \EndIf
        \EndFor
    \EndIf
\EndFor
\item[]
\Return ($\Tilde{\Gamma}$, $\mathcal{M}$)
\end{algorithmic}
\end{algorithm}

\section{Preprocessing Generated INR of plane}
\label{section:preprocessinggenai}
\begin{figure}[tp]
    \centering
    \includegraphics[width=0.4\linewidth,trim=20 40 20 100, clip]{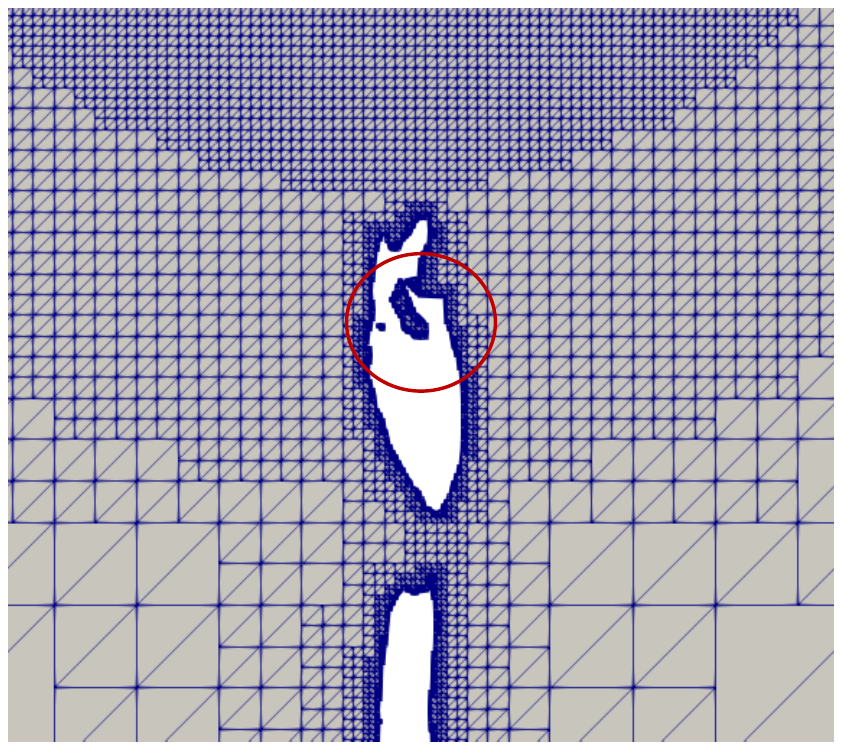}
    \caption{Unconnected component during carving of generative model.}
    \label{fig:diffusion_unwanted}
\end{figure}

\begin{figure}[t!]
    \centering
    \begin{subfigure}[t]{0.45\textwidth}
        \centering
        \includegraphics[width=0.95\textwidth, trim=10 20 10 20, clip]{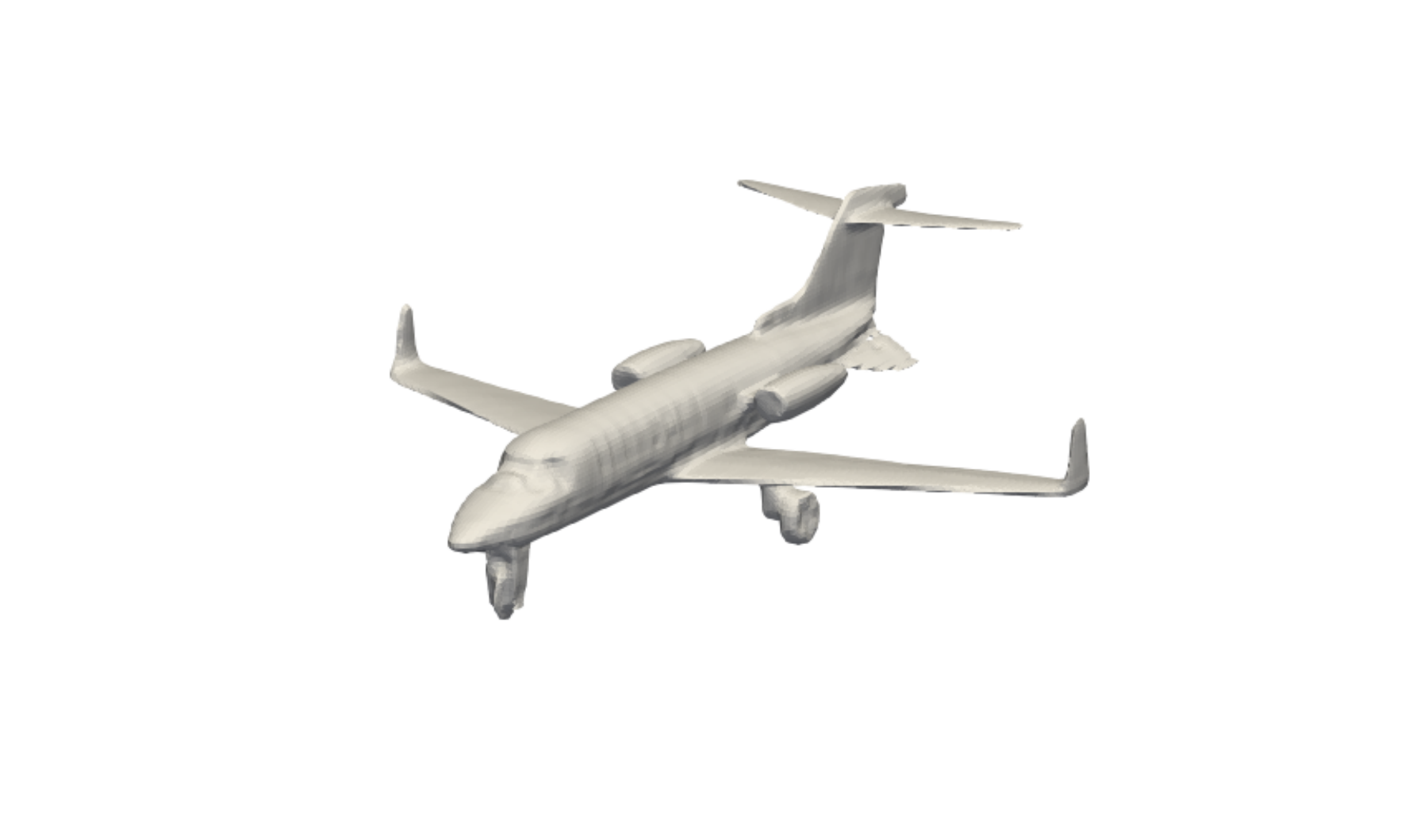}
        \caption{First Design}
    \end{subfigure}
    \hfill
    \begin{subfigure}[t]{0.45\textwidth}
        \centering
        \includegraphics[width=0.95\textwidth, trim=15 25 15 25, clip]{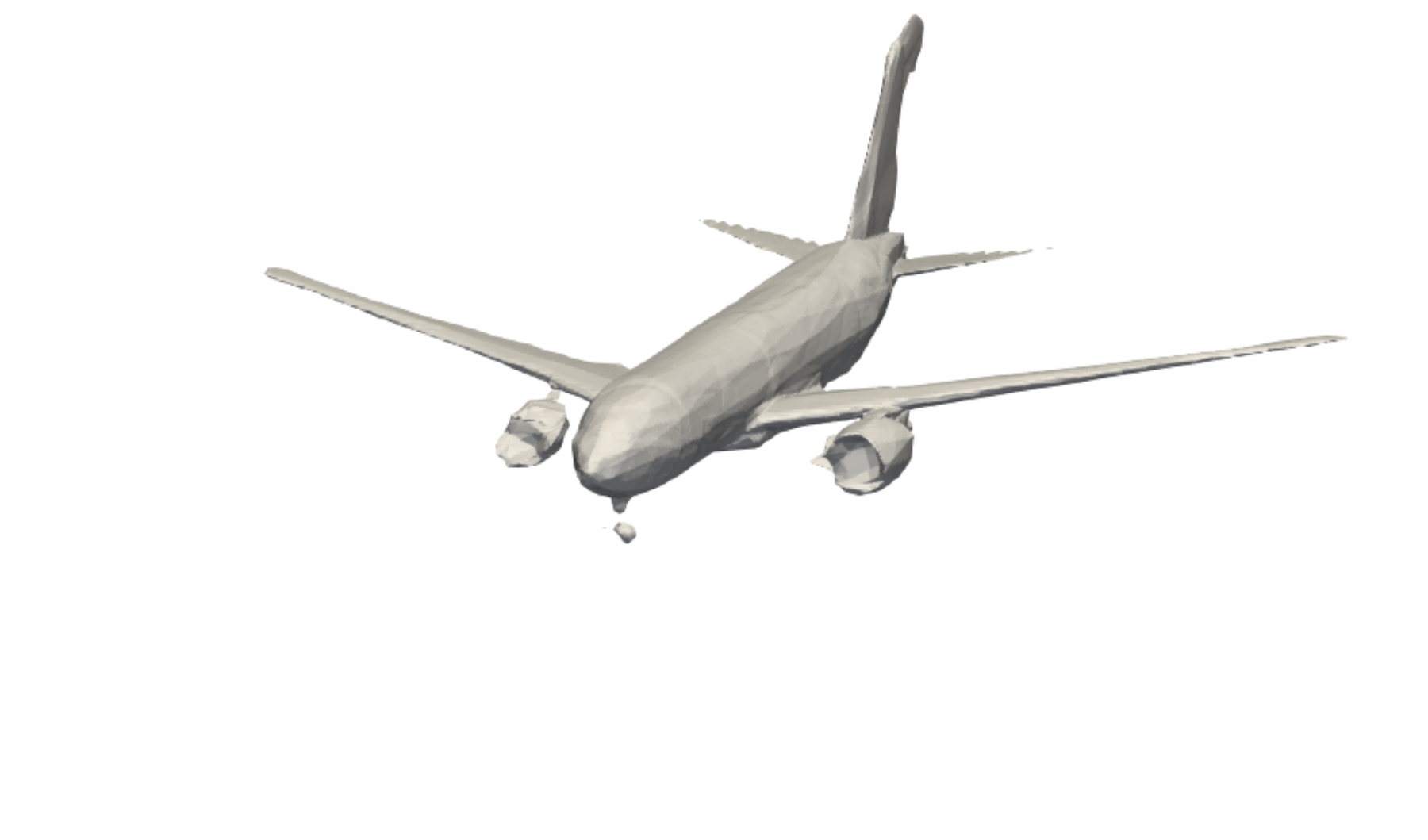}
        \caption{Second Design}
    \end{subfigure}
    
    \vspace{10pt}
    
    \begin{subfigure}[t]{0.45\textwidth}
        \centering
        \includegraphics[width=0.95\textwidth, trim=10 15 10 15, clip]{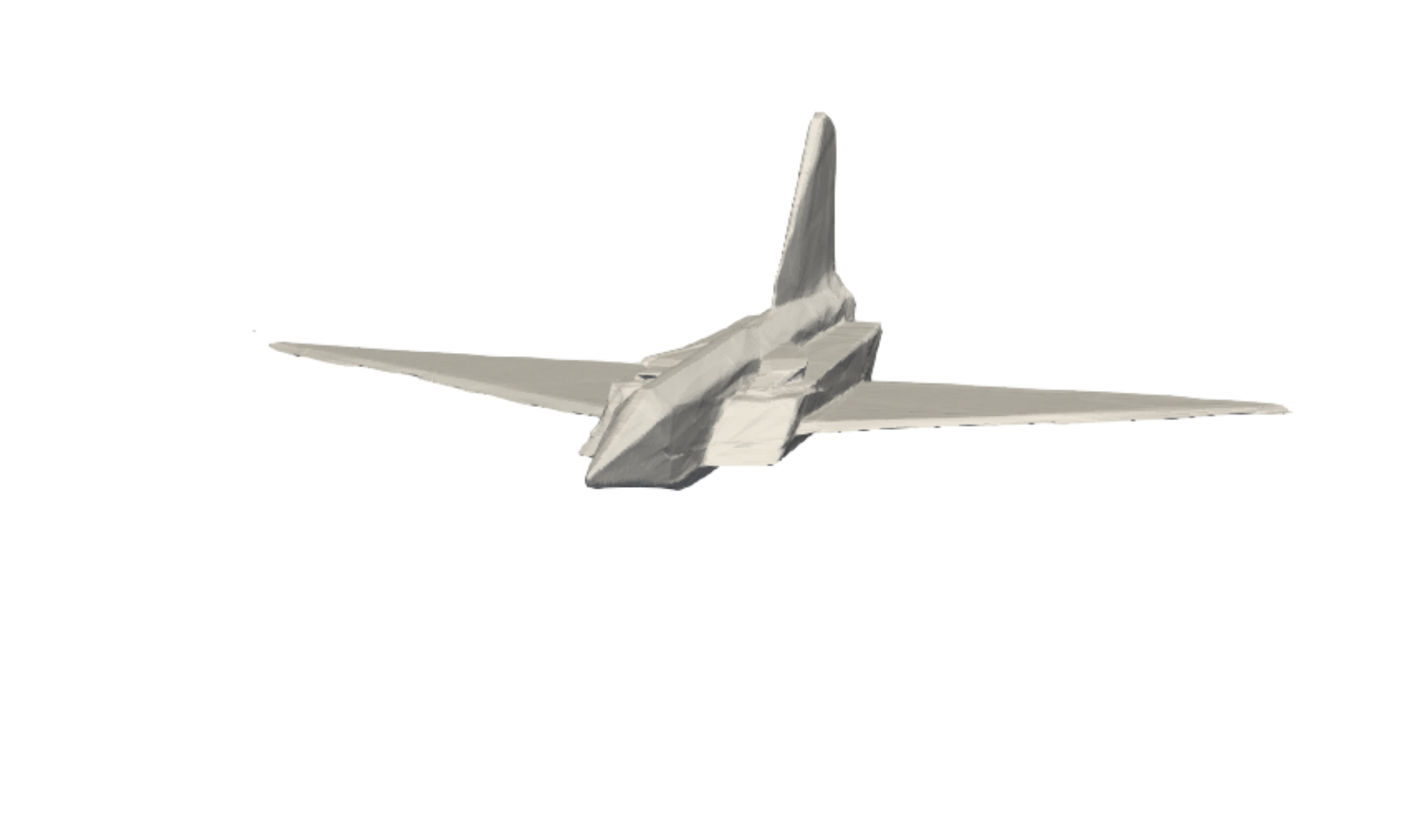}
        \caption{Third Design}
    \end{subfigure}
    \hfill
    \begin{subfigure}[t]{0.45\textwidth}
        \centering
        \includegraphics[width=0.95\textwidth, trim=20 30 20 30, clip]{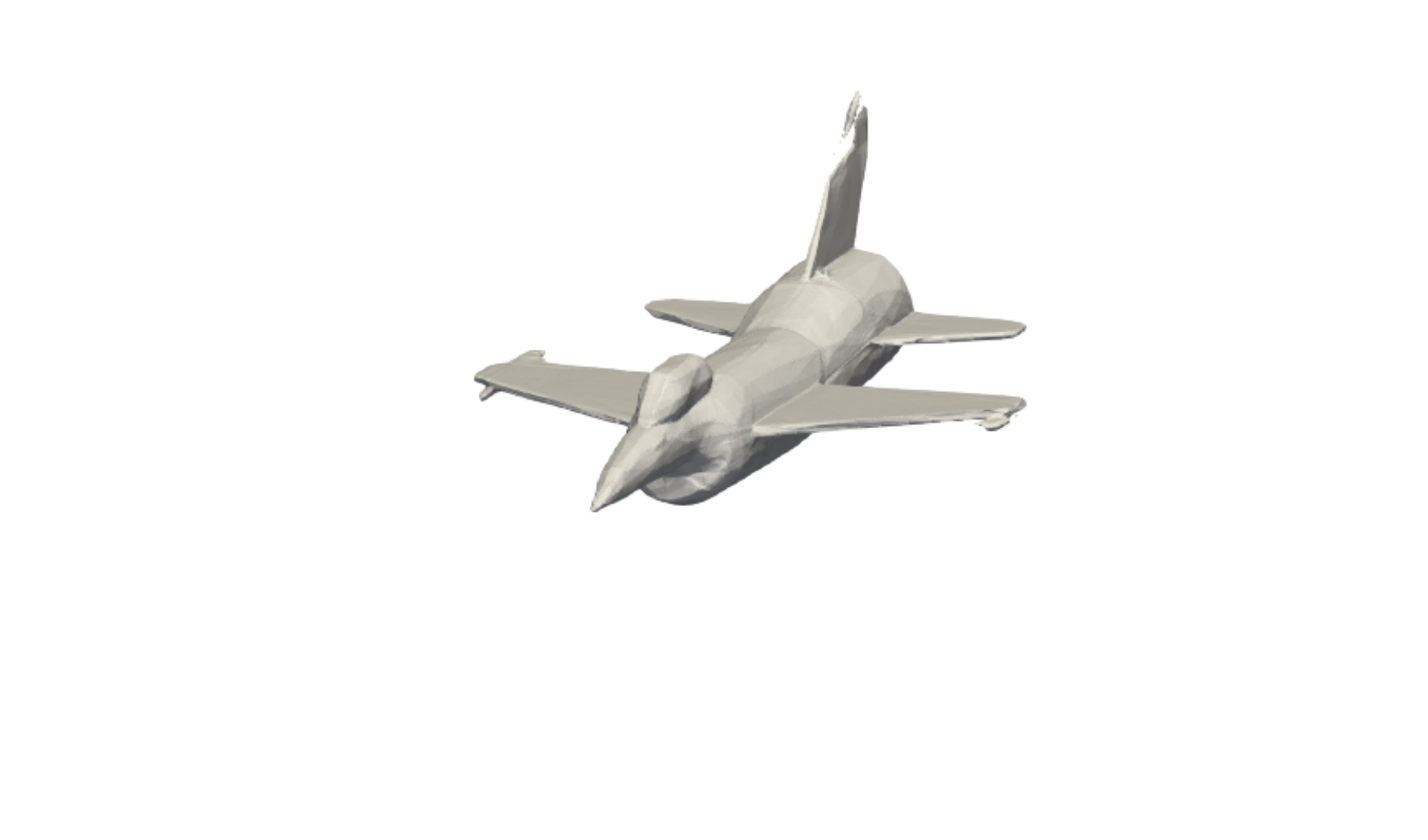}
        \caption{Fourth Design}
    \end{subfigure}
    
    \caption{Generated Plane from hyperdiffusion model \citep{erkocc2023hyperdiffusion}}
    \label{fig:generative_designs}
\end{figure}

\figref{fig:generative_designs} shows the generated INRs obtained from \citet{erkocc2023hyperdiffusion}. INR obtained directly from the generative model when used to generate carved-out regions for flow analysis contained disconnected components in unnecessary regions, as presented in \figref{fig:diffusion_unwanted}. To get rid of this as a preprocessing step, the obtained polygonal mesh obtained by performing marching cube over the Implicit Representation is used to obtain Implicit Representation using method described in \appendixref{appendix:Neural Implicit}.



\end{document}